\documentclass[aps,prl,superscriptaddress,twocolumn,showpacs,longbibliography,amsmath,amssymb,nofootinbib]{revtex4-2}
\usepackage[utf8]{inputenc}
\usepackage[T1]{fontenc}
\usepackage{microtype}
\usepackage{graphicx}
\usepackage{mathbbol}
\usepackage{amssymb}
\usepackage{comment}
\usepackage[usenames,dvipsnames,svgnames,table]{xcolor}
\usepackage[colorlinks,linkcolor=blue,citecolor=blue,urlcolor=blue]{hyperref}
\usepackage{siunitx}
\usepackage{symbol-defs}
\usepackage{placeins}
\usepackage[caption=false]{subfig}
\newcommand{\phantomsubfloat}[1]{
    {
        \captionsetup[subfigure]{labelformat=empty}
        \subfloat[][]{#1}
    }
}

\def\equationautorefname~#1\null{Eq.~(#1)\null}

\makeatletter
\newcommand{\supplementtitleformat}{\frontmatter@title@format}
\newcommand{\supplementauthorformat}{\normalfont\normalsize}
\newcommand{\supplementaffiliationformat}{\normalfont\frontmatter@affiliationfont}
\makeatother

\begin{document}
	\title{Chiral thermal fluctuations and enhanced refrigeration in a nonreciprocal nanomechanical system}
	\author{Jesse J. Slim}
	\thanks{J. J. S. and J. d. P. contributed equally to this work.}
    \affiliation{Center for Nanophotonics, AMOLF, Science Park 104, 1098 XG Amsterdam, The Netherlands}
    \affiliation{Australian Research Council Centre of Excellence in Quantum Biotechnology (QUBIC), School of Mathematics and Physics, The University of Queensland, St Lucia, Queensland 4072, Australia}
	\author{Javier del Pino}
	\thanks{J. J. S. and J. d. P. contributed equally to this work.}
	\affiliation{Departamento de Física Teórica de la Materia Condensada and Condensed Matter Physics Center (IFIMAC), Universidad Autónoma de Madrid, E-28049 Madrid, Spain}
	\affiliation{Center for Nanophotonics, AMOLF, Science Park 104, 1098 XG Amsterdam, The Netherlands}
    \author{Sander A. Mann}
    \affiliation{Institute of Physics, University of Amsterdam, Amsterdam, The Netherlands}
	\author{Ewold Verhagen}
	\email{verhagen@amolf.nl}
	\affiliation{Center for Nanophotonics, AMOLF, Science Park 104, 1098 XG Amsterdam, The Netherlands}

\begin{abstract}
Understanding how the breaking of reciprocity influences thermal flows and microscopic thermodynamic processes such as refrigeration and energy conversion is of broad current interest. 
We report experimental measurements of thermal flows and refrigeration in nanomechanical resonator networks in which nonreciprocity is controlled through an optically-induced synthetic magnetic flux. 
Time-modulated optomechanical interactions allow controlled coupling, gauge fields, and refrigeration processes through nanomechanical frequency conversion.
We quantify nonequilibrium heat flows between resonators coupled to dissipative baths of different occupation and image heat and effective temperature in networks through measuring correlations of fluctuations.
Synthetic magnetism is shown to imprint chirality on thermal fluctuations in a loop of resonators, leading to pronounced chiral flows with different handedness in distinct frequency bands.
We find that the heat flows in a non-equilibrium system are tuned by the synthetic magnetic flux, which redistributes energy in the thermal steady-state. 
Specifically, we illustrate how nonreciprocity enhances the refrigeration of a resonator in the strong coupling regime, reducing its temperature below the bound that applies to time-reversal symmetric networks. 
These results experimentally demonstrate the impact of nonreciprocity on thermodynamic machines, and provide new methods to characterize them at the microscopic level.

\end{abstract}

	\maketitle
	\DeclareGraphicsExtensions{.pdf,.png,.jpg}
Controlling heat flow at the nanoscale is central to testing microscopic thermodynamics, the study of fluctuations in low-dimensional, well-controlled open systems, both in classical stochastic thermodynamics~\cite{Seifert2012stochastic, Ciliberto2017experiments} and its quantum generalization~\cite{Parrondo2015thermodynamics,Rivas2020}.
Key directions include nonequilibrium fluctuations \cite{Rademacher2022nonequilibrium}, transport of heat and other fluctuation currents~\cite{Bermudez2013controlling,Huber2019active}, heat-to-work conversion in nanoscale engines~\cite{Benenti2017fundamental,Serra-Garcia2016mechanical} and efficient refrigeration or cooling schemes~\cite{Buddhiraju2020photonic}. 
The role of time-reversal symmetry (TRS) in microscopic thermodynamics has gained growing attention.
TRS breaking can produce various manifestations of \emph{nonreciprocal} dynamics. Understanding how nonreciprocity shapes thermodynamic behavior --- and, particularly, how it impacts the performance of microscopic machines including heat engines and refrigerators --- remains an open challenge.
Nonreciprocity enables directional energy transport \cite{Giazotto2012josephson,Zhu2014nearcomplete,Ege2025exploring,Biehs2023breakdown,Fernandez-Alcazar2021extreme}, leading to thermal diodes with thermal rectification~\cite{Seif2018thermal} and currents in topological insulators~\cite{Rivas2017topological,Silveirinha2017topological}. 
When effective magnetic fields break reciprocity, persistent circulating heat or probability currents have been predicted, allowing chiral flows even in equilibrium~\cite{Zhu2016persistent,Denis2020permanent,Biehs2023nonreciprocal,Biehs2025onpersistent}, e.g. in magneto-optic photonic systems~\cite{Zhu2016persistent} and optomechanical lattices~\cite{Denis2020permanent}. 

\begin{figure*}[ht]
		\includegraphics[width=\linewidth]{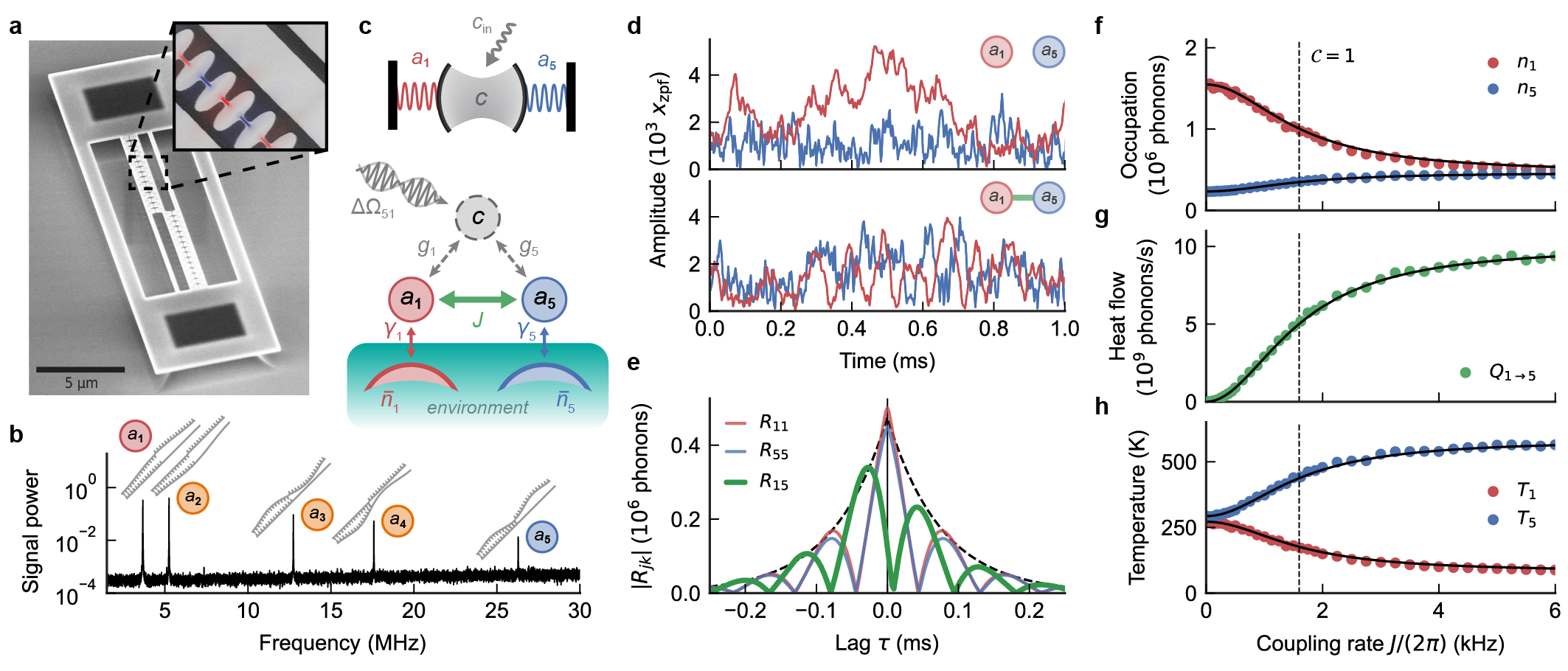}
		\caption{\textbf{Imaging heat flows in a nano-optomechanical cavity.} 
        \subidc{a} Experimental setup consisting of a sliced nanobeam nanocavity coupled to multiple mechanical modes.
        \subidc{b} Its thermomechanical spectrum reveals multiple optomechanically active modes (`resonators') that span a synthetic dimension. Flexural mode profiles of the first five modes are shown. 
        \subidc{c} Resonators are coupled to a room-temperature bath at rates $\gamma_j$ and can be optically coupled via time-modulated radiation pressure. This couples resonators with different frequency $\Omega_j$ and corresponding bath occupancy $\nres{j}$. Experiments in this figure concern a `hot' resonator $a_1$ with frequency $\Omega_1/(2\pi)=3.66$~MHz, decay rate $\gamma_1/(2\pi) = 1.5$~kHz, and `cold' resonator $a_5$ at $\Omega_5/(2\pi)=26.22$~MHz and $\gamma_5/(2\pi) = 6.9$~kHz. 
        \subidc{d} Thermal Brownian motion time traces for uncoupled and coupled resonators $a_{1,5}$ ($J/(2\pi)=6$ kHz) (top and bottom respectively). 
        \subidc{e} Absolute value of temporal auto- and cross-correlation functions involving the hot and cold resonators in the steady-state. Correlations are enveloped by the average resonator amplitude decay rate $\bar\gamma/2=(\gamma_1+\gamma_5)/4$ (dashed).
        \subidc{f} Phononic occupation of the resonators for increasing coupling rate, 
        \subidc{g} thermal heat flow $Q_{1\rightarrow5}$ to the cold resonator, and
        \subidc{h} effective temperature in steady state. Dashed lines in (f-h) represent the coupling rate corresponding to unit mechanical cooperativity $\mathcal{C} = 4J^2/(\gamma_1\gamma_5) = 1$ in the system. 
        } \label{fig:1}

	\end{figure*}

In systems with nonreciprocal interactions, i.e., non-Hermitian systems that microscopically violate Newton's third law, entropy production and cold-to-hot heat transfer have been predicted~\cite{Loos2020irreversibility,Zhang2023entropy,Loos2023nonreciprocal}. 
Lorentz nonreciprocity could also enable violating Kirchhoff's radiation law, impacting thermal radiation, radiative cooling and energy harvesting~\cite{Yang2024nonreciprocal, Shayegan2023direct}. 
A widely debated open question in stochastic thermodynamics is whether breaking TRS can boost heat-engine efficiency at finite power~\cite{Benenti2011thermodynamic, Brandner2013strong, Luo2020onsager, Saryal2022bounds}.
Conversely, nonreciprocity has been proposed to improve cooling and refrigeration mechanisms~\cite{Lai2020nonreciprocal,Messina2021radiative,Wang2022superior,Wang2024enhancement,Chakraborty2024quantum}.
While the role of nonreciprocity in thermodynamics has been explored theoretically, experimental investigations remain limited.
Parametric couplings in time-modulated systems offer useful testbeds~\cite{Roushan2017chiral}, and cavity optomechanical systems provide the control and measurement needed to study heat transport and thermodynamic phenomena in well-defined microscopic systems~\cite{Xuereb2015heat,Barzanjeh2018manipulating,Fong2019phonon,Yang2020phonon,Barzanjeh2022optomechanics,Sheng2023nonequilibrium,Reisenbauer2024nonhermitian}.
An inspiring experiment showed that breaking reciprocity can affect laser sideband cooling of two mechanical modes~\cite{Xu2019nonreciprocal}. 
Laser-mediated couplings between multiple mechanical modes can be induced through temporal modulation of the optical spring, enabling synthetic magnetic fluxes that break reciprocity and induce chiral transport in nanomechanical resonator networks~\cite{Mathew2020synthetic,MercierDeLepinay2020nonreciprocal,DelPino2022nonhermitian,Slim2025programmable}.
This provides a powerful platform for studying thermal flows in nonreciprocal systems.
In this work, we harness cavity-mediated phonon transport and artificial magnetic gauge fields to control thermal transport in a multimode nanomechanical loop of three modes with distinct frequencies and bath thermal occupations, and demonstrate how nonreciprocity introduces chiral heat flows and enhances refrigeration performance. 
We quantify and image cavity-mediated heat transfer in networks where optical coupling between detuned resonators with unequal bath occupations establishes heat pumps. 
In a loop threaded by a synthetic magnetic flux~\cite{Roushan2017chiral, DelPino2022nonhermitian}, we observe chiral circulation of thermal fluctuations against the occupation gradient.
We reveal that these consist of spectrally separated chiral-flow contributions with opposite handedness, which persist in equilibrium.
In the strong coupling regime, the flux biases heat pumping and suppresses dark hybrid modes~\cite{Lai2020nonreciprocal}, a known bottleneck in multimode sideband cooling~\cite{Ockeloen-Korppi2019sideband,Huang2022multimode}, thereby boosting nanomechanical refrigeration, cooling a resonator further than allowed by time-reversal-symmetric bounds. 
These findings show how synthetic magnetism can steer nanoscale heat flows and enable experimental studies of improved thermodynamic processes in low-dimensional systems. 

\section{Harnessing Nanomechanical Thermal Fluctuations via Radiation Pressure}

We study thermal motion in a room-temperature multimode nanomechanical system. This sliced nanobeam optomechanical resonator, shown in Fig.~\ref{fig:1}(a), hosts multiple MHz-frequency mechanical flexural modes, each parametrically coupled to a single photonic crystal nanocavity~\cite{DelPino2022nonhermitian}.
Lasers driving this cavity at different detunings allow both measurement of thermomechanical motion and control of intermode mechanical couplings~\cite{Mathew2020synthetic}.
Hereafter, we refer to the undriven normal mechanical modes as `resonators', to distinguish them from the eigenmodes of coupled networks. 
The thermally driven displacement fluctuations $x_j(t)$ of mechanical resonator $j$ with  ($j,k=\{1,2,\cdots,N\}$) are imprinted linearly on the reflected intensity signal $z(t)$ of a laser detuned by multiple cavity linewidths.  
The power spectrum of $z(t)$, shown in Fig.~\ref{fig:1}(b), reveals five high-$Q$ mechanical resonators with distinct MHz resonance frequencies $\Omega_j$ and comparable kHz damping rates $\gamma_j$.
A second laser, tuned to the cavity-resonance slope, generates an optical spring effect on all resonators, which is time-modulated through electro-optic modulation of the laser intensity. 
Parametric modulation of the cross-mode optical spring effect at any of the frequency splittings $\Delta\Omega_{jk}=\Omega_j-\Omega_k$ couples resonators $j$ and $k$ (Fig.~\ref{fig:1}(c)), yielding the interaction Hamiltonian $H_{jk}=J_{jk}\cos(\Delta\Omega_{jk}t+\phi_{jk})x_jx_k$ with coupling amplitudes $J_{jk}$ and phases $\phi_{jk}$ proportional to the electro-optic modulation depth and phase, respectively \cite{Mathew2020synthetic}. 
Each resonator experiences stochastic thermal forces $F_j(t)$ from its local dissipative environment at temperature $T\approx300$~K, with thermal occupation $\bar{n}^{\mathrm{th}}_j = \kb T / (\hbar \Omega_j)$ for $\kb T\gg\hbar\Omega_j$ (Fig.~\ref{fig:1}(c) bottom).
We analyse the resonators' time-dependent motion by comparing oscillations in $z(t)$ with fixed electronic local oscillators (LOs) at the mechanical resonance frequencies $\Omega_j$. 
Specifically, we demodulate $z(t)$ for all $N$ resonators in parallel at frequencies $\Omega_j$. 
After low-pass filtering, this yields complex envelopes $a_j=(x_j+i p_j/(m_j\Omega_j))e^{i\Omega_j t}/(2x_{\mathrm{zpf},j})$, which describe the slowly-evolving amplitude and phase of each resonator in a frame rotating at its natural frequency.
Since the resonators have high quality factors $\Omega_j/\gamma_j\sim10^3$ and well-separated frequencies, their vibrations are individually resolvable.
In the rotating-wave approximation, valid for slow coupling and damping compared with the mechanical oscillation and intermode splittings, the dynamics are described by~\cite{Mathew2020synthetic},
\begin{equation}
\dot a_j=-\frac{\gamma_j}{2}a_j+i\sum_{k\neq j}J_{jk}e^{-i\phi_{jk}}a_k+\xi_j^a .
\label{ctd:eq:resonator-a-langevin}
\end{equation}
Here $\xi_j^a$ denotes the thermal noise in the rotating frame (Methods). We model the baths as Markovian and independent, with $\langle \xi_j^a(t)\xi_j^{a*}(t-t^\prime)\rangle=\bar n_j^{\mathrm{th}}\gamma_j\delta(t-t^\prime)$~\cite{Aranas2017quantum} and $\langle \xi_j^a(t)\xi_{k\neq j}^{a*}(t-t^\prime)\rangle=0$.
Under these assumptions, each bath occupation is extracted experimentally from the variance of the corresponding uncoupled resonator $(J_{jk}=0)$ via $\langle x_j^2\rangle=2x_{\mathrm{zpf},j}^2\bar n_j^{\mathrm{th}}$~\cite{supmat}.
The coherent part of Eq.~\eqref{ctd:eq:resonator-a-langevin} is equivalently described by $H\approx \sum_{j,k}J_{jk}e^{-i\phi_{jk}}a_j^\ast a_k+\mathrm{H.c.}$ with a $U(1)$ gauge freedom, so only the gauge-invariant fluxes around closed loops, $\triangle$, namely $\Phi=\sum_{j,k\in\triangle}\phi_{jk}$, are physically relevant~\cite{Koch2010timereversalsymmetry}.

\section{Measuring heat flow through thermal correlations}

We study the network's thermal state and energy flows between resonators through statistical analysis of their measured Brownian motion. 
We will first consider a network of only two resonators: Figure~\ref{fig:1}(d) shows representative traces of fluctuating amplitudes $|a_j(t)|$ for two uncoupled resonators ($J_{15}=0$, top) and coupled resonators ($J_{15}\neq 0$, bottom).
To quantify fluctuation decay and exchange, we evaluate time-resolved stationary-state correlation functions
\begin{align}
R_{jk}(\tau) = \left\langle a_j(t + \tau) a_k^\ast(t) \right\rangle_{t\rightarrow\infty}\,, \label{ctd:eq:crosscorr-def}
\end{align}
where $\tau=t'-t$ is the time lag. 
For uncoupled resonators, auto-correlations $R_{jj}(\tau)$ decay at rate $\gamma_j/2$ while cross-correlations $R_{15}(\tau)\approx0$ are negligible, as expected for independently fluctuating resonators coupled to separate thermal baths. 
For a strongly coupled dimer, $2J_{15}\gg \{\gamma_1,\gamma_5$\}, the auto-correlations exhibit dips and revivals within an overall decay envelope $\widebar{\gamma}/2=(\gamma_1+\gamma_5)/4$ (Fig.~\ref{fig:1}(e)), matching the decay of the hybrid modes~\cite{supmat}.
The large cross-correlations at these revivals reveal Rabi-like exchange of thermal fluctuations between the resonators, previously observed only for equal-frequency resonators driven by artificial thermal fluctuations~\cite{Yang2020phonon}. 
Moreover, $R_{15}(\tau)$ is asymmetric in time lag and finite at $\tau=0$. This asymmetry arises from the occupation gradient $\nres1>\nres5$, which drives the coupled system out of equilibrium.
By contrast, in equilibrium all baths would have equal thermal occupation, and the correlators would obey the fluctuation-dissipation relation and detailed balance. 
Because the modulated spring couples two resonators with different bath occupations, it drives a \textit{net heat flow} from the low- to the high-frequency resonator.
From the continuity equation for stationary mode occupations~\cite{Rivas2017topological,supmat}, this flow is determined by the off-diagonal stationary covariance matrix elements, i.e., by the zero-lag cross-correlation through
\begin{equation}
    Q_{j\mapsto k}=2\mathrm{Im}\left[J_{jk}e^{i\phi_{jk}}R_{jk}(0)\right].
\label{ctd:eq:heat-flow-def}
\end{equation}
The heat flow directly modifies the resonator occupations. 
Figure~\ref{fig:1}(f,g) shows the measured thermal occupations of the strongly coupled $a_1$--$a_5$ dimer versus coupling strength, together with the measured heat flow $Q_{1\mapsto5}$. 
Occupations and heat flow are related by $n_{1,5}=\langle |a_{1,5}|^2\rangle=\nuth_{1,5}\mp Q_{1\mapsto5}/\gamma_{1,5}$. 
The data agree well with the steady-state solution of Eq.~\eqref{ctd:eq:resonator-a-langevin}~\cite{supmat}. In particular,
\begin{equation}
Q_{1\mapsto5}
=
\frac{\mathcal{C}}{1+\mathcal{C}}\frac{\gamma_1\gamma_5\,\Delta \nuth}{\gamma_1+\gamma_5},
\end{equation}
where $\Delta \nuth=\nuth_1-\nuth_5$ is the bath occupation imbalance and $\mathcal{C}=4J_{15}^2/(\gamma_1\gamma_5)$ the cooperativity of the mechanical dimer. 
For $J=0$, the imbalance, $\Delta \nuth$, simply reflects that of the baths. 
As $J$ increases, $a_1$ is sympathetically cooled while $a_5$ heats, consistent with a positive current $Q_{1\mapsto5}>0$ and heat-pump behaviour. 
Stronger coupling accelerates thermal fluctuation exchange, so excitations flow more efficiently from the more populated low-frequency mode $a_1$ to the less populated high-frequency mode $a_5$, increasing the heat current.
Meanwhile, local baths inject and remove excitations; since $\gamma_5>\gamma_1$, resonator $a_5$ evacuates incoming heat more efficiently.
The trend saturates once inter-resonator exchange $2J_{15} \gg \gamma_1,\gamma_5$ outpaces dissipation, in which case the occupations approach 
$n^{\rm max}_{1,5}=(\gamma_1 \nres1+\gamma_5 \nres5)/(\gamma_1+\gamma_5)$. 
Expressed as effective temperatures, $T_j=\nres{j}\hbar\Omega_j/\kb$, the heat-pump action is clear in Fig.~\ref{fig:1}(h): increasing $J$ drives $T_5$ above the laboratory temperature $T_{\rm lab}\approx295$~K while lowering $T_1$ below it, saturating at aforementioned cooling limit in the strong-coupling regime.
The refrigeration work is supplied by the modulated drive laser, analogous to optomechanical sideband cooling. This proof-of-concept experiment shows how heat and temperature in a microscopic refrigerator can be controlled and quantified through optomechanical correlation measurements.

\begin{figure}
	\includegraphics[width=\linewidth]{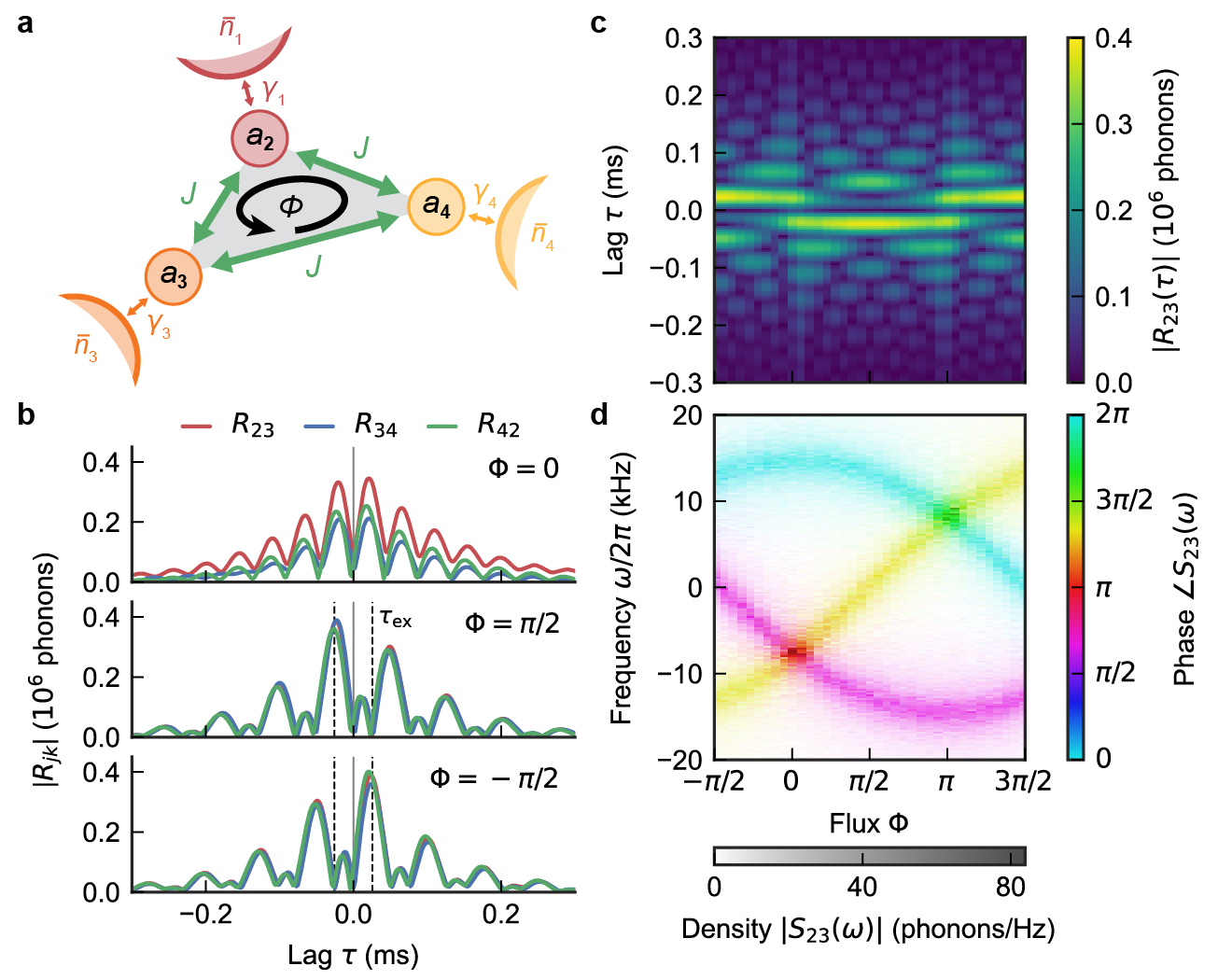}
	\phantomsubfloat{\label{ctd:fig:loop-fluctuations:loop-diagram}} 
	\phantomsubfloat{\label{ctd:fig:loop-fluctuations:correlation-linecuts}} 
	\phantomsubfloat{\label{ctd:fig:loop-fluctuations:correlation-2d}} 
	\phantomsubfloat{\label{ctd:fig:loop-fluctuations:correlation-fft-2d}} 
    \vspace{-2\baselineskip}
	\caption{\textbf{Chiral thermal fluctuations in a nanomechanical loop.} \subidc{a} Three resonators $a_2, a_3, a_4$ form a loop (coupling strength $J/(2\pi)=7.5$ kHz) threaded by a synthetic flux $\Phi$, and interact with independent phonon baths with occupations 
    $\nres{2,3,4} = \{0.70, 0.64, 0.32\} \times 10^6$. 
    \subidc{b} Cross-correlations $R_{jk}(\tau)$  measured for $\Phi=0$ (time-reversal symmetric) and $\Phi = \pm \pi/2$ (broken TRS). For $\Phi = \pm\pi/2$, the intermode exchange time $\tau_\text{ex} = \pm 2\pi/(3J\sqrt{3})$ is indicated (dashed lines). 
    \subidc{c} Measured cross-correlation magnitude $|R_{23}(\tau,\Phi)|$ with varying flux $\Phi$.
    \subidc{d} Magnitude and phase of the cross-spectral density $S_{23}(\omega,\Phi)$, obtained by Fourier transform of $R_{23}(\tau,\Phi)$. In the rotationally symmetric gauge (see text) $S_{34}(\omega)$ and $S_{42}(\omega)$ look similar and exhibit the same relative phases.
	}
		\label{ctd:fig:loop-fluctuations}

\end{figure}

\begin{figure*}
    \centering
    \includegraphics[width=\linewidth]{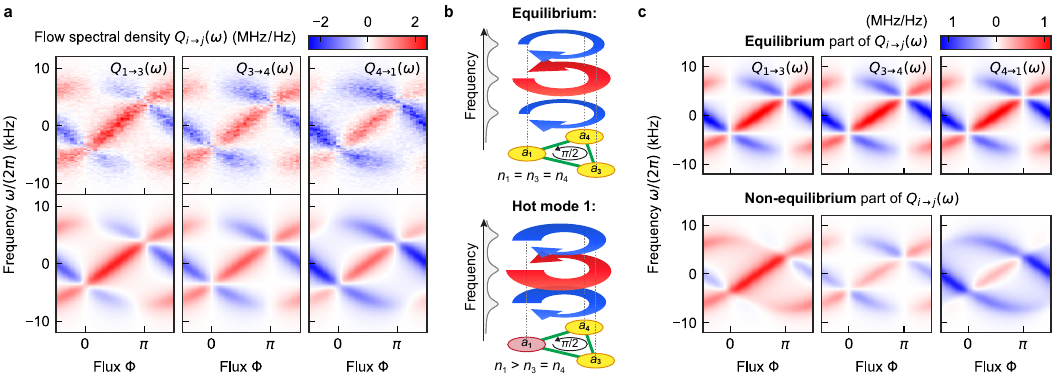}
    \phantomsubfloat{\label{ctd:fig:circulator-flows:2d-flows}}
    \phantomsubfloat{\label{ctd:fig:circulator-flows:chiral-diagram}}
    \phantomsubfloat{\label{ctd:fig:circulator-flows:eq-noneq}}
    \vspace{-2\baselineskip}
    \caption{\textbf{Spectrally resolved chiral heat flows.} 
    \subidc{a} Flux $\Phi$ tunes the spectral densities $Q_{i\to j}(\omega)$ of the flows between resonators $a_1, a_3, a_4$ (bath occupations $\nuth_{1,3,4} = \{1.51, 0.45, 0.36\} \times 10^6$) connected in a loop. All three coupling strengths are $J/(2\pi) = 3.6$ kHz, comparable to the optothermally matched decay rates $\gamma_3/(2\pi)\approx\gamma_4/(2\pi)=3.8$ kHz, while $\gamma_1/(2\pi)=1.5$ kHz. Experiment (top) is compared to theory (bottom). 
    \subidc{b} At TRS-breaking $\Phi = \pi/2$,  the three hybridized modes at different frequencies carry chiral currents with opposite handedness. In equilibrium (top), these spectrally resolved currents integrate to zero net flow. Out of equilibrium, with resonator 1 hotter than the other two, the same chiral flows carry heat from mode 1 through the network (bottom).
    \subidc{c} The total flow spectral density is decomposed into an equilibrium part (top) for equal $\nres{j} = \bar{n} =  0.4 \times 10^6$ and a non-equilibrium part that is driven by the occupancy difference $\nres{1} - \bar{n} = 1.1 \times 10^6$ between the hot mode bath and the two colder baths, here set to equal $\nres{3}=\nres{4}=\bar{n}$ for simplicity. Comparable plots for experimental $\nres{j}$ (Extended Data Fig.~\ref{fig:ed1:net_flows}) appear similar.
	}
	\label{ctd:fig:circulator-flows}

\end{figure*}

\section{Chiral Thermal Fluctuations via Synthetic Magnetic Fields}

We now consider a loop of resonators $a_2$, $a_3$, and $a_4$ with equal couplings $J_{jk}=J$, each coupled to its own thermal bath (Fig.~\ref{ctd:fig:loop-fluctuations:loop-diagram}).
The gauge-invariant flux $\Phi=\phi_{23}+\phi_{34}+\phi_{42}$ controls the Hamiltonian’s TRS, and we choose a gauge that distributes the flux equally over the three links. 
Using cross-correlations between adjacent resonators, we probe fluctuation dynamics versus time lag $\tau$ and flux $\Phi$ at strong coupling, $J/(2\pi)=7.5$~kHz $>\gamma_j/(2\pi)$ (Fig.~\ref{ctd:fig:loop-fluctuations:correlation-linecuts}).
For $\Phi=0$, $|R_{jk}(\tau)|$ is nearly symmetric in $\tau$, with only a weak asymmetry from the thermal gradient, $\nres2 \gtrsim \nres3 > \nres4$.
In contrast, for $\Phi=\pm\pi/2$, it becomes strongly asymmetric, evidencing TRS breaking with a handedness set by the magnetic flux.
At $\Phi=\pi/2$, for instance, $R_{23}(\tau)$ peaks at multiples of $\tau_\text{ex}=2\pi/(3J\sqrt{3})$, associated with excitation hopping in that direction, while transfer in the opposite direction is minimal, noting $R_{kj}(\tau)=R_{jk}^*(-\tau)$ (dashed lines in Fig.~\ref{ctd:fig:loop-fluctuations:correlation-linecuts}).
Reversing $\Phi$ reverses the circulation chirality, which is continuously tunable with flux (Fig.~\ref{ctd:fig:loop-fluctuations:correlation-2d}).
To identify the modes underlying this chiral dynamics, we Fourier transform $R_{jk}(\tau)$ to obtain the complex cross-spectral density $S_{jk}(\omega)=\mathcal{F}[R_{jk}(\tau)]$, which captures fluctuation frequencies and relative phases.
Its three bands in Fig.~\ref{ctd:fig:loop-fluctuations:correlation-fft-2d} follow the flux-dependent eigenfrequencies of the angular-momentum modes $\tilde{a}_l=\sum_{j=2}^{4}e^{-i2\pi lj/3}a_j/\sqrt{3}$, with $l={-1,0,1}$, while color encodes the phase difference between resonators $a_2$ and $a_3$ (see Methods).
The light-blue band is the symmetric in-phase mode, whereas the other two are phase-chiral, with inter-resonator phase lags $\pm2\pi/3$. At the TRS-preserving fluxes $\Phi_0={0,\pi}$, two modes become degenerate and can be combined into real modes with trivial phase delays.
Accordingly, we see that then $\arg(S_{23})\in\{0,\pi\}$ for all frequencies after a global phase transformation, whereas this is never true for broken TRS.
Thus, thermal cross-correlation phases reveal gauge-independent chirality.
Through Eq.~\ref{ctd:eq:heat-flow-def}, cross-correlations quantify heat flows $Q_{j\rightarrow k}$ in the network. The effect of magnetic flux on thermal flows becomes especially clear when resolving their spectral content. We define the \emph{flow spectral density}
\begin{equation}\label{eq:sp_flow_dens}
Q_{j\rightarrow k}(\omega)=2\mathrm{Im}[J_{jk} e^{i \phi_{jk}}S_{jk}(\omega)],
\end{equation}
whose frequency integral gives the total heat flow $Q_{j \rightarrow k}$. 
Figure~\ref{ctd:fig:circulator-flows:2d-flows} (top) shows the measured flow spectral density in all links of another loop comprising the `hottest' resonator $a_1$, and colder $a_3$ and $a_4$.
We observe that a magnetic field creates circulating currents of opposite handedness in different loop eigenmodes: for $\Phi=\pi/2$, the middle-frequency mode carries clockwise flow $1\rightarrow3\rightarrow4\rightarrow1$, whereas the higher- and lower-frequency modes carry counterclockwise flow $1\rightarrow4\rightarrow3\rightarrow1$. For $\Phi=-\pi/2$, all flows reverse. 
The measured flow spectral density agrees well with theory (Fig.~\ref{ctd:fig:circulator-flows:2d-flows}, bottom). 
Notably, the observed chiral flows are not primarily driven by thermal gradients, and persist even in equilibrium (Fig.~\ref{ctd:fig:circulator-flows:chiral-diagram}). 
As the system is linear, we can separate the heat flow into the equilibrium and non-equilibrium contributions. 
Figure~\ref{ctd:fig:circulator-flows:eq-noneq} decomposes the flow spectral densities for a loop with one hot resonator of bath occupation $\nres{1}$ and two colder resonators with $\nres{3}=\nres{4}=\bar{n}<\nres{1}$, close to the conditions in Fig.~\ref{ctd:fig:circulator-flows:2d-flows} (see Extended Data Fig.~\ref{fig:ed1} for the exact experimental case. Note that optothermal backaction changes the effective bath occupations compared to the experiment shown in Fig.~\ref{ctd:fig:loop-fluctuations}, such that here $\nres{3}\approx\nres{4}$). 
The equilibrium contribution (top) is calculated for three equal bath occupations $\bar{n}$, while the non-equilibrium contribution (bottom) has resonator 1 at bath occupation $\nres{1}-\bar{n}$ and the other two at zero. 
We recognize that the persistent spectrally resolved chiral currents observed in Fig.~\ref{ctd:fig:circulator-flows:2d-flows} are dominated by the equilibrium, counterpropagating flows of Fig.~\ref{ctd:fig:circulator-flows:eq-noneq} (top).
The equilibrium flows vanish only for $\Phi\in\{0,\pi\}$ and cancel to zero \emph{net} integrated heat flow for all $\Phi$~\cite{supmat}.
Additionally, the thermal gradient $\nres{1}-\bar{n}$ produces the non-equilibrium flow contribution in Fig.~\ref{ctd:fig:circulator-flows:eq-noneq} (bottom), carrying heat from resonator 1 to the colder baths of 3 and 4.
Crucially, when TRS is broken, these flows still respect the eigenmode chiralities: the middle eigenmode carries heat clockwise from resonator 1 through the loop $1\rightarrow3\rightarrow4$ for $\Phi=\pi/2$, while the top and bottom modes carry counterclockwise flows. In fact, the non-equilibrium flow spectral density can even be counter to the occupation gradient, such as the $4\rightarrow1$ flow for zero frequency. 
Thus, the synthetic flux turns chirality into a control knob for heat transport, inducing strong equilibrium spectral flows and allowing thermal currents to be spectrally routed.

\section{Nonreciprocity-enhanced refrigeration}

\begin{figure*}
    \centering
    \includegraphics{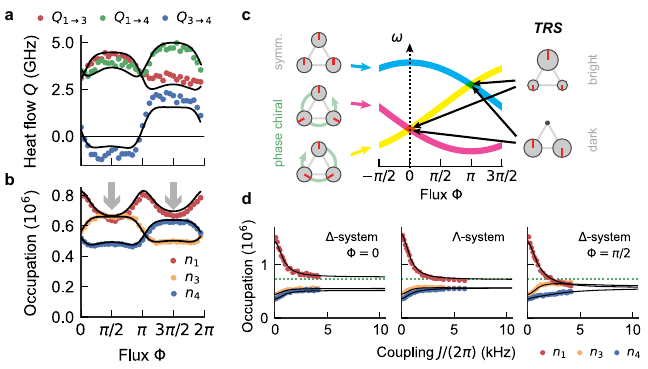}
    \phantomsubfloat{\label{ctd:fig:chiral-equilibration:lc-flows}}
    \phantomsubfloat{\label{ctd:fig:chiral-equilibration:lc-energies}}
    \phantomsubfloat{\label{ctd:fig:chiral-equilibration:dark-modes}}
    \phantomsubfloat{\label{ctd:fig:chiral-equilibration:J-sweep}}
    \vspace{-2\baselineskip}
    \caption{\textbf{Broken TRS enhances refrigeration.} 
    \subidc{a} Thermal heat flows $Q_{j\to k}$ and 
    \subidc{b} average resonator occupations $n_j$ as a function of flux $\Phi$, for the network presented in Fig.~\ref{ctd:fig:circulator-flows}.
    Black lines are calculated by integrating \eqref{eq:meth:spectral_density_matrix}. Grey arrows indicate the fluxes for which optimal refrigeration is reached. 
    \subidc{c} Spectral structure of a three-mode loop. Its three eigenmodes can always be written as flux-independent angular momentum states (left) that tune with $\Phi$ (colored bands, cf. Fig.~\ref{ctd:fig:loop-fluctuations:correlation-fft-2d}), two of which carry chiral phase advance. The chiral modes become degenerate at TRS-preserving $\Phi=0$, where they can be combined into bright and dark modes (right). The same holds at TRS-preserving flux $\Phi=\pi$, after gauge transformation.
    \subidc{d} Thermal equilibration in chiral and achiral systems. Occupations $n_j$ are shown for increasing coupling strength $J$ in the three-mode `$\Delta$' loop $a_1, a_3, a_4$ at $\Phi = 0$ (left) and $\Phi=\pi/2$ (right), and in a three-mode `$\Lambda$' network where the cold link $a_3-a_4$ is severed (middle). In both TRS-preserving cases (left, middle), the hot resonator effectively equilibrates with a \emph{single} cold resonator with average bath occupation $(\nres{3}+\nres{4})/2$ as $J\to\infty$, setting a limit on the maximum cooling (green dotted line). In the TRS-broken loop (right), the hot resonator fully equilibrates with \emph{both} cold resonators, allowing for cooling below the TRS limit.
    }
    \label{fig:chiral-equilibration}

\end{figure*}

Out of equilibrium, the spectrally resolved flows no longer cancel upon integration, yielding the net heat flows in Fig.~\ref{ctd:fig:chiral-equilibration:lc-flows}.
The magnetic field affects all flows and even reverses the net current on link $3\rightarrow4$, to run opposite to the thermal gradient~\cite{supmat}.
It is therefore natural to ask whether broken TRS impacts heating and cooling performance.
Indeed, the chiral spectral flows reported above suggest the possibility for improved refrigeration: we recall that in the two-resonator system, cooling was limited at strong coupling because fluctuations were exchanged between hot and cold resonators faster than they could dissipate into the cold bath.
The same limitation occurs in a TRS loop, where the hot resonator exchanges heat with \emph{either} cold resonator.
With broken TRS, however, chiral currents carry fluctuations along \emph{both} cold resonators, maximizing the possibility for dissipation.

Experimentally, Fig.~\ref{ctd:fig:chiral-equilibration:lc-flows} shows that the heat flow from $a_1$ to $a_3$ and $a_4$ peaks for $\Phi\in\{\pi/2,-\pi/2\}$.
This flow cools the hot resonator below its bath occupation, $n_1<\nres{1}$, while heating the colder resonators above theirs, $n_{3,4}>\nres{3,4}$ (Fig.~\ref{ctd:fig:chiral-equilibration:lc-energies}).
Indeed, the total heat flow out of the hot resonator is significantly enhanced at nontrivial fluxes, as shown in Extended Data Fig.~\ref{fig:ed1:net_flows}. Consequently, the cooled population $n_1$ is significantly smaller than for trivial fluxes $\Phi\in\{0,\pi\}$, as indicated by grey arrows in Fig.~\ref{ctd:fig:chiral-equilibration:lc-energies}. 
The observed enhancement of refrigeration through symmetry breaking, evidenced through the measured heat flows and resonator temperature, can be understood from the network's modal structure.
Fluctuations  from the hot bath are injected into the eigenmodes, which distribute them across all three sites. 
Optimal refrigeration imposes two conditions on the eigenmodes. Importantly, fluctuations from the hot resonator must excite all eigenmodes equally: any eigenmode with little or no weight on the hot resonator removes heat from it inefficiently.
Additionally, the eigenmodes must be maximally separated in frequency: if two modes are degenerate, their relative phase is conserved, allowing formation of `bright' and `dark' modes (see Fig.~\ref{ctd:fig:chiral-equilibration:dark-modes}).
Since the dark mode has no weight on the hot resonator, this reduces the number of modes available for cooling~\cite{Lai2020nonreciprocal}.
The eigenmode picture explains the refrigeration performance of three network configurations shown in Fig.~\ref{ctd:fig:chiral-equilibration:J-sweep}: the trivial $\Delta$-configuration, a TRS-preserving loop; the $\Lambda$-configuration, where the link between the cold resonators is removed; and the TRS-broken flux loop.
In the $\Delta$-configuration each eigenmode in principle has equal weight on the hot resonator, but the two chiral modes are degenerate and form a dark mode decoupled from it, preventing effective refrigeration and equilibration of populations $n_j$ even as $J \rightarrow \infty$ (Fig.~\ref{ctd:fig:chiral-equilibration:J-sweep}, left)~\cite{Lai2020nonreciprocal}.
The $\Lambda$-configuration lifts this degeneracy, with eigenfrequencies $(-\sqrt{2}J,0,\sqrt{2}J)$ maximally separated for a given $J$ ~\cite{supmat}. However, incomplete mixing persists as $J \rightarrow \infty$ (Fig.~\ref{ctd:fig:chiral-equilibration:J-sweep}, middle), because the eigenvectors no longer have equal weight on each resonator: the hot resonator contributes only to two modes while the third anti-symmetric mode has zero weight on it.
The TRS-broken flux loop (Fig.~\ref{ctd:fig:chiral-equilibration:J-sweep}, right), satisfies both requirements for optimal refrigeration: at $\Phi = \pm \pi/2$, the eigenfrequencies are maximally separated, as in the $\Lambda$ configuration, while the hot resonator contributes equally to all eigenmodes, with weight $1/3$. TRS breaking thus enables the strongest possible refrigeration for a given coupling strength $J$, by simultaneously lifting all degeneracies maximally \emph{and} equalizing each eigenmode's weight on the hot resonator.
These conditions cannot be met simultaneously in TRS-preserving three-resonator systems~\cite{supmat}. As a result, their occupation is never smaller than the general TRS cooling limit, shown as the green dotted line in Fig.~\ref{ctd:fig:chiral-equilibration:J-sweep}. In contrast, for the nonreciprocal loop at $\Phi=\pi/2$, the cooled occupation $n_1$ fully equilibrates with $n_3, n_4$ for large $J$ and asymptotes significantly below the TRS limit.

\section*{Outlook}

In conclusion, we have used tunable, light-mediated mechanical interactions in a multi-mode cavity optomechanical system to manipulate and image heat transport in situ, revealing how controlled time-reversal-symmetry breaking affects chiral heat currents and refrigeration performance. 
Cross-correlation analysis shows chirality in both amplitudes and phases of thermal fluctuations in a nonreciprocal system. Synthetic magnetism induces spectrally resolved chiral heat flows and steers net thermal flows out of equilibrium, enhancing refrigeration in specific multimode settings beyond the limits set by reciprocity.
The control and analysis methods developed here provide tools for studying nonequilibrium fluctuations~\cite{Horowitz2020thermodynamic}.
In particular, heat-flow spectral densities offer a powerful way to study heat currents in and out of equilibrium.
In the presence of dissipative couplings, these persistent currents would sum to net heat flows even in equilibrium~\cite{Zhu2016persistent,Denis2020permanent,Biehs2023nonreciprocal,Biehs2025onpersistent}.
While we demonstrated TRS-breaking-induced chirality and enhanced refrigeration in a specific network, it will be important to understand how these mechanisms generalize to different network scales and topologies.
Promising directions include heating and cooling in anti-parity-time symmetric systems~\cite{Jiang2020energylevelattraction} and heat transport in topological insulators~\cite{Rivas2017topological,Slim2025programmable}.
The nonreciprocal routing and efficient cooling we study may also aid the development of photonic or microwave components where suppressing thermal noise in selected output channels is important~\cite{Peterson2017demonstration, Bernier2017nonreciprocal}.
Since the non-degenerate circulator acts as a heat pump driven by modulated light, natural next steps would be quantifying the work supplied by the drive, the refrigeration efficiency, and entropy production. 
This perspective remains unexplored in light-mediated nonreciprocal phonon transport~\cite{Xu2019nonreciprocal, Seif2018thermal, Yang2020phonon} and connects to ongoing debates on finite-power efficiency bounds for heat engines with broken TRS~\cite{Benenti2011thermodynamic, Brandner2013strong, Luo2020onsager, Saryal2022bounds}.
Finally, similar principles could be applied to mechanical systems with quantum fluctuations~\cite{Barzanjeh2022optomechanics}, opening new opportunities to study classical and quantum thermodynamics in well-controlled microscopic systems with broken symmetries.

\clearpage

\section{Methods}

\subsection{Theory of engineered gauge fields in nano-optomechanical networks}

We model the device as a network of $N$ nanomechanical resonators with annihilation operators $c_j$ and bare frequencies $\omega_j$.
The resonators are coupled through an optical spring generated by an amplitude-modulated cavity field, which enables the programming of hopping interactions, synthetic gauge phases, and non-Hermitian nanomechanics~\cite{Mathew2020synthetic,DelPino2022nonhermitian,Slim2025programmable,Slim2024optomechanical}.
Linearizing the optomechanical interaction and eliminating the cavity in the large-detuning, large-bandwidth limit, with $\kappa\ll\omega_j$, gives
\begin{equation}
H_{\rm os}(t)=-\frac{\Delta\,\bar n(t)}{\Delta^2+(\kappa/2)^2}\left[\sum_j g_0^{(j)}(c_j+c_j^\dagger)\right]^2 ,
\end{equation}
where $\Delta$ and $\kappa$ are the cavity detuning and linewidth, $g_0^{(j)}$ is the single-photon coupling to resonator $j$, and $\bar n(t)$ is the intracavity photon number.
The optical intensity contains a static component, which shifts the resonator frequencies, and pairwise modulations, which activate the selected links,
\begin{equation}
\bar n(t)=\bar n_{\rm dc}+\sum_{j<k}\bar n_{jk}\cos(\Delta\Omega_{jk}t+\phi_{jk}) .
\end{equation}
Here $\Delta\Omega_{jk}=\Omega_j-\Omega_k$. The dc component gives the optical spring shift
\begin{equation}
\Omega_j=\omega_j-\frac{2g_0^{(j)2}\Delta\,\bar n_{\rm dc}}{\Delta^2+(\kappa/2)^2},
\end{equation}
where we used $(c_j+c_j^\dagger)^2\simeq 2c_j^\dagger c_j+1$ and omitted the constant shift. The modulated components generate phonon hopping on the selected links.
We next move to a frame rotating with each mechanical resonance, defining $a_j=e^{i\Omega_j t}c_j$. Since the modulation frequencies are chosen close to the mechanical splittings, $\Delta\Omega_{jk}=\Omega_j-\Omega_k$, the rotating-wave approximation keeps only the near-resonant hopping processes $c_j^\dagger c_k$ driven by the corresponding modulation tone. This yields the time-independent Hamiltonian
\begin{equation}
H=-\sum_{j<k}\left[J_{jk}e^{-i\phi_{jk}}a_j^\dagger a_k+{\rm H.c.}\right],
\end{equation}
valid for $J_{jk}\ll\Omega_j$, with
\begin{equation}
J_{jk}=\frac{g_0^{(j)}g_0^{(k)}\Delta\,\bar n_{jk}}{\Delta^2+(\kappa/2)^2}.
\end{equation}
Any overall sign of $J_{jk}$ can be absorbed into $\phi_{jk}$. The programmable phases $\phi_{jk}$ define synthetic fluxes through gauge-invariant sums around closed loops, while the Hamiltonian conserves total phonon number.
Dissipation and fluctuations are described by independent Markovian thermal baths, justified by the spectral separation of the mechanical resonators. In the rotating frame,
\begin{equation}
\dot{\mathbf a}(t)=-i\left(\boldsymbol{\mathcal A}-i\frac{\boldsymbol{\gamma}}{2}\right)\mathbf a(t)+\mathbf a_{\rm in}(t),
\end{equation}
where $\mathbf a=(a_1,\ldots,a_N)^T$, $\boldsymbol{\gamma}={\rm diag}(\gamma_1,\ldots,\gamma_N)$, and the coupling matrix is defined by $\mathcal A_{jk}=-J_{jk}e^{-i\phi_{jk}}$ for $j<k$, with $\mathcal A_{kj}=\mathcal A_{jk}^\ast$ and $\mathcal A_{jj}=0$. The vector $\mathbf a_{\rm in}$ collects the bath inputs, whose only non-zero correlations are
\begin{align}
\langle a_{\rm in}^{(j)}(t)a_{\rm in}^{(k)\dagger}(t')\rangle&=\gamma_j(\nres{j}+1)\delta_{jk}\delta(t-t'),\\
\langle a_{\rm in}^{(j)\dagger}(t)a_{\rm in}^{(k)}(t')\rangle&=\gamma_j\nres{j}\delta_{jk}\delta(t-t').
\end{align}
Here $\nres{j}=\bar n(\Omega_j)\simeq \kb T_j/(\hbar\Omega_j)$ in the high-temperature limit. The equivalent Lindblad master equation and cavity elimination are given in Supplementary Sections~I.A and~I.B.

\subsection{Correlation and spectral-density formalism}

We extract mode occupations, cross correlations and spectra from the linear Langevin dynamics introduced above. The central object is the steady-state two-time correlation matrix
\begin{equation}
\mathbf R(\tau)
=
\lim_{t\to\infty}
\left\langle
\mathbf a(t+\tau)\mathbf a^\dagger(t)
\right\rangle ,
\end{equation}
whose elements $R_{ij}(\tau)=\langle a_i(t+\tau)a_j^\dagger(t)\rangle$ contain both auto-correlations and cross correlations between mechanical modes. For $\tau>0$, linearity implies that these correlations regress with the same drift matrix as the mean amplitudes,
\begin{equation}
\frac{{\rm d}}{{\rm d}\tau}\mathbf R(\tau)
=
\boldsymbol{\mathcal M}\mathbf R(\tau),
\qquad
\boldsymbol{\mathcal M}
=
-i\boldsymbol{\mathcal A}
-\frac{\boldsymbol{\gamma}}{2}.
\end{equation}
This is the quantum regression theorem for the present linear open system, and reduces to the classical Onsager regression principle in the high-temperature limit.
The equal-time correlator $\mathbf R_0=\mathbf R(0)$ is fixed by a Lyapunov equation~\cite{Meystre2007elements}. Using
\begin{equation}
\left\langle
\mathbf a_{\rm in}(t)\mathbf a_{\rm in}^\dagger(t')
\right\rangle
=
\mathbf D\,\delta(t-t'),
\end{equation}
with $\mathbf D  = {\rm diag}\!\left[\gamma_j(\bar n_j^{\rm th}+1)
\right]$, one obtains
\begin{equation}
\boldsymbol{\mathcal M}\mathbf R_0
+
\mathbf R_0\boldsymbol{\mathcal M}^\dagger
+
\mathbf D
=
0 .
\end{equation}
This equation determines all steady-state equal-time correlations. The normally ordered correlator is
$\mathbf C_0=\langle \mathbf a^\dagger\mathbf a\rangle=(\mathbf R_0-\mathbb 1)^T$. In the classical high-temperature regime considered here, the vacuum contribution is negligible, so the measured occupation is $n_j\simeq R_{jj}(0)$.
The frequency-resolved correlations are defined from the Fourier transform of the two-time correlator,
\begin{equation}\label{eq:sp_density_aa}
\boldsymbol{\mathcal S}_{\mathbf a\mathbf a^\dagger}(\omega)
=
\int_{-\infty}^{\infty}
{\rm d}\tau\,
e^{-i\omega\tau}\mathbf R(\tau).
\end{equation}
Equivalently, solving the Langevin equation in frequency space gives
\begin{equation}
\mathbf a(\omega)
=
\boldsymbol{\chi}(\omega)\mathbf a_{\rm in}(\omega),
\qquad
\boldsymbol{\chi}(\omega)
=
\left(
-i\omega\mathbb 1
+i\boldsymbol{\mathcal A}
+\frac{\boldsymbol{\gamma}}{2}
\right)^{-1},
\end{equation}
and therefore
\begin{equation}
\boldsymbol{\mathcal S}_{\mathbf a\mathbf a^\dagger}(\omega)
=
\boldsymbol{\chi}(\omega)
\mathbf D
\boldsymbol{\chi}^\dagger(\omega).
\label{eq:meth:spectral_density_matrix}
\end{equation}
The diagonal elements give the displacement spectra of the individual modes, while the off-diagonal elements
\begin{equation}
S_{ij}(\omega)
=
\left[
\boldsymbol{\mathcal S}_{\mathbf a\mathbf a^\dagger}(\omega)
\right]_{ij}
\end{equation}
give the complex cross spectra. Their phases provide a frequency-resolved probe of the phase structure of the hybridized mechanical modes and of the synthetic gauge flux. Formal solution details are given in Supplementary Section~I.B.

\subsection{Heat currents and occupation continuity equations}

The heat-flow analysis follows from the local occupation dynamics of the number-conserving rotating-frame Hamiltonian. For each mode we define $n_i=a_i^\dagger a_i$. The exchange with the local reservoir is
\begin{equation}
Q_i^{\rm bath}
=
\gamma_i\left(\bar n_i^{\rm th}-\langle n_i\rangle\right),
\end{equation}
which is positive when the bath injects energy into mode $i$ and negative when mode $i$ releases energy to its bath.
For each ordered link, we use the phase convention of the measured zero-lag correlations,
\begin{equation}
R_{ij}(0)=\langle a_i a_j^\dagger\rangle=C_{ji}.
\end{equation}
The coherent heat current from mode $i$ to mode $j$ is then
\begin{equation}\label{eq:heat_ij}
Q_{i\to j}
=
2\,{\rm Im}\!\left\{\mathcal A_{ij}R_{ij}(0)\right\}
=
2\,{\rm Im}\!\left\{\mathcal A_{ij}C_{ji}\right\},
\end{equation}
which reads, in matrix form, as $\mathbf Q_0= 2\,{\rm Im}\!\left(
\boldsymbol{\mathcal A}\circ\mathbf C_0^T\right)$, where $\circ$ denotes element-wise multiplication and the diagonal entries are set to zero.
Thus, heat currents in Eq.~\eqref{eq:heat_ij} are fixed by equal-time coherences ($\tau=0$). In frequency space, these coherences accumulate contributions from the full fluctuation spectrum. From Eq.~\eqref{eq:sp_density_aa}, we have the spectral decomposition
\begin{equation}
    R_{ij}(0)=\int_{-\infty}^{\infty}\frac{{\rm d}\omega}{2\pi}S_{ij}(\omega).
\end{equation}
We thus rewrite Eq.~\eqref{eq:heat_ij} as $Q_{i\to j}  =\int_{-\infty}^{\infty}\frac{{\rm d}\omega}{2\pi}\, Q_{i\to j}(\omega)$, where 
\begin{equation}
    Q_{i\to j}(\omega)=2\,{\rm Im}\!\left\{\mathcal A_{ij}S_{ij}(\omega) \right\},
\end{equation}
is the flow spectral density in Eq.~\eqref{eq:sp_flow_dens}. These frequency-resolved flows decompose the bond currents of Eq.~\eqref{eq:heat_ij} into spectral contributions.
The local continuity equation is
\begin{equation}
\partial_t\langle n_i\rangle
=
Q_i^{\rm bath}
-
\sum_{j\neq i}\langle Q_{i\to j}\rangle .
\end{equation}
In the steady state, it simply reads as
\begin{equation}
0
=
\gamma_i\left(\bar n_i^{\rm th}-C_{ii,0}\right)
-
\sum_{j\neq i}
2\,{\rm Im}\!\left\{\mathcal A_{ij}C_{ji,0}\right\}.
\end{equation}
This relation identifies the coherent bond terms as internal heat currents: any deviation of a mode occupation from its local thermal value is sustained by a compensating flow through the mechanically coupled network.
Crucially, the continuity equation constrains only the frequency-integrated currents; finite frequency-resolved flows may counterpropagate in different spectral bands and cancel in the net current.

\subsection{Experimental platform and measurement protocol}

The experiment was performed on a sliced photonic-crystal nanobeam cavity at room temperature in vacuum, at about $2\times10^{-6}$ mbar. The device was rotated by $45^\circ$ relative to the incoming polarization for cross-polarized detection. Several high-$Q$ MHz flexural modes, all coupled to the same optical cavity, were selected to realize the rotating-frame mechanical network introduced above.
Two optical fields were used. A modulated drive laser generated the time-dependent optical spring that mediates programmable couplings, while a second, far-detuned laser provided displacement readout. The reflected detection light was separated from the drive by polarization and spectral filtering, then measured on a fast photodetector. Its intensity fluctuations encode the thermomechanical motion. The network was programmed with a multi-frequency modulation of the drive intensity. Each tone was placed near a selected mechanical frequency difference, activating the corresponding resonant beam-splitter coupling. Its amplitude sets the coupling strength, and its electronic phase sets the Peierls phase of $\mathcal A_{jk}$. One tone realizes a dimer, while three tones realize a closed loop with tunable synthetic flux.
The detection signal was demodulated at the selected mechanical frequencies to obtain the complex rotating-frame amplitudes $a_j(t)$. From these traces we extract occupations, equal-time correlations, spectra and heat currents. Bath occupations and damping rates are calibrated from uncoupled thermomechanical spectra and variances. Dimer measurements benchmark heat transfer, while loop measurements probe flux-dependent correlations, thermal-flow redistribution and refrigeration.

\subsection{Signal demodulation and calibration}

The reflected detection signal $z(t)$ contains the thermomechanical motion of the selected resonators. Mode $j$ is isolated by demodulating at its measured resonance frequency $\Omega_j$ and low-pass filtering,
\begin{equation}
    \tilde z_j(t)=\mathrm{LPF}\!\left[z(t)e^{i\Omega_j t}\right],
\end{equation}
using a lock-in amplifier (LIA; Zurich Instruments UHFLI).
The filter bandwidth ($50$~kHz at the 3-dB point) is chosen larger than the linewidths and coupling rates, but below the intermode separation, yielding the slowly varying quadratures of mode $j$.
After correcting for detector gain and mode-dependent transduction, the calibrated signal is
\begin{equation}
    a_j(t)
    =
    \frac{x_j(t)+i p_j(t)/(m_j\Omega_j)}
    {2x_{\mathrm{zpf},j}}\,
    e^{i\Omega_j t}.
\end{equation}
The scale is fixed from thermally driven uncoupled spectra and variances, and the phase is referenced to the demodulation LO and corrected for the signal propagation time through the set-up.
Damping rates $\gamma_j$ are obtained from Lorentzian fits to uncoupled spectra. Bath occupations follow from
\begin{equation}
    \langle |a_j|^2\rangle_{J=0}=\bar n_j^{\rm th},
\end{equation}
or from integrated displacement spectra, and are kept fixed in the network analysis. 
Notably, the effective linewidths and bath occupations were tuned by optothermal backaction~\cite{DelPino2022nonhermitian}, which depends on drive laser power and detuning. As a result, even though modes $a_3$ and $a_4$ featured in both loops studied, the relevant bath occupations $\nres{3}$ and $\nres{4}$ were different in Fig.~\ref{ctd:fig:loop-fluctuations} compared to the experiments presented in later Figures. Optothermal backaction was also responsible for the different mode temperatures observed in Fig.~\ref{fig:1}(h) at zero coupling.
Couplings are calibrated link by link. For each pair $(j,k)$, a tone near $\Omega_j-\Omega_k$ activates the beam-splitter interaction. Its frequency is adjusted to the shifted resonances, and its amplitude is calibrated from normal-mode splitting, coherent exchange, or correlation dynamics, yielding $J_{jk}$. The hopping phase is set by the tone phase, which is electronically referenced to the LO phases of the mode pair it couples~\cite{Slim2024optomechanical}. Since LO phases define a local gauge, we fix a common convention and identify the physical loop phase with
\begin{equation}
    \Phi=\sum_{(j,k)\in\mathrm{loop}}\phi_{jk}.
\end{equation}
In loop measurements, $\Phi$ is swept by changing one tone phase while keeping all coupling magnitudes fixed.

\subsection{Noise treatment and data analysis}

The demodulated traces contain thermomechanical motion together with detector and electronic noise. The lock-in low-pass filter gives otherwise broadband detector noise a finite correlation time, producing a short-time peak in the measured auto-correlations. This contribution is measured independently and subtracted from the auto-correlations used to extract occupations. Cross-correlations between different mechanical frequencies are much less affected, since the detector noise is uncorrelated between demodulation channels.
All observables are extracted from stationary time windows of the calibrated rotating-frame traces. We first compute the two-time correlations and their Fourier transforms, then obtain occupations from the corrected equal-time auto-correlations and heat currents from the zero-lag cross-correlations using the expressions given above. Spectral flow densities are obtained analogously from the complex cross spectra.
Data are averaged over repeated traces or independent stationary windows. Uncertainties are estimated from the statistical spread between these independent averages and propagated through the same analysis pipeline for occupations, spectra and heat currents.
\FloatBarrier
\setcounter{figure}{0}
\renewcommand{\figurename}{Extended Data Figure}
\renewcommand{\theHfigure}{extended-data.\arabic{figure}}

\subsection{Flux-threaded trimer eigenmode picture}

For a uniform three-mode loop with hopping strength $J$ and total flux $\Phi$, we choose a gauge in which each link carries phase $\Phi/3$,
\begin{equation}\label{eq:trimer-coupling-matrix}
    \mathcal A_{j,j+1}=J e^{i\Phi/3},
    \qquad
    \mathcal A_{j+1,j}=J e^{-i\Phi/3},
\end{equation}
with $j+1$ understood modulo three. The normal modes are angular-momentum modes,
\begin{equation}\label{eq:trimer-angular-momentum-modes}
    \tilde a_\mu=\frac{1}{\sqrt{3}}\sum_j e^{-ik_\mu j}a_j,
\end{equation}
with wavenumbers $k_\mu=2\pi\mu/3$ ($\mu=0,\pm1$), and flux-dependent eigenfrequencies
\begin{equation}\label{eq:trimer-eigenfrequencies}
    \omega_\mu=2J\cos\!\left(k_\mu+\frac{\Phi}{3}\right).
\end{equation}
The modes $\mu=\pm1$ carry opposite phase winding around the loop, while $\mu=0$ is the symmetric mode. At time-reversal-symmetric fluxes, such as $\Phi=0$ or $\pi$, degeneracies allow the chiral modes to be recombined into real standing-wave modes. Away from these points the eigenmodes have intrinsically complex phase patterns.
This phase structure is directly visible in the complex cross spectra. In the symmetric limit of equal damping and equal bath occupation (equilibrium condition), the off-diagonal spectrum reduces to
\begin{equation}\label{eq:trimer-cross-spectrum}
    S_{ij}(\omega)
    =
    \frac{\gamma(\bar n+1)}{3}
    \sum_{\mu=0,\pm1}
    \frac{e^{ik_\mu(i-j)}}
    {(\omega-\omega_\mu)^2+\gamma^2/4}.
\end{equation}
Thus each spectral peak carries the relative phase of the corresponding angular-momentum eigenmode. Measuring the magnitude and phase of $S_{ij}(\omega)$ therefore provides a frequency-resolved probe of the chiral eigenmodes and of the synthetic flux threading the loop. A more general derivation, including the full eigenmode expansion of $S_{ij}(\omega)$ and the steady-state trimer correlations for non-uniform baths and damping rates, is given in Supplementary Sections~II.B and~II.C.
The corresponding eigenmode decomposition of the heat flow follows by substituting Eq.~\eqref{eq:trimer-cross-spectrum} into the flow spectral density defined in Eq.~\eqref{eq:sp_flow_dens},
\begin{equation}
Q_{i\to j}(\omega) = \frac{2J\gamma(\bar n+1)}{3} \sum_{\mu=0,\pm1} \frac{{\rm Im}\!\left\{e^{i\phi_{ij}}e^{ik_\mu(i-j)} \right\} }{(\omega-\omega_\mu)^2+\gamma^2/4}.
\label{eq:trimer-modal-flow-spectrum}
\end{equation}
Thus, each spectral peak represents the bond current carried by one trimer eigenmode, with its direction fixed by the phase winding of that mode across the link. For a single link $j\to j+1$, Eqs.~\eqref{eq:trimer-coupling-matrix} and~\eqref{eq:trimer-modal-flow-spectrum} give
\begin{equation}
Q_{j\to j+1}(\omega)
=
\frac{2J\gamma(\bar n+1)}{3}
\sum_{\mu=0,\pm1}
\frac{
\sin\!\left(\frac{\Phi}{3}-k_\mu\right)
}{
(\omega-\omega_\mu)^2+\gamma^2/4
}.
\label{eq:trimer-nearest-neighbour-flow}
\end{equation}
Eigenmodes with different phase winding can therefore carry currents in opposite directions at different frequencies.
Since each Lorentzian in Eq.~\eqref{eq:trimer-nearest-neighbour-flow} has the same integrated weight, the contribution of eigenmode $\mu$ to the total bond current is proportional to $\sin(\Phi/3-k_\mu)$.
Since
\begin{equation}
\sum_{\mu=0,\pm1} \sin\!\left(\frac{\Phi}{3}-k_\mu\right) = 0,
\label{eq:trimer-flow-cancellation}
\end{equation}
the spectral densities exactly cancel after frequency integration. Thus, an equilibrium loop supports finite counterpropagating spectral flows but no net bond current. 
Out of equilibrium, unequal bath occupations produce unequal and generally correlated modal fluctuations, so this cancellation is generally incomplete and a finite integrated heat current can emerge.

\subsection*{Data and code availability}

All codes and data supporting this study are available in the public repository \#.

\subsection*{Contributions}

J.J.S. designed and fabricated the samples and carried out the experiments with input from E.V. and J.d.P. J.J.S. and J.d.P. analyzed the data. J.d.P., J.J.S., and S.M. developed the theory with input from E.V.. All authors contributed to the interpretation of the results and the writing of the manuscript. E.V. supervised the project.

\subsection*{Acknowledgments}

We thank Uro\v{s} Deli\'{c} and Rafael Sánchez for fruitful discussions. This work is part of the research programme of the Netherlands Organisation for Scientific Research (NWO). It is supported by the European Union's Horizon 2020 research and innovation programme under grant agreement No 732894 (FET-Proactive HOT) and the European Research Council (ERC Grants~759644 (TOPP) and 101088055 (Q-MEME)). Views and opinions expressed are however those of the authors only and do not necessarily reflect those of the European Union or the European Research Council Executive Agency. Neither the European Union nor the granting authority can be held responsible for them.  JdP acknowledges funding from the Ram\'on y Cajal program (RYC2023-043827-I), funded by MICIU/AEI (10.13039/501100011033) and FSE+, and from the ``Mar\'ia de Maeztu'' Programme for Units of Excellence in R\&D (CEX2023-001316-M), funded by the Spanish Ministry of Science, Innovation and Universities. J.d.P. also acknowledges support from the Proyectos de Generaci\'on de Conocimiento program, Grant No.~PID2024-158923NA-I00, funded by MICIU/AEI (10.13039/501100011033) and FEDER, UE.

\subsection*{Competing interests}

The authors declare no competing interest.

\begin{figure*}
    \centering
    \includegraphics{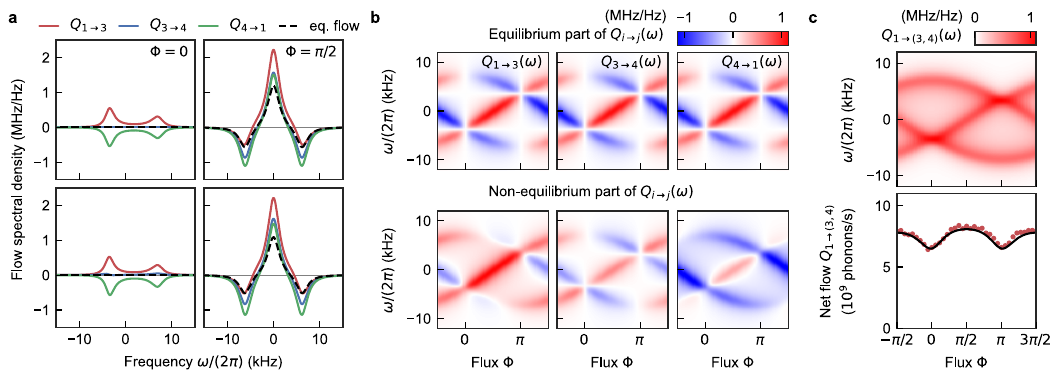}
    \phantomsubfloat{\label{fig:ed1:linecuts}}
    \phantomsubfloat{\label{fig:ed1:density_plots}}
    \phantomsubfloat{\label{fig:ed1:net_flows}}
    \vspace{-2\baselineskip}
    \caption{\textbf{Flow spectral densities for the loop comprising $a_1,a_3,a_4$.} 
    \subidc{a} Theoretical flow spectral densities $Q_{i \to j}(\omega)$ at flux $\Phi = 0$ (left) and $\Phi = \pi/2$ (right) for the network presented in Fig.~\ref{ctd:fig:circulator-flows}, in the idealized case $\nres{3}=\nres{4}$ (top) and the exact experimental conditions where $\nres{3}\approx\nres{4}$
    \subidc{b} Density plots of the equilibrium ($\nres{j} = \bar{n} = 0.36 \times 10^6$) and non-equilibrium ($\Delta\nres{1,3,4} = \{1.15, 0.09, 0\} \times 10^6$) flow spectral densities for the network's exact experimental conditions.
    \subidc{c} Net flow $Q_{1\to3}(\omega) + Q_{1\to4}(\omega)$ out of the `hottest' resonator $a_1$, spectrally resolved (top) and integrated (bottom). Measured flows (red circles) agree well across all fluxes $\Phi$.
	}
	\label{fig:ed1}

\end{figure*}

\clearpage

\onecolumngrid

\setcounter{section}{0}
\setcounter{subsection}{0}
\setcounter{equation}{0}
\setcounter{figure}{0}
\setcounter{secnumdepth}{2}
\renewcommand{\thesection}{\Roman{section}}
\renewcommand{\thesubsection}{\Alph{subsection}}
\renewcommand{\theequation}{S\arabic{equation}}
\renewcommand{\thefigure}{S\arabic{figure}}
\renewcommand{\theHsection}{supplement.\Roman{section}}
\renewcommand{\theHsubsection}{supplement.\Roman{section}.\Alph{subsection}}
\renewcommand{\theHequation}{supplement.\arabic{equation}}
\renewcommand{\theHfigure}{supplement.\arabic{figure}}
\renewcommand{\figurename}{Supplementary Figure}
\begin{center}
{\supplementtitleformat
Supplementary Information for\\[0.7em]
Chiral thermal fluctuations and enhanced refrigeration in a nonreciprocal nanomechanical system\par}
\vspace{1.2em}
{\supplementauthorformat
Jesse J. Slim,\textsuperscript{1,2,*}
Javier del Pino,\textsuperscript{1,3,*}
Sander A. Mann,\textsuperscript{4}
and Ewold Verhagen\textsuperscript{1,$\dagger$}\par}
\vspace{0.9em}
{\supplementaffiliationformat
\textsuperscript{1}Center for Nanophotonics, AMOLF, Science Park 104, 1098 XG Amsterdam, The Netherlands\par
\textsuperscript{2}Australian Research Council Centre of Excellence in Quantum Biotechnology (QUBIC), School of Mathematics and Physics, The University of Queensland, St Lucia, Queensland 4072, Australia\par
\textsuperscript{3}Departamento de Física Teórica de la Materia Condensada and Condensed Matter Physics Center (IFIMAC), Universidad Autónoma de Madrid, E-28049 Madrid, Spain\par
\textsuperscript{4}Institute of Physics, University of Amsterdam, Amsterdam, The Netherlands\par}
\vspace{0.7em}
{\supplementaffiliationformat
\textsuperscript{*}These authors contributed equally to this work.\par
\textsuperscript{$\dagger$}\href{mailto:verhagen@amolf.nl}{verhagen@amolf.nl}\par}
\end{center}

\section*{Contents}

\begingroup
\setlength{\parindent}{0pt}
\newcommand{\supptocsection}[3]{
  \textbf{#1\quad #2}\dotfill\pageref{#3}\par}
\newcommand{\supptocsubsection}[3]{
  \hspace*{2em}#1\quad #2\dotfill\pageref{#3}\par}
\supptocsection{I}{Theoretical formalism}{supp:sec:formalism}
\supptocsubsection{A}{Effective nanomechanical Hamiltonian}{supp:subsec:hamiltonian}
\supptocsubsection{B}{Coupling to fluctuating-dissipative thermal bath}{subsec:therma_reg}
\supptocsubsection{C}{Steady-state occupations and currents}{supp:subsec:steady-state}
\supptocsection{II}{Supporting theoretical results}{supp:sec:theory-results}
\supptocsubsection{A}{Dimer with bath occupation imbalance}{supp:subsec:dimer}
\supptocsubsection{B}{Flux-threaded loop: Cross correlations}{supp:subsec:cross-correlations}
\supptocsubsection{C}{Flux-threaded loop: stationary occupations}{supp:subsec:stationary-occupations}
\supptocsubsection{D}{Absence of persistent directional heat currents}{supp:subsec:persistent-currents}
\supptocsubsection{E}{Nonreciprocity for optimal mixing}{supp:subsec:mixing}
\supptocsubsection{F}{Counterflow between cold resonators}{supp:subsec:counterflow}
\supptocsection{III}{Supporting experimental results}{supp:sec:experimental}
\supptocsubsection{A}{Supporting measurements: Single resonator}{supp:subsec:single-resonator}
\supptocsubsection{B}{Time-resolved thermalization measurements}{supp:subsec:thermalization}
\supptocsubsection{C}{Filtered instrument noise in demodulated correlations}{supp:subsec:noise}
\endgroup

	\section{Theoretical formalism}\label{supp:sec:formalism}

	In this section we detail the derivation of the effective Hamiltonian employed in the main text, the incorporation of dissipative coupling to thermal baths, and the calculation of the system's response to thermal and coherent fluctuations. For convenience we express the mechanical modes using quantum creation and annihilation operators, but the dynamics discussed here and in the main text is fully classical: all operators can be consistently replaced by complex amplitudes, recovering the semiclassical equations used throughout.

	\subsection{Effective nanomechanical Hamiltonian}\label{supp:subsec:hamiltonian}

    The main text considers a nanomechanical network of $N$ modes $b_j$ (label $j=1,2,\cdots N$) whose couplings are tuned by cavity modes $c_l$ through optomechanical interactions $g_{0}^{(i,l)}$. Cavity modes are coherently driven by time-dependent inputs $C^{(l)}_{\mathrm{in}}(t)$. The full Hamiltonian reads $H_s(t)+H_\mathrm{in}(t)$ ($\hbar=1$), where the network dynamics obey the multimode optomechanical Hamiltonian
    \begin{equation}
    H_s(t)=\sum_{j}\omega_j\, b_{j}^{\dagger}b_{j}-\sum_{l}c_{l}^{\dagger}c_{l}\left[\Delta_l+\sum_{j}g_{0}^{(j,l)}(b_{j}+b_{j}^{\dagger})\right],\label{eq:H}
    \end{equation}
    and the inputs enter through the in-coupling rates $\kappa_l$ as
    \begin{equation}
    H_\mathrm{in}(t)=\sum_{l}i\sqrt{\kappa_l}\,c^{\dagger}_l\,C^{(l)}_\mathrm{in}(t) + \mathrm{H.c.}\, .
    \end{equation}
    The driving laser intensity incident on the device is modulated with frequency $\omega_l$ and phase $\phi_l$. In the limit of large detuning and bandwidth ($|\Delta_l|,\kappa_l\gg\omega_i$), cavity fields follow the drives quasi-instantaneously, so that the coherent intracavity amplitudes $\bar c_l(t)$ and populations $\bar n_l(t)=|\bar c_l(t)|^2$ track the input envelopes. In particular, the total population $\bar n_c(t)=\sum_l \bar n_l(t)$ takes the form
    \begin{equation}\label{eq:n(t)}
    \bar{n}_c(t)
    =\sum_{l} |C^{(l)}_\mathrm{in}|^2\left(1+c_l\cos(\omega_l t+\phi_l)\right),
    \end{equation}
    where $c_l$ is the modulation depth of tone $l$.
    Finally, since the single-photon optomechanical couplings $g_{0}^{(i,l)}$ are weak, we linearize the optomechanical interaction around the coherent cavity fields by writing $c_l(t)=\bar c_l(t)+d_l$ and keeping only terms up to first order in the fluctuations $d_l$, which yields
    \begin{equation}
    H_{\rm lin}(t)
    =\sum_j \omega_i\, b_j^\dagger b_j
    -\sum_l \Delta_l\, d_l^\dagger d_l
    -\sum_{j,l}\Big(G_{j,l}(t)\, d_l+G_{j,l}^*(t)\, d_l^\dagger\Big)\,(b_i+b_i^\dagger),
    \end{equation}
    with $G_{j,l}(t)=g_{0}^{(j,l)}\,\bar c_l(t)$ (so $|G_{j,l}(t)|=g_{0}^{(j,l)}\sqrt{\bar n_l(t)}$). We have dropped constant energy shifts and absorbed static radiation-pressure offsets into the definitions of detunings and equilibrium positions.
    Within the same limit, the cavity modes can be adiabatically eliminated. We first evaluate the coherent cavity field in the absence of optomechanical backaction,
    $c_l\rightarrow \bar c_{0,l}(t)+\delta c_l$, with
    \begin{equation}
    \bar c_{0,l}(t)\simeq \sqrt{\kappa_l}\,\chi_l(0)\,C^{(l)}_{\mathrm{in}}(t),
    \qquad
    \chi_l(\omega)=\left(\frac{\kappa_l}{2}-i(\omega+\Delta_l)\right)^{-1},
    \end{equation}
    where $\chi_l(\omega)$ is the cavity susceptibility. Under these assumptions, the mechanical frequencies acquire an optical-spring shift,
    \begin{equation}
    \omega_j\mapsto\Omega_j=\omega_j-\sum_l \frac{2\,(g_{0}^{(j,l)})^{2}\,\Delta_l}{\Delta_l^{2}+(\kappa_l/2)^2}\,\bar n_{l}^{(\mathrm{dc})},
    \end{equation}
    set by the DC component $\bar n_{l}^{(\mathrm{dc})}$ of the intracavity population in Eq.~\eqref{eq:n(t)}. Eliminating the cavity fluctuations yields effective phonon--phonon interactions of the form
    \begin{equation}\label{eq:Hbint}
    H_b^{\mathrm{int}}(t)\simeq
    -\sum_l \frac{\Delta_l}{\Delta_l^{2}+(\kappa_l/2)^2}\,
    \left|\bar c_{0,l}(t)\right|^{2}
    \left(\sum_{j}g_{0}^{(j,l)}(b_{j}+b_{j}^{\dagger})\right)^{2},
    \end{equation}
    where we kept the leading dispersive contribution relevant for mechanically mediated couplings.
    We focus on the experimental regime with a single relevant cavity mode and drop the index $l$. Under moderate effective couplings $J_{jk}\ll\Omega_i$, the rotating-wave approximation applies: fast-oscillating terms average out, leaving only resonant contributions. Choosing modulation tones at the mechanical frequency differences $\omega_l\simeq \Omega_i-\Omega_j$ and assuming the mechanical frequencies are incommensurate, the effective Hamiltonian becomes
    \begin{equation}\label{eq:Ht}
    H_b(t)\simeq\sum_{j}\Omega_{j}\,b_{j}^{\dagger}b_{j}
    +\sum_{l\in\mathrm{resonant}}\sum_{j\neq k}
    \left(J_{jk}^{(l)}\, b_{j}^{\dagger}b_{k}\,e^{i\omega_l t}+\mathrm{H.c.}\right),
    \end{equation}
    with hopping amplitudes proportional to the modulation depth and dispersive kernel, e.g.
    \begin{equation}
    J_{jk}^{(l)}= \frac{c_l\, g_{0}^{(j)} g_{0}^{(k)}\Delta}{\Delta^{2}+(\kappa/2)^2}.
    \end{equation}
    For tones tuned to $\omega_l=\tilde\omega_j-\tilde\omega_k$, the Hamiltonian \eqref{eq:Ht} becomes time independent in the rotating frame $\tilde a_j=e^{-i\tilde\omega_j t}b_i$, implemented via $\hat U_F=\exp\!\left(-it\sum_j \tilde\omega_j\, b_j^\dagger b_j\right)$. One obtains
    \begin{equation}\label{eq:H_rot}
    H_a\approx \sum_{j,k}J_{jk}e^{-\iu\phi_{jk}}a_j^\dagger a_k + \mathrm{H.c.},
    \end{equation}
    which conserves the phonon number $\sum_{j} a_j^{\dagger} a_j$. For notational simplicity, in the main text and the following we later drop the subscript ``$a$''.

	\subsection{Coupling to fluctuating-dissipative thermal bath}\label{subsec:therma_reg}

	To complete the description we include coupling to thermally fluctuating baths. We assume that each mechanical mode interacts with an \textit{independent} thermal bath at temperature $T$, justified by two facts: (i) the mode frequencies differ by amounts far exceeding their linewidths, and (ii) each mode occupies a very narrow spectral band. Under these conditions the environmental noise seen by different modes is effectively uncorrelated. The master equation for the mechanical density matrix $\hat\rho$ then follows standard open-quantum-systems arguments~\cite{Gardiner2004}. Each mode has thermal occupation $\bar{n}^{\mathrm{th}}_i=\bar n(\tilde\omega_i)\simeq k_\mathrm{B}T/\tilde\omega_i$, and $T=0$ Markovian losses $\gamma_i$. The mechanical density matrix obeys:
	\begin{equation}\label{eq:masterE}
	\dot{\rho}=-i[\rho,H]+\sum_j \gamma_j(\bar{n}^{\mathrm{th}}_j+1)\mathcal{D}_{ a_j}[\rho]+
	\gamma_i\bar{n}^{\mathrm{th}}_j\mathcal{D}_{ a^{\dagger}_j}[\rho], 
	\end{equation}
	where $\mathcal{D}_{X}[\rho]=X\rho X^{\dagger}-\frac{1}{2}\{X^{\dagger}X,\rho\}$ is a Lindblad dissipator and $H$ given by \eqref{eq:H_rot}. Note that adiabatic elimination of the cavity also generates small induced losses and dissipative couplings between the mechanical modes, but these scale as $\gamma_m^{(i)}/\Delta\ll1$\cite{Reiter2012effective} and can be safely ignored within the current setting.
    The Heisenberg--Langevin formulation is fully equivalent to the master-equation description, but it is more convenient for evaluating correlations because the noise inputs appear explicitly and the resulting linear dynamics can be solved in closed form. We therefore work in this picture. In the rotating frame introduced in the main text, the mechanical mode amplitudes obey
    \begin{equation}\label{eq:HEOM_evo}
    \dot{\mathbf{a}}(t)=-i\!\left(\boldsymbol{\mathcal{A}}-i\frac{\boldsymbol{\gamma}}{2}\right)\mathbf{a}(t)+\mathbf{a}_{\mathrm{in}}(t),
    \end{equation}
    where $\mathbf{a}(t)=(a_1(t),\ldots,a_N(t))^{T}$ collects the mode amplitudes, $\boldsymbol{\mathcal{A}}$ is the hopping matrix with elements $\mathcal{A}_{jk}=J_{jk}e^{i\phi_{jk}}$, dissipation is encoded by $\boldsymbol{\gamma}=\mathrm{diag}(\gamma_1,\ldots,\gamma_N)$, and $\mathbf{a}_{\mathrm{in}}(t)=(a^{(1)}_{\mathrm{in}}(t),\ldots,a^{(N)}_{\mathrm{in}}(t))^{T}$ denotes the bath input operators. Assuming independent Markovian thermal baths, the only non-vanishing noise correlators are
    \begin{equation}\label{eq:noise_corr}
    \big\langle a^{(j)}_{\mathrm{in}}(t)\,a^{(k)\dagger}_{\mathrm{in}}(t')\big\rangle
    =\gamma_j(\bar n_j+1)\,\delta_{jk}\,\delta(t-t'),
    \end{equation}
    or, in compact matrix form, $\langle \mathbf{a}_{\mathrm{in}}(t)\mathbf{a}_{\mathrm{in}}^{\dagger}(t')\rangle=\mathbf{D}\,\delta(t-t')$ with diffusion matrix
    $\mathbf{D}=\mathrm{diag}\big(\gamma_1(\bar n_1+1),\ldots,\gamma_N(\bar n_N+1)\big)$.
    In Fourier space, Eq.~\eqref{eq:HEOM_evo} can be solved in the steady state to give $\mathbf{a}(\omega)=\boldsymbol{\chi}(\omega)\,\mathbf{a}_{\mathrm{in}}(\omega)$ with susceptibility matrix
    \begin{equation}\label{eq:chi_matrix}
    \boldsymbol{\chi}(\omega)=\left[-i\omega\,\mathbb{1}+i\boldsymbol{\mathcal{A}}+\frac{\boldsymbol{\gamma}}{2}\right]^{-1}.
    \end{equation}
    Although the thermal force is $\delta$-correlated in time, the mechanical inertia generates finite-time correlations in the amplitudes.
    For linear open systems (i.e.\ those governed by a quadratic Hamiltonian with linear damping), the corresponding quantum statement is the \emph{quantum regression theorem}~\cite{Gardiner2004quantum}. In the classical limit, replacing operators by their expectation values reproduces the Onsager regression principle~\cite{Onsager1931reciprocal2}.
    Let us consider the vector of annihilation operators $\mathbf{a}(t)=(a_1(t),\ldots,a_N(t))^{T}$ evolving under the linear Heisenberg--Langevin equation \eqref{eq:HEOM_evo}. The central object is the (reduced) two-time correlation matrix
    \begin{equation}\label{ctd:eq:R-def}
    \mathbf{R}(t+\tau,t)\equiv \big\langle \mathbf{a}(t+\tau)\,\mathbf{a}^{\dagger}(t)\big\rangle .
    \end{equation}
    In the steady state, the quantum regression theorem states that for $\tau>0$ the correlation with positive time lag obeys the same drift equation as $\langle\mathbf{a}\rangle$~\cite{Meystre2007}:
    \begin{equation}\label{ctd:eq:R-regression}
    \frac{\dd}{\dd \tau}\mathbf{R}(t+\tau,t)
    =-i\!\left(\boldsymbol{\mathcal{A}}-i\frac{\boldsymbol{\gamma}}{2}\right)\mathbf{R}(t+\tau,t),
    \end{equation}
    with formal solution
    \begin{equation}\label{ctd:eq:R-solution}
    \mathbf{R}(t+\tau,t)=e^{-i\left(\boldsymbol{\mathcal{A}}-i\frac{\boldsymbol{\gamma}}{2}\right)\tau}\mathbf{R}(t,t).
    \end{equation}
    The equal-time correlator $\mathbf{R}(t,t)$ follows from a Lyapunov equation obtained by combining the drift in \eqref{eq:HEOM_evo} with the diffusion encoded in \eqref{eq:noise_corr},
    \begin{equation}\label{ctd:eq:lyapunov}
    \dot{\mathbf{R}}(t)
    =-i\!\left(\boldsymbol{\mathcal{A}}-i\frac{\boldsymbol{\gamma}}{2}\right)\mathbf{R}(t,t)
    +i\,\mathbf{R}(t,t)\!\left(\boldsymbol{\mathcal{A}}+i\frac{\boldsymbol{\gamma}}{2}\right)
    +\mathbf{D}.
    \end{equation}
    In the steady state ($\dot{\mathbf{R}}(t,t)=0$, $t\rightarrow\infty$) the equal-time correlation settles into a value $\mathbf{R}_0$ satisfying
    \begin{equation}\label{ctd:eq:lyapunov_ss}
    i\!\left(\boldsymbol{\mathcal{A}}-i\frac{\boldsymbol{\gamma}}{2}\right)\mathbf{R}_0
    -i\,\mathbf{R}_0\!\left(\boldsymbol{\mathcal{A}}+i\frac{\boldsymbol{\gamma}}{2}\right)
    =\mathbf{D},
    \end{equation}
    which can be formally solved in terms of the drift propagator.\footnote{Defining $\boldsymbol{\mathcal{M}}=-i\boldsymbol{\mathcal{A}}-\boldsymbol{\gamma}/2$, Eq.~\eqref{ctd:eq:lyapunov_ss} is equivalent to $\boldsymbol{\mathcal{M}}\mathbf{R}_0+\mathbf{R}_0\boldsymbol{\mathcal{M}}^\dagger+\mathbf{D}=0$. Provided that all eigenvalues of $\boldsymbol{\mathcal{M}}$ have negative real parts, i.e. $\boldsymbol{\mathcal{M}}$ is dynamically stable, the solution for $\mathbf{R}_0$
    \begin{equation}
        \mathbf{R}_0=\int_{0}^{\infty}\!e^{\boldsymbol{\mathcal{M}}t}\mathbf{D}e^{\boldsymbol{\mathcal{M}}^\dagger t}\,\dd t.
    \end{equation}
    Eq.~\eqref{ctd:eq:lyapunov_ss} can equivalently be solved by vectorizing the unknown matrix, using $\mathrm{vec}(AXB)=(B^{T}\!\otimes A)\mathrm{vec}(X)$ to reduce it to a linear system for $\mathrm{vec}(\mathbf{R}_0)$.
    }
    The two-time correlation matrix only depends on the delay $\tau$,
    \begin{equation}\label{ctd:eq:R-solution_tau}
    \mathbf{R}(\tau)\equiv\lim_{t\rightarrow\infty}\mathbf{R}(t+\tau,t)
    =e^{-i\left(\boldsymbol{\mathcal{A}}-i\frac{\boldsymbol{\gamma}}{2}\right)\tau}\mathbf{R}_0\, .
    \end{equation}
    This result, combined with the Wiener--Khinchin theorem yields the spectral density matrix in the steady state in a simple closed form
    \begin{equation}\label{ctd:eq:spectral-matrix}
    \boldsymbol{\mathcal{S}}_{\mathbf{a}\mathbf{a}^\dagger}(\omega)
    =\int_{-\infty}^{\infty}e^{-i\omega\tau}\,\mathbf{R}(\tau)\,\dd\tau
    =\boldsymbol{\chi}(\omega)\,\mathbf{D}\,\boldsymbol{\chi}^{\dagger}(\omega),
    \end{equation}
    where $\boldsymbol{\chi}(\omega)$ is given in Eq.~\eqref{eq:chi_matrix}.

    \subsection{Steady-state occupations and currents}\label{supp:subsec:steady-state}

    Starting from the Lindblad master equation~\eqref{eq:masterE}, we derive the continuity equation for the mode occupations and connect it to equal-time correlators. We consider a number-conserving Hamiltonian in the rotating frame,
    \begin{equation}\label{eq:H_tb}
    H=\sum_{i,j}\mathcal{A}_{ij}\,a_i^\dagger a_j,
    \end{equation}
    and define the mode occupation operator $n_i\equiv a_i^\dagger a_i$, with $\langle n_i\rangle=\mathrm{Tr}(n_i\rho)$. Taking the trace of $\dot\rho$ against $n_i$ gives
    \begin{equation}\label{eq:pop_dyn_start}
    \partial_t\langle n_i\rangle=\mathrm{Tr}\!\left(n_i\dot\rho\right)
    =-i\langle[n_i,H]\rangle+\mathrm{Tr}\!\left(n_i\,\mathcal{L}_{\rm diss}[\rho]\right),
    \end{equation}
    where $\mathcal{L}_{\rm diss}$ denotes the dissipative part of~\eqref{eq:masterE}.
    Evaluating the commutator using~\eqref{eq:H_tb} yields
    \begin{equation}\label{eq:commutator_part}
    -i\,\langle[n_i,H]\rangle
    =-i\sum_{j}\Big(\mathcal{A}_{ij}\langle a_i^\dagger a_j\rangle-\mathcal{A}_{ji}\langle a_j^\dagger a_i\rangle\Big)
    =\sum_{j\neq i}\langle Q_{i\rightarrow j}\rangle,
    \end{equation}
    which identifies the \emph{internal bond current operator} from $i$ to $j$ as
    \begin{equation}\label{eq:current_operator_def}
    Q_{i\rightarrow j}
    \equiv i\!\left(\mathcal{A}_{ij}\,a_i^\dagger a_j-\mathcal{A}_{ij}^*\,a_j^\dagger a_i\right)
    =2\,\mathrm{Im}\!\left\{\mathcal{A}_{ij}\,a_i^\dagger a_j\right\}.
    \end{equation}
    In steady state, these currents are therefore fixed by the equal-time coherences $\langle a_i^\dagger a_j\rangle$.
    On the other hand, the dissipators in~\eqref{eq:masterE} yield the net exchange with the local bath coupled to mode $i$,
    \begin{equation}\label{eq:dissipator_part}
    \mathrm{Tr}\!\left(n_i\,\mathcal{L}_{\rm diss}[\rho]\right)
    =\gamma_i\big(\bar n_i^{\rm th}-\langle n_i\rangle\big)
    \equiv Q^{\rm bath}_i,
    \end{equation}
    which we interpret as the \emph{external (reservoir) current} injected into mode $i$ by its bath.
    Combining \eqref{eq:commutator_part} and \eqref{eq:dissipator_part}, the population dynamics takes the continuity form
    \begin{equation}\label{eq:continuity_full}
    \partial_t\langle n_i\rangle
    =Q^{\rm bath}_i+\sum_{j\neq i}\langle Q_{i\rightarrow j}\rangle
    =\gamma_i\big(\bar n_i^{\rm th}-\langle n_i\rangle\big)+\sum_{j\neq i}\langle Q_{ij}\rangle.
    \end{equation}
    In steady state, $\partial_t\langle n_i\rangle=0$, so any deviation of $\langle n_i\rangle$ from the local bath value $\bar n_i^{\rm th}$ is sustained by nonzero internal bond currents.
    The continuity equation \eqref{eq:continuity_full} can be closed after introducing the normally ordered equal-time correlator matrix
    \begin{equation}\label{eq:C_def}
    C_{ij}(t)=\langle a_i^\dagger(t)a_j(t)\rangle,
    \qquad
    \mathbf{C}_0\equiv\lim_{t\to\infty}\mathbf{C}(t).
    \end{equation}
    Then the steady-state occupations are simply $\langle n_i\rangle_{t\to\infty}=C_{ii,0}$, and the steady-state internal currents follow from \eqref{eq:current_operator_def} as
    \begin{equation}\label{eq:current_from_C}
    \langle Q_{i\rightarrow j}\rangle_{t\to\infty}
    =2\,\mathrm{Im}\!\left\{\mathcal{A}_{ij}\,C_{ij,0}\right\}.
    \end{equation}
    Introducing the steadystate current matrix $\boldsymbol{Q_0}$ with elements $\langle  Q_{i\rightarrow j}\big\rangle_{t\to\infty}$ and $\mathrm{diag}(\boldsymbol{Q}_0)=0$, Eq. \eqref{eq:current_from_C} can be compactly written in terms of the linkwise product $\circ$:
    \begin{equation}\label{eq:I_matrix_hadamard}
        \boldsymbol{Q_0}
        =2\,\mathrm{Im}\!\left(\boldsymbol{\mathcal{A}}\circ \mathbf{C}_0^{T}\right),
    \end{equation}
    Therefore, the full set of occupations and internal currents is determined once $\mathbf{C}_0$ is known. Note that in the regression section \ref{subsec:therma_reg} we instead used the anti-normally ordered equal-time correlator $\mathbf{R}_0$; using $a_i a_j^\dagger=\delta_{ij}+a_j^\dagger a_i$, one obtains $\mathbf{C}_0=\big(\mathbf{R}_0-\mathbb{1}\big)^{T}$ and the steadystate current matrix  reads
    \begin{equation}
    \boldsymbol{Q}
    =2\,\mathrm{Im}\!\left(\boldsymbol{\mathcal{A}}\circ \mathbf{R}_0^{T}\right).
    \end{equation}

    \section{Supporting theoretical results}\label{supp:sec:theory-results}

    \subsection{Dimer with bath occupation imbalance}\label{supp:subsec:dimer}

    Here we solve the thermal steady state of the minimal two-mode problem within the same correlator notation used above. We consider two modes $(a_1,a_2)$ with hopping matrix
    \begin{equation}\label{eq:dimer_A}
    \boldsymbol{\mathcal{A}}=
    \begin{pmatrix}
    0 & J e^{i\phi}\\
    J e^{-i\phi} & 0
    \end{pmatrix},
    \qquad
    \boldsymbol{\gamma}=\mathrm{diag}(\gamma_1,\gamma_2),
    \qquad
    \mathbf{D}=\mathrm{diag}\!\big(\gamma_1(\bar n_1+1),\gamma_2(\bar n_2+1)\big),
    \end{equation}
    The equal-time correlator is the $2\times 2$ matrix
    \begin{equation}\label{eq:R0_def_dimer}
    \mathbf{R}_0 \equiv \langle \mathbf{a}\mathbf{a}^\dagger\rangle_{t\to\infty}
    =
    \begin{pmatrix}
    R_{11} & R_{12}\\
    R_{21} & R_{22}
    \end{pmatrix}
    =
    \begin{pmatrix}
    \langle a_1 a_1^\dagger\rangle & \langle a_1 a_2^\dagger\rangle\\
    \langle a_2 a_1^\dagger\rangle & \langle a_2 a_2^\dagger\rangle
    \end{pmatrix}_{t\to\infty}.
    \end{equation}
    In steady state, $\mathbf{R}_0$ is determined from Eq.\eqref{ctd:eq:lyapunov_ss}. Writing this equation as a linear system for the vector
    \begin{equation}\label{eq:v_def_dimer}
    \mathbf{v}\equiv\big(R_{11},\,R_{22},\,R_{12},\,R_{21}\big)^T,
    \end{equation}
    Eq.~\eqref{eq:pop_dyn_start} becomes $\boldsymbol{\mathcal{N}}\,\mathbf{v}=\mathbf{h}$ with
    \begin{equation}\label{eq:N_matrix_dimer}
    \boldsymbol{\mathcal{N}}=
    \begin{pmatrix}
    \gamma_1 & 0 & -iJ e^{-i\phi} & \ \ iJ e^{i\phi} \\
    0 & \gamma_2 & \ \ iJ e^{-i\phi} & -iJ e^{i\phi} \\
    -iJ e^{i\phi} & \ \ iJ e^{i\phi} & \frac{\gamma_1+\gamma_2}{2} & 0 \\
    \ \ iJ e^{-i\phi} & -iJ e^{-i\phi} & 0 & \frac{\gamma_1+\gamma_2}{2}
    \end{pmatrix},
    \end{equation}
    and $\mathbf{h}=\big(D_1,\,D_2,\,0,\,0\big)^T$.    Hence the steady state follows directly as $\mathbf{v}_{t\rightarrow\infty}=\boldsymbol{\mathcal{N}}^{-1}\mathbf{h}$. The steady-state heat current can be written in terms of the bath imbalance $\Delta \bar n\equiv \bar n_1-\bar n_2$. Introducing the cooperativity $\mathcal{C}\equiv \frac{4J^2}{\gamma_1\gamma_2}$, the dimer current takes the compact form
    \begin{equation}\label{eq:Q_dimer}
    Q_{1\mapsto2}=
    \frac{\mathcal{C}}{1+\mathcal{C}}\,
    \frac{\gamma_1\gamma_2}{\gamma_1+\gamma_2}\,
    \Delta \bar n,
    \end{equation}
    showing explicitly how the coupling-induced hybridization (through $\mathcal{C}$) controls the efficiency of heat transfer between baths.

    \subsection{Flux-threaded loop: Cross correlations}\label{supp:subsec:cross-correlations}	

    The chiral flow of thermal excitations underlying the thermal redistributions discussed in the previous section is directly tied to complex (gauge-invariant) phase patterns in the eigenmodes of the hopping matrix $\boldsymbol{\mathcal{A}}$. In practice, these phases can be accessed from steady-state cross correlations and their associated cross spectral densities.
    The cross spectral densities are defined as the off-diagonal $(i,j)$ elements of the spectral matrix in Eq.~\eqref{ctd:eq:spectral-matrix},
    \begin{equation}\label{eq:Sij_def}
    S_{i,j}(\omega)\equiv \big[\boldsymbol{\mathcal{S}}_{\mathbf{a}\mathbf{a}^\dagger}(\omega)\big]_{ij}
    =\int_{-\infty}^{\infty}e^{-i\omega\tau}\,R_{ij}(\tau)\,\dd\tau,
    \end{equation}
    and can be evaluated directly from the closed-form expression \eqref{ctd:eq:spectral-matrix}. 
    We rewrite $S_{i,j}(\omega)$ explicitly in the eigenmode basis of matrix $\boldsymbol{\mathcal{M}}\equiv -i\boldsymbol{\mathcal{A}}-\frac{\boldsymbol{\gamma}}{2}$. In terms of right/left eigenvectors $\mathbf{r}_\mu$, $\mathbf{l}_\mu$ and eigenvalues $\lambda_\mu$ of $\boldsymbol{\mathcal{M}}$, such that $\boldsymbol{\mathcal{M}}\mathbf{r}_\mu=\lambda_\mu\mathbf{r}_\mu$, $\mathbf{l}_\mu^\dagger\boldsymbol{\mathcal{M}}=\lambda_\mu\mathbf{l}_\mu^\dagger$ and $\mathbf{l}_\mu^\dagger\mathbf{r}_\nu=\delta_{\mu\nu}$, the susceptibility admits the expansion
    \begin{equation}\label{eq:chi_modeexp}
    \boldsymbol{\chi}(\omega)=\sum_\mu \frac{\mathbf{r}_\mu\mathbf{l}_\mu^\dagger}{-i\omega-\lambda_\mu},
    \qquad
    \boldsymbol{\chi}^\dagger(\omega)=\sum_\nu \frac{\mathbf{l}_\nu\mathbf{r}_\nu^\dagger}{i\omega-\lambda_\nu^*},
    \end{equation}
    Therefore, Eq.~\eqref{ctd:eq:spectral-matrix} gives the spectral matrix in the form
    \begin{equation}\label{eq:S_modeexp}
    \boldsymbol{\mathcal{S}}_{\mathbf{a}\mathbf{a}^\dagger}(\omega)
    =\sum_{\mu,\nu}\frac{\mathbf{r}_\mu\;\big(\mathbf{l}_\mu^\dagger\mathbf{D}\mathbf{l}_\nu\big)\;\mathbf{r}_\nu^\dagger}
    {\big(-i\omega-\lambda_\mu\big)\big(i\omega-\lambda_\nu^*\big)}.
    \end{equation}
    Taking matrix elements yields
    \begin{equation}\label{eq:Sij_modeexp}
    S_{i,j}(\omega)
    =\sum_{\mu,\nu}\frac{r_{\mu,i}\,r_{\nu,j}^*}{\big(-i\omega-\lambda_\mu\big)\big(i\omega-\lambda_\nu^*\big)}
    \sum_{m} D_{mm}\,(\ell_{\mu,m})^*\,\ell_{\nu,m},
    \end{equation}
    where $r_{\mu,i}$ is the $i$-th component of $\mathbf{r}_\mu$, $\ell_{\mu,m}$ the $m$-th component of $\mathbf{l}_\mu$ and $D_{mm}=\gamma_m(\bar n_m+1)$. The phases of the off-diagonal elements $S_{i,j}(\omega)$ provide a frequency-resolved probe of the complex phase distribution of the eigenmodes of $\boldsymbol{\mathcal{A}}$, and thereby of the chirality responsible for the directed thermal transport discussed above.
    In the equal damping case $\gamma_i=\gamma$ (so $\boldsymbol{\gamma}=\gamma\mathbb{1}$), $\boldsymbol{\mathcal{M}}$ is normal whenever $\boldsymbol{\mathcal{A}}$ is Hermitian, and one can choose $\mathbf{l}_\mu=\mathbf{r}_\mu\equiv \mathbf{u}_\mu$ with real eigenfrequencies $\omega_\mu$ of $\boldsymbol{\mathcal{A}}$ and drift eigenvalues $\lambda_\mu=-i\omega_\mu-\gamma/2$. If, moreover, all baths have the same occupation $\bar n_m=\bar n$, then $\mathbf{D}=\gamma(\bar n+1)\mathbb{1}$ and \eqref{eq:Sij_modeexp} simplifies to the diagonal mode sum
    \begin{equation}\label{eq:Sij_simplified}
    S_{i,j}(\omega)
    =\gamma(\bar n+1)\sum_{\mu}\frac{u_{\mu,i}\,u_{\mu,j}^*}{(\omega-\omega_\mu)^2+(\gamma/2)^2},
    \end{equation}
    making explicit that the phases of $S_{i,j}(\omega)$ track the phase structure of the eigenmodes $\mathbf{u}_\mu$.
    For the three-site ring with uniform hopping $J$ and Peierls phases $\pm\Phi/3$ on each bond,
    \begin{equation}\label{eq:A_trimer}
    \mathcal{A}_{j,j+1}=J\,e^{+i\Phi/3},\qquad
    \mathcal{A}_{j+1,j}=J\,e^{-i\Phi/3},
    \qquad (j\ \mathrm{mod}\ 3),
    \end{equation}
    the matrix $\boldsymbol{\mathcal{A}}$ is diagonalized by the discrete Fourier transform. Writing the quasi-momenta as
    \begin{equation}
    k_\mu=\frac{2\pi}{3}\mu,\qquad \mu\in\{-1,0,1\},
    \end{equation}
    the normalized eigenvectors can be chosen as plane waves
    \begin{equation}\label{eq:u_trimer}
    u_{\mu,j}=\frac{1}{\sqrt{3}}\,e^{ik_\mu j},
    \qquad j\in\{-1,0,1\},
    \end{equation}
    with dispersion $\omega_\mu=2J\cos\!\left(k_\mu+\frac{\Phi}{3}\right)$.
    Inserting \eqref{eq:u_trimer} into \eqref{eq:Sij_simplified} yields
    \begin{equation}\label{eq:Sij_trimer}
    S_{i,j}(\omega)
    =\frac{\gamma(\bar n+1)}{3}\sum_{\mu=\{-1,0,1\}}
    \frac{e^{i k_\mu (i-j)}}{(\omega-\omega_\mu)^2+\frac{\gamma^2}{4}},
    \end{equation}
    so that the relative phase between neighbouring sites is set by the Fourier factors $e^{\pm i2\pi/3}$ weighted by the Lorentzian contributions of each chiral eigenmode.
    Time-reversal symmetry is effectively restored at parameter values where the eigenmodes can be chosen real up to a gauge (or where degeneracies allow such a choice), so that the cross phases become trivial. Away from these points, the cross spectral densities retain nontrivial relative phases (e.g.\ $\pm 2\pi/3$ for nearest neighbours in a three-site ring), reflecting the underlying chiral eigenmodes.

\subsection{Flux-threaded loop: stationary occupations}\label{supp:subsec:stationary-occupations}

    In this section we solve for the steady-state normally ordered correlators
    \begin{equation}
    C_{ij}=\langle a_i^\dagger a_j\rangle_{t\rightarrow\infty}
    \end{equation}
    in the three-site ring, and use them to obtain the stationary occupations and internal bond currents. For a uniform ring with site labels $j\in\{1,2,3\}$, hopping amplitude $J$, and total synthetic flux $\Phi$, the hopping matrix is
    \begin{equation}
    \mathcal A_{j,j+1}=J e^{i\Phi/3},\qquad
    \mathcal A_{j+1,j}=J e^{-i\Phi/3},
    \end{equation}
    where $j+1$ is understood modulo $3$. The steady-state current from site $i$ to site $j$ is then
    \begin{equation}
    \langle Q_{i\to j}\rangle_{t\rightarrow\infty}
    =2\,\mathrm{Im}\!\left\{\mathcal A_{ij} C_{ij}\right\}.
    \end{equation}
	In the high-temperature limit, where $\bar{n}_{i}\simeq k_{B}T/\Omega_{i}$ and $\sum_{i}\gamma_i\langle a_{i}^{\dagger}a_{i}\rangle_{t\rightarrow\infty}=\sum_{i}\gamma_ik_{B}T/\Omega_{i}$,  which does not neccessarily ensures that, individually, $\Omega_{i}\langle a_{i}^{\dagger}a_{i}\rangle_{t\rightarrow\infty}=k_{B}T$. When the currents vanish ($J=0$) modes become independent, and the average energy of the modes is equal to $k_BT$ (equipartition theorem):
	$\langle a_{i}^{\dagger}a_{i}\rangle_{t\rightarrow\infty}=k_{B}T/\Omega_{i}$. 
    The steady-state equations for the correlators close within the vectorized basis
    \begin{equation}
    \mathbf v
    =
    \big(
    C_{21},\,C_{32},\,C_{13},\,C_{23},\,C_{31},\,C_{12},\,C_{11},\,C_{22},\,C_{33}
    \big)^T,
    \end{equation}
    which obeys 
    \begin{equation}
    \dot{\mathbf v}=\boldsymbol{\mathcal{M}}\,\mathbf v+\mathbf f.
    \label{eq:thermalization_evo}
    \end{equation} Here
	\begin{align}
	\boldsymbol{\mathcal{M}}=\left(\begin{array}{ccccccccc}
	-\frac{\gamma_1}{2}-\frac{\gamma_2}{2} & 0 & 0 & -iJe^{\frac{i\Phi}{3}} & iJe^{\frac{i\Phi}{3}} & 0 & iJe^{-\frac{i\Phi}{3}} & -iJe^{-\frac{i\Phi}{3}} & 0\\
	0 & -\frac{\gamma_2}{2}-\frac{\gamma_3}{2} & 0 & 0 & -iJe^{\frac{i\Phi}{3}} & iJe^{\frac{i\Phi}{3}} & 0 & iJe^{-\frac{i\Phi}{3}} & -iJe^{-\frac{i\Phi}{3}}\\
	0 & 0 & -\frac{\gamma_1}{2}-\frac{\gamma_3}{2} & iJe^{\frac{i\Phi}{3}} & 0 & -iJe^{\frac{i\Phi}{3}} & -iJe^{-\frac{i\Phi}{3}} & 0 & iJe^{-\frac{i\Phi}{3}}\\
	-iJe^{-\frac{i\Phi}{3}} & 0 & iJe^{-\frac{i\Phi}{3}} & -\frac{\gamma_2}{2}-\frac{\gamma_3}{2} & 0 & 0 & 0 & -iJe^{\frac{i\Phi}{3}} & iJe^{\frac{i\Phi}{3}}\\
	iJe^{-\frac{i\Phi}{3}} & -iJe^{-\frac{i\Phi}{3}} & 0 & 0 & -\frac{\gamma_1}{2}-\frac{\gamma_3}{2} & 0 & iJe^{\frac{i\Phi}{3}} & 0 & -iJe^{\frac{i\Phi}{3}}\\
	0 & iJe^{-\frac{i\Phi}{3}} & -iJe^{-\frac{i\Phi}{3}} & 0 & 0 & -\frac{\gamma_1}{2}-\frac{\gamma_2}{2} & -iJe^{\frac{i\Phi}{3}} & iJe^{\frac{i\Phi}{3}} & 0\\
	iJe^{\frac{i\Phi}{3}} & 0 & -iJe^{\frac{i\Phi}{3}} & 0 & iJe^{-\frac{i\Phi}{3}} & -iJe^{-\frac{i\Phi}{3}} & -\gamma_1 & 0 & 0\\
	-iJe^{\frac{i\Phi}{3}} & iJe^{\frac{i\Phi}{3}} & 0 & -iJe^{-\frac{i\Phi}{3}} & 0 & iJe^{-\frac{i\Phi}{3}} & 0 & -\gamma_2 & 0\\
	0 & -iJe^{\frac{i\Phi}{3}} & iJe^{\frac{i\Phi}{3}} & iJe^{-\frac{i\Phi}{3}} & -iJe^{-\frac{i\Phi}{3}} & 0 & 0 & 0 & -\gamma_3
	\end{array}\right),
	\end{align}
	and the driving vector due to the independent thermal baths is $\mathbf f=(0, 0, 0, 0, 0, 0, \gamma_1\bar{n}_{1} , \gamma_2\bar{n}_{2}, \gamma_3\bar{n}_{3})^{T}$. 
    The steady-state solution is therefore
    \begin{equation}
    \mathbf v_{t\rightarrow\infty}=-\boldsymbol{\mathcal{M}}^{-1}\mathbf f.
    \end{equation}
    Although in general both the damping rates $\gamma_j$ and bath occupations $\bar n_j$ differ from site to site in the experiment, useful analytical expressions are obtained in the symmetric case $\gamma_j=\gamma$. One then finds
    \begin{subequations}\label{eq:solution_eql_gamma_ring}:
		\begin{align}\label{eq:solution_eql_gamma}
		\langle n_1\rangle_{t\rightarrow\infty}
        ={}&\bar{n}_{1}+\frac{2J^{2}\left[\left(9J^{4}\cos(2\Phi)-\left(\gamma^{4}+9J^{4}+9\gamma^{2}J^{2}\right)\right)(2\text{\ensuremath{\bar{n}_{1}}}-\text{\ensuremath{\bar{n}_{2}}}-\text{\ensuremath{\bar{n}_{3}}})+3\gamma J\left(\gamma^{2}+3J^{2}\right)(\text{\ensuremath{\bar{n}_{2}}}-\text{\ensuremath{\bar{n}_{3}}})\sin(\Phi)\right]}{\gamma^{6}+54J^{6}\left(1-\cos(2\Phi)\right)+81\gamma^{2}J^{4}+18\gamma^{4}J^{2}},\hspace{-2mm}\\
		\langle n_2\rangle_{t\rightarrow\infty}
        ={}&\bar{n}_{2}-\frac{2J^{2}\left[\left(9J^{4}\cos(2\Phi)-\left(\gamma^{4}+9J^{4}+9\gamma^{2}J^{2}\right)\right)(\text{\ensuremath{\bar{n}_{1}}}-2\text{\ensuremath{\bar{n}_{2}}}+\text{\ensuremath{\bar{n}_{3}}})+3\gamma J\left(\gamma^{2}+3J^{2}\right)(\text{\ensuremath{\bar{n}_{1}}}-\text{\ensuremath{\bar{n}_{3}}})\sin(\Phi)\right]}{\gamma^{6}+54J^{6}\left(1-\cos(2\Phi)\right)+81\gamma^{2}J^{4}+18\gamma^{4}J^{2}},\hspace{-2mm}
		\end{align}
		\end{subequations}
		The third occupation follows from particle-number balance,
        \begin{equation}
        \langle n_3\rangle_{t\rightarrow\infty}
        =
        (\bar n_1+\bar n_2+\bar n_3)-(\langle n_1\rangle_{t\rightarrow\infty}+\langle n_2\rangle_{t\rightarrow\infty}).
        \end{equation}
        The corresponding steady-state bond currents are
        \begin{subequations}\label{eq:ring_currents_ss}
    		\begin{align}
    		\langle Q_{2\to1}\rangle_{t\rightarrow\infty}=&\frac{2 \gamma  J^2 \left(9 J^4 (\bar{n}_2-\bar{n}_1) \cos (2 \Phi )-\gamma  J \left(\gamma ^2+3 J^2\right) \sin (\Phi ) (\bar{n}_1+\bar{n}_2-2 \bar{n}_3)+\left(\gamma ^4+9 J^4+9 \gamma ^2 J^2\right) (\bar{n}_1-\bar{n}_2)\right)}{54 J^6 \cos (2 \Phi )-\left(\gamma ^2+6 J^2\right) \left(\gamma ^4+9 J^4+12 \gamma ^2 J^2\right)},\\
    		\langle Q_{1\to3}\rangle_{t\rightarrow\infty}=&\frac{2 \gamma  J^2 \left((\bar{n}_1-\bar{n}_3) \left(-\gamma ^4+9 J^4 \cos (2 \Phi )-9 J^4-9 \gamma ^2 J^2\right)-\gamma  J \left(\gamma ^2+3 J^2\right) \sin (\Phi ) (\bar{n}_1-2 \bar{n}_2+\bar{n}_3)\right)}{54 J^6 \cos (2 \Phi )-\left(\gamma ^2+6 J^2\right) \left(\gamma ^4+9 J^4+12 \gamma ^2 J^2\right)},\\
    		\langle Q_{3\to2}\rangle_{t\rightarrow\infty}=&-\frac{2 \gamma  J^2 \left(\gamma  J \left(\gamma ^2+3 J^2\right) \sin (\Phi ) (-2 \bar{n}_1+\bar{n}_2+\bar{n}_3)+(\bar{n}_2-\bar{n}_3) \left(-\gamma ^4+9 J^4 \cos (2 \Phi )-9 J^4-9 \gamma ^2 J^2\right)\right)}{54 J^6 \cos (2 \Phi )-\left(\gamma ^2+6 J^2\right) \left(\gamma ^4+9 J^4+12 \gamma ^2 J^2\right)}.
    		\end{align}
    	\end{subequations}
        These expressions satisfy the steady-state continuity equations,
        \begin{align}
        \gamma \bar n_1+\langle Q_{2\to1}\rangle_{t\rightarrow\infty}-\langle Q_{1\to3}\rangle_{t\rightarrow\infty} &= \gamma\langle n_1\rangle_{t\rightarrow\infty},
        \\
        \gamma \bar n_2+\langle Q_{3\to2}\rangle_{t\rightarrow\infty}-\langle Q_{2\to1}\rangle_{t\rightarrow\infty} &= \gamma\langle n_2\rangle_{t\rightarrow\infty},
        \\
        \gamma \bar n_3+\langle Q_{1\to3}\rangle_{t\rightarrow\infty}-\langle Q_{3\to2}\rangle_{t\rightarrow\infty} &= \gamma\langle n_3\rangle_{t\rightarrow\infty},
        \end{align}
        in agreement with Eq.~\eqref{eq:continuity_full}.
	   For zero flux, \(\Phi=0\), the occupations simplify to
    \begin{subequations}\label{eq:eqs_th_sim}
    \begin{align}
    \langle n_1\rangle_{t\rightarrow\infty}^{(\Phi=0)}
    &=
    \bar n_1-\frac{2J^2}{\gamma^2+9J^2}\big[(\bar n_1-\bar n_2)+(\bar n_1-\bar n_3)\big],
    \\
    \langle n_2\rangle_{t\rightarrow\infty}^{(\Phi=0)}
    &=
    \bar n_2-\frac{2J^2}{\gamma^2+9J^2}\big[(\bar n_2-\bar n_1)+(\bar n_2-\bar n_3)\big],
    \end{align}
    \end{subequations}
    while $\langle n_3\rangle_{t\rightarrow\infty}^{(\Phi=0)}$ follows by cyclic permutation. For $\bar n_1>\bar n_2>\bar n_3$, Eq.~\eqref{eq:eqs_th_sim} shows that heat flows from higher- to lower-occupation baths, with a magnitude set by the occupation imbalance.

    \subsection{Absence of persistent directional heat currents}\label{supp:subsec:persistent-currents}\noindent

    The dynamics of the three-resonator system is governed by the Heisenberg--Langevin equation
    \begin{equation}
    \dot{\mathbf a}(t)
    =
    -i\!\left(\boldsymbol{\mathcal A}-i\frac{\boldsymbol{\gamma}}{2}\right)\mathbf a(t)
    +\mathbf a_{\mathrm{in}}(t),
    \end{equation}
    where $\boldsymbol{\mathcal A}=\boldsymbol{\mathcal A}^\dagger$ is the coherent hopping matrix and $\boldsymbol{\gamma}=\gamma \mathbf I$ corresponds to identical local losses into independent baths. The equal-time anti-normally ordered correlator
    \begin{equation}
    \mathbf R_0=\langle \mathbf a(t)\mathbf a^\dagger(t)\rangle_{t\to\infty}
    \end{equation}
    then satisfies the steady-state Eq.~\eqref{ctd:eq:lyapunov_ss}
    \begin{equation}\label{ctd:eq:lyapunov_ss_3_modes}
    i\!\left(\boldsymbol{\mathcal A}-i\frac{\boldsymbol{\gamma}}{2}\right)\mathbf R_0
    -i\,\mathbf R_0\!\left(\boldsymbol{\mathcal A}+i\frac{\boldsymbol{\gamma}}{2}\right)
    =
    \mathbf D,
    \end{equation}
    with diffusion matrix $\mathbf D=\mathrm{diag}\bigl(\gamma_1(\bar n_1^{\rm th}+1),\gamma_2(\bar n_2^{\rm th}+1),\gamma_3(\bar n_3^{\rm th}+1)\bigr)$.
    In equilibrium all baths have the same occupation, $\bar n_i^{\rm th}=\bar n^{\rm th}$, and since we also have chosen $\gamma_i=\gamma$, one has
    \begin{equation}
    \mathbf D=\gamma(\bar n^{\rm th}+1)\mathbf I.
    \end{equation}
    By direct substitution one verifies that $\mathbf R_0=(\bar n^{\rm th}+1)\mathbf I$  solves Eq.~\eqref{ctd:eq:lyapunov_ss_3_modes}, since
    \begin{equation}
    i\!\left(\boldsymbol{\mathcal A}-i\frac{\boldsymbol{\gamma}}{2}\right)\mathbf R_0
    -i\,\mathbf R_0\!\left(\boldsymbol{\mathcal A}+i\frac{\boldsymbol{\gamma}}{2}\right)
    =
    i(\bar n^{\rm th}+1)\left[
    \left(\boldsymbol{\mathcal A}-i\frac{\boldsymbol{\gamma}}{2}\right)
    -
    \left(\boldsymbol{\mathcal A}+i\frac{\boldsymbol{\gamma}}{2}\right)
    \right]
    =
    (\bar n^{\rm th}+1)\boldsymbol{\gamma}
    =
    \mathbf D.
    \end{equation}
    Since $\boldsymbol{\gamma}$ is positive definite, the dynamics are stable and this steady state is unique \cite{Purkayastha2022}.
    The heat current along a bond $j\to k$ is determined by the corresponding steady-state coherence,
    \begin{equation}
    \langle Q_{j\to k}\rangle_{t\rightarrow\infty}
    =
    2\,\mathrm{Im}\!\left\{\mathcal A_{jk} C_{jk,0}\right\},
    \end{equation}
    where $C_{jk,0}=\langle a_j^\dagger a_k\rangle_{t\to\infty}$. Equivalently, for $j\neq k$, this may be written in terms of $\mathbf R_0$, since $C_{jk,0}=R_{kj,0}$. 
    Because $\mathbf R_0$ is diagonal in equilibrium, all off-diagonal coherences vanish, and therefore no persistent directional heat current can exist.
    Instead, a persistent equilibrium directional heat current requires non-local dissipation, for instance dissipative coupling between resonators and one or more shared baths, such that the damping and diffusion matrices, $\boldsymbol{\gamma}$ and $\mathbf D$, acquire off-diagonal elements. In that case the Eq.~\eqref{ctd:eq:lyapunov_ss}, can sustain non-vanishing off-diagonal coherences. More specifically, the off-diagonal $(j,k)$ component of Eq.~\eqref{ctd:eq:lyapunov_ss}, with $j\neq k$, reads
    \begin{equation}
    -i\,\mathcal A_{jk}\bigl((R_0)_{kk}-(R_0)_{jj}\bigr)
    -\frac{\gamma_{jk}}{2}\bigl((R_0)_{kk}+(R_0)_{jj}\bigr)
    +\langle Q_{j\to k}\rangle_{t\rightarrow\infty}
    =
    D_{jk},
    \end{equation}
    where
    \begin{equation}
    \langle Q_{j\to k}\rangle_{t\rightarrow\infty}
    =
    2\,\mathrm{Im}\!\left\{\mathcal A_{jk}(C_0)_{jk}\right\},
    \qquad
    (C_0)_{jk}=\langle a_j^\dagger a_k\rangle_{t\to\infty}.
    \end{equation}
    For independent local baths one has $\gamma_{jk}=0$ and $D_{jk}=0$ for $j\neq k$, so in equilibrium the only solution is again vanishing off-diagonal coherence. By contrast, shared baths or dissipative inter-mode couplings generate off-diagonal terms in $\boldsymbol{\gamma}$ and $\mathbf D$, which can support finite steady-state coherences and thereby persistent equilibrium directional currents.

\subsection{Nonreciprocity for optimal mixing}\label{supp:subsec:mixing}\noindent

Let us now consider the impact of the coupling phases on the mixing rate, \emph{i.e.} how well the temperatures of the different resonators are equilibrated due to coupling. To be specific, we are interested in the average bath occupation $\bar{R}$ and deviation $\Delta_i$ given by
\begin{equation}
\bar{R} = \frac{1}{3} \sum R_{jj}, \quad \Delta_j = R_{jj} - \bar{R}
\end{equation}
Let us now move to the eigenmode basis by diagonalizing the Lyapunov equation. We write $\mathbf{U}^\dagger \boldsymbol{\mathcal A} \mathbf{U} = \mathbf{\Lambda}$, which yields
\begin{equation}\label{ctd:eq:lyapunov_ss_3_modes_diag}
    i\!\left(\mathbf{\Lambda}  -i\frac{\gamma}{2} \mathbf{I}  \right) \tilde{\mathbf R}_0
    -i\, \tilde{\mathbf R}_0\!\left( \mathbf{\Lambda} + i\frac{\gamma}{2} \mathbf{I} \right)
    =
    \tilde{\mathbf D},
    \end{equation}
where $\tilde{\mathbf D} = \mathbf{U}^\dagger \mathbf{DU}$, $\tilde{\mathbf R}_0 = \mathbf{U}^\dagger \mathbf{R}_0 \mathbf{U}$ and $\mathbf{\Lambda} = \text{diag}(\omega_\mu)$. If we now write this for a single element of $\tilde{\mathbf R}_0$, we find
\begin{equation}
i(\omega_\mu - i\frac{\gamma}{2})\tilde{R}_{\mu\nu} -i \tilde{R}_{\mu \nu} (\omega_\nu - i\frac{\gamma}{2}) = \tilde{D}_{\mu \nu},
\end{equation}
which results in
\begin{equation}
\tilde{R}_{\mu \nu} = \frac{\tilde{D}_{\mu \nu}}{\gamma + i (\omega_\mu-\omega_\nu)}.
\end{equation}
We are interested in the temperatures of the resonances $R_{ii}$, which are related to the eigenmodes as
\begin{equation}
R_{jj} = \sum_{\mu,\nu} u_{\mu,j} \tilde{R}_{\mu \nu} u_{\nu,j}^*
\end{equation}
via the inverse transformation. We can split this into its diagonal and off-diagonal elements,
\begin{equation}
R_{jj} = \sum_\mu |u_{\mu,j}|^2 \tilde{R}_{\mu \mu} + \sum_{\mu \neq \nu} u_{\mu,j} \tilde{R}_{\mu \nu} u_{\nu,j}^*, 
\end{equation}
where the first term is due to the thermal occupation of the eigenmodes of the system, while the second is due to thermal coherence between the different eigenmodes.
The eigenmode occupations are given by, using $\mathbf D=\gamma \, \mathrm{diag}\bigl(\bar n_1^{\rm th}+1,\bar n_2^{\rm th}+1,\bar n_3^{\rm th}+1\bigr)$,
\begin{equation}
\tilde{R}_{\mu \mu} = \frac{\tilde{D}_{\mu \mu}}{\gamma} = \sum_j |u_{\mu,j}|^2 ( \bar n_{j}^{\rm th}+1)
\end{equation}
For the loop with equal coupling between each site, we previously gave $u_{\mu,j}=\frac{1}{\sqrt{3}}\,e^{ik_\mu j}$, with $k_\mu = 2\pi \mu/3 $ and $j\in\{-1,0,1\}$ (Eq.~\ref{eq:u_trimer}), so $|u_{\mu,j}|^2 = 1/3$. We thus find
\begin{equation}
\tilde{R}_{\mu \mu} = \bar{n}^{th}_{av} + 1 = \bar{R}
\end{equation}
where $\bar{n}^{th}_{av}$ is the average occupation, for all $p$: each eigenmode has the same thermal occupation. This is of course equal to the average occupation $\bar{R}$ since the diagonalization is unitary. Hence, we find for the resonance occupations
\begin{equation}
R_{jj} = \bar{n}^{th}_{av} + 1 + \sum_{\mu \neq \nu}   u_{\mu,j} u_{\nu,j}^* \tilde{R}_{\mu\nu}.
\end{equation}
This yields for the difference in occupation
\begin{equation}
\Delta_j = \sum_{\mu \neq \nu} u_{\mu,j} u_{\nu,j}^* \tilde{R}_{\mu \nu}.
\end{equation}
Hence, any difference in resonator temperature is due to coherence between the eigenmodes of the system. We can write $u_{\mu,j} u_{\nu,j}^* = e^{i(k_\mu -k_\nu)j}/3$, which yields
\begin{equation}
\Delta_j = \sum_{\mu \neq \nu} \frac{ e^{i(k_\mu -k_\nu)j}}{3} \tilde{R}_{\mu \nu}.
\end{equation}
This lets us simplify to
\begin{equation}
\Delta_j = \frac{2}{3}  \Re \bigl(  e^{-i\frac{2\pi }{3}j}\Xi \bigr), \qquad \Xi = \tilde{R}_{-1, 0} + \tilde{R}_{0, 1} + \tilde{R}_{-1, 1},
\end{equation}
where we have used that each term in $\Xi$ has the same phase factor $\frac{ e^{i(k_\mu -k_\nu)j}}{3} = e^{-i\frac{2\pi }{3}j}$, and that the terms $\tilde{R}_{0,-1}$ etc. are conjugate. Defining $\Xi = |\Xi|e^{i\xi}$, we can further simplify to 
\begin{equation}
\Delta_j = \frac{2}{3} | \Xi | \cos \biggl( \xi - \frac{2\pi}{3}j \biggr).
\end{equation}
Using $\cos(\xi)^2 + \cos(\xi-2\pi/3)^2 + \cos(\xi+2\pi/3)^2 = 3/2$, we can write for the rms occupation difference:
\begin{equation}
\frac{1}{3} \sum \Delta_j^2 = \frac{2}{9} |\Xi|^2 = \frac{2}{9} \biggl| \frac{\tilde{D}_{-1,0}}{\gamma + i (\omega_{-1}-\omega_0)} + \frac{\tilde{D}_{0,1}}{\gamma + i (\omega_0-\omega_1)} + \frac{\tilde{D}_{1,-1}}{\gamma + i (\omega_1-\omega_{-1})} \biggr|^2
\end{equation}
This expression already indicates that, if there are two degenerate modes independent of the coupling rate $J$, then total equilibration is impossible since one of the terms would become $\tilde{D}_{\mu , \nu}/\gamma$. Let's consider the explicit eigenvalues $\omega_\mu = 2 J \cos(k_\mu + \Phi/3)$ for the three-site ring with uniform $J$, and using that 
\begin{equation}
\tilde{D}_{\mu,\nu} = (\gamma/3)  \sum (\bar{n}_j +1)e^{i (k_\mu -k_\nu)j}
\end{equation}
We will consider one resonator to be coupled to a hot bath, with $j=0$ and $\bar{n}_0 = \bar{n}_h$. The other two resonators are coupled to a cold bath, $\bar{n}_{-1}=\bar{n}_{1}=\bar{n}_{c}$. This particular choice of hot bath location yields a simplified expression
\begin{equation}
\tilde{D}_{\mu,\nu} = \frac{\gamma (\bar{n}_h - \bar{n}_c) }{3}  
\end{equation}
We can use this to further simplify the rms mixing difference as a function of the phase $\Phi$
\begin{equation}
\frac{1}{3} \sum \Delta_j^2 = \frac{2\gamma^2(\bar{n}_h - \bar{n}_c) ^2}{81} \biggl| \frac{1}{\gamma + i \delta_{-1,0}} + \frac{1}{\gamma + i \delta_{0,1}} + \frac{1}{\gamma + i \delta_{1,-1}} \biggr|^2
\end{equation}
with $\delta_{-1,0} = \omega_{-1} - \omega_0 = 2 J \bigl( \cos(-2\pi/3 + \Phi/3) - \cos(\Phi/3) \bigr) = J\bigl( \sqrt{3} \sin(\Phi/3) - \cos(\Phi/3)\bigr)$, $\delta_{0,1} = J\bigl( \sqrt{3} \sin(\Phi/3) + \cos(\Phi/3)\bigr)$, and $\delta_{1,-1} = -2\sqrt{3}J \sin(\Phi/3)$. 
The largest splitting of eigenmodes for a given $J$ yields the most efficient mixing. We can see that $\Phi=0$ yields $\delta_{1,-1} = 0$, which results in incomplete mixing for any $J$ since coherence always persists between the clockwise and counterclockwise modes. Breaking time-reversal symmetry with $\Phi \neq 0$ lifts the degeneracy, and the smallest $\delta$ is largest for $\Phi=\pi/2$, which results in optimal mixing/cooling for a given $J$.
One can also lift the degeneracy by making one of the links different, but in this case the eigenmodes do not achieve the same temperature, as the eigenvectors are not plane waves with equal weight on each site. As a result, even in the limit of $J\rightarrow \infty$, the resonators will not equilibrate. This is evident in the $\Lambda$-structure in the main text: the frequency splitting is the same as in the $\pi/2$ flux loop, but $\tilde{R}_{\mu \mu}$ is not equal for all $\mu$, and an occupation difference therefore remains.

\subsection{Counterflow between cold resonators}\label{supp:subsec:counterflow}\noindent

Considering again the situation with one hot and two cold reservoirs, as in the previous subsection. For simplicity, we will consider the cold baths to be at zero temperature, and for the bond current we can ignore the vacuum contribution to the occupation. This lets us simply consider the analogous case of a circulator driven by a single input port, which yields for the amplitudes of modes $a_3$ and $a_4$ as described in the main text
\begin{equation}
a_3(\omega) = \chi_{31}(\omega)\xi_1^a(\omega), \qquad a_4(\omega) = \chi_{41}(\omega)\xi_1^a(\omega),
\end{equation}
where $\xi_1^a(\omega)$ is the stochastic input of resonator $a_1$. The bond current between resonators 3 and 4 is given by
\begin{equation}
Q_{3\rightarrow4}=2J \Im ( e^{-i\Phi/3} a_3^* a_4) =2J \Im ( e^{-i\Phi/3} \chi_{31}^* \chi_{41}) \langle |\xi_1^a(\omega)|^2 \rangle.
\end{equation}
We can now write the susceptibilities/Green's functions, as a sum of possible paths, but truncating at the direct and first indirect path:
\begin{equation}
\chi_{31}(\omega) = d e^{i\Phi/3} + l e^{-2i\Phi/3} + \dots, \qquad \chi_{41}(\omega) = d e^{-i\Phi/3} + l e^{2i\Phi/3}  + \dots
\end{equation}
where $d$ and $l$ are the hopping amplitudes via the short side or long side of the loop, and we have ignored all higher order walks. Inserting this into the bond current, we find
\begin{align}
Q_{3\rightarrow4} & \approx 2J \Im \bigl( e^{-i\Phi/3} (d^* e^{-i\Phi/3} + l^* e^{2i\Phi/3})(d e^{-i\Phi/3} + l e^{2i\Phi/3} )\bigr) \langle |\xi_1^a|^2 \rangle \\
& = 2J \langle |\xi_1^a|^2 \rangle \Im \bigl( |d|^2 e^{-i\Phi} + |l|^2 e^{i\Phi} + 2 \Re(d^* l) \bigr) = 2J \langle |\xi_1^a|^2 \rangle \bigl( |l|^2 -|d|^2  \bigr) \sin(\Phi)
\end{align}
It is safe to assume that generally $|l(\omega)|^2 \leq |d(\omega)|^2$, which means that for positive $\Phi$, for which the direction is $1 \rightarrow 3 \rightarrow 4 \rightarrow 1$, we observe a negative bond current between 3 and 4, due to interference between the direct and indirect paths.

\section{Supporting Experimental results}\label{supp:sec:experimental}

\subsection{Supporting measurements: Single resonator}\label{supp:subsec:single-resonator}

\begin{figure*}
	\centering
	\includegraphics[width=\textwidth]{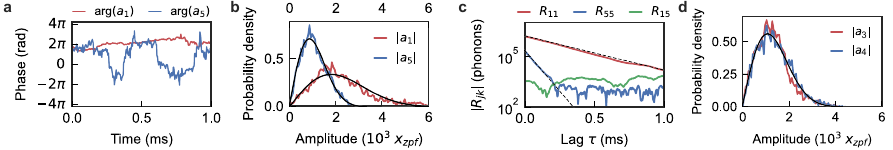}
    \phantomsubfloat{\label{ctd:fig:brownian-uncoupled:spectrum}} 
    \phantomsubfloat{\label{ctd:fig:brownian-uncoupled:time-traces-ampl}} 
    \phantomsubfloat{\label{ctd:fig:brownian-uncoupled:time-traces-phase}} 
    \phantomsubfloat{\label{ctd:fig:brownian-uncoupled:histograms}} 
    \phantomsubfloat{\label{ctd:fig:brownian-uncoupled:correlations}} 
    \vspace{-2\baselineskip}
	\caption{\textbf{Brownian motion of uncoupled nano-optomechanical resonators.}
	\subidc{a} Phase fluctuations $\arg(a_j)$ corresponding to main text Fig.~1\subidc{d} top.
	\subidc{b} Histograms of the amplitudes $|a_j|$ collected over a duration of $10$ ms. Black lines represent the probability densities $f_\text{MB}$ of the corresponding Maxwell-Boltzmann distribution \eqref{ctd:eq:mb-distribution}.
	\subidc{c} Auto-correlations $R_{11}(\tau)$ and $R_{22}(\tau)$ and cross-correlations $R_{12}(\tau)$ as a function of time lag $\tau$. Black dashed lines represent exponentially decaying amplitude correlations with rates $\gamma_j/2$ equal to half the spectral linewidth of the resonances. \subidc{d} Same as \subidc{b}, reconstructed from modes $a_3$ and $a_4$, with similar thermal occupation.
	}
	\label{ctd:fig:brownian-uncoupled}

\end{figure*}

We first characterize the Brownian motion of individual mechanical resonators, following standard optomechanical practice. Representative time traces for the lowest- and highest-frequency modes, $a_1$ and $a_2$, are shown in Fig.~\ref{ctd:fig:brownian-uncoupled:time-traces-ampl} and Fig.~\ref{ctd:fig:brownian-uncoupled:time-traces-phase} for amplitude and phase, respectively. Two features stand out: $a_1$ has the larger average amplitude, while $a_2$ fluctuates faster. The first follows from the frequency-dependent thermal variance,
\begin{equation}
a_{\mathrm{rms},j}^2=\langle x_j^2\rangle
=
2x_{\mathrm{zpf},j}^2 n_{\mathrm{th},j},
\qquad
n_{\mathrm{th},j}=\frac{k_{\mathrm B}T}{\hbar\Omega_j},
\end{equation}
where $T$ is the bath temperature. The second reflects their different coupling strengths to the environment, quantified by the damping rates $\gamma_1/(2\pi)=1.5$ kHz and $\gamma_2/(2\pi)=6.9$ kHz.
A histogram of the amplitudes $|a_j|$ in a time trace, shown in Fig.~\ref{ctd:fig:brownian-uncoupled:histograms}, recovers the thermal Boltzmann distribution $\propto \exp(-E_j/\kb T)$, with
\begin{equation}
E_j=\frac{1}{2}m_j\Omega_j^2 x_j^2+\frac{1}{2}m_j v_j^2,
\end{equation}
where $x_j$, $v_j$, and $m_j$ are the displacement, velocity, and mass of mode $j$. Equivalently, the probability density for the normalized oscillation amplitude $r_j=|a_j|/a_{j,\mathrm{rms}}$, with $a_{j,\mathrm{rms}}^2=\langle |a_j|^2\rangle$, is
\begin{equation}
f_{\mathrm{MB}}(r_j)=2r_j e^{-r_j^2}.
\label{ctd:eq:mb-distribution}
\end{equation}
The sizeable deviations of the experimental histograms from the expected densities, shown as black lines in Fig.~\ref{ctd:fig:brownian-uncoupled:histograms}, arise from the short acquisition time of $10$ ms, only about fifty times longer than the dissipation times $1/\gamma_j$, which are on the order of $0.2$ ms.

\subsection{Time-resolved thermalization measurements}\label{supp:subsec:thermalization}

\begin{figure}
	\centering
	\includegraphics{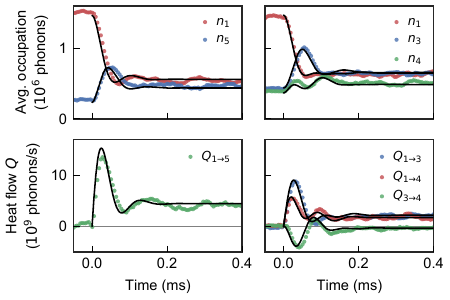}
	\caption{\textbf{Chirality in time-resolved thermalization.} Time-resolved ensemble averages ($1000$ runs) of the resonator occupations (upper row) and heat flows (lower row) in the dimer $a_1\text{-}a_2$ (left column, $J/(2\pi)=5$ kHz) and circulator $a_1\text{-}a_3\text{-}a_4$ (right column, $J/(2\pi) = 3.6$ kHz, $\Phi = \pi/2$). The resonators are in thermal equilibrium with their baths $\nres{j}$ until at $t = 0$ the interactions are switched on and the correlators thermalize following Eq.~\eqref{eq:thermalization_evo} to a new steady-state. The evolution of the occupations and flows reflect the competition between energy transfer with time-scale $1/J$ and thermal relaxation with time-scales $1/\gamma_j$. In the circulator, the heat flows $Q_{3 \mapsto 4}$ between the cold resonators is seen to slosh back and forth until a steady state is reached.
	}
	\label{ctd:fig:time-resolved-thermalization}

\end{figure}

We probe the competition between thermal relaxation and beamsplitter-mediated energy transfer by tracking time-resolved occupations and flows~\cite{Clos2016timeresolved}. In the ensemble-averaged measurements of Fig.~\ref{ctd:fig:time-resolved-thermalization}, the resonators are initially uncoupled and thermalized with their individual baths. At $t=0$, the interactions are switched on, and the system relaxes toward the coupled steady state.
For the dimer $a_1$-$a_2$, the initial occupation imbalance produces a large transient energy current that overshoots the final steady-state flow. The cold resonator $a_2$ even becomes temporarily more occupied than the hot resonator $a_1$. In the limit of large $J$, a pulsed interaction of duration $\pi/J$ would fully swap the resonator states \cite{Verhagen2012quantumcoherent}. Here, relaxation on the scale $1/\gamma_j$ instead drives the system to a steady state with continuous heat flow from $a_1$ to $a_2$.
Similar dynamics appear in the circulator $a_1$-$a_3$-$a_4$ with TRS-breaking flux $\Phi=\pi/2$: thermal energy sloshes between the cold resonators $a_3$ and $a_4$ before reaching a steady-state flow.

\subsection{Filtered instrument noise in demodulated correlations}\label{supp:subsec:noise}

\begin{figure*}
	\centering
	\includegraphics{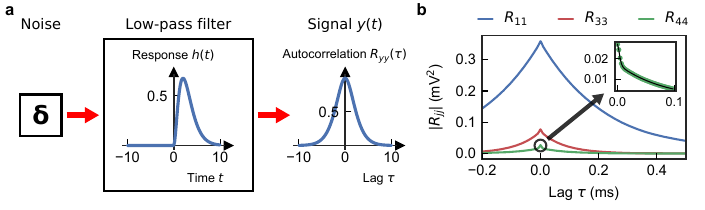}
	\phantomsubfloat{\label{ctd:fig:noise-autocorr:a}} 
	\phantomsubfloat{\label{ctd:fig:noise-autocorr:b}} 
    \vspace{-2\baselineskip}
	\caption{\textbf{Instrument noise in auto-correlations.} \subidc{a} In the lock-in amplifier used to analyze thermomechanical signals, $\delta$-correlated instrument noise is filtered by a third-order low-pass filter (response $h(t)$ shown for unity time constant) in the demodulation process. This introduces time-correlations in the demodulated signal $y(t)$ and results in a peaked auto-correlation $R_{yy}(\tau) \propto h(\tau) \star h(-\tau) $ given by the convolution of $h(t)$ with itself. \subidc{b} Measured auto-correlation voltages $|R_{jj}(\tau)|$ for three uncoupled resonators. On top of the long-time exponential thermal decay of amplitude correlations, there is a short-time contribution of filtered noise correlations. The inset zooms in on $|R_{44}(\tau)|$ at the short time scale and shows that it is described well by the (uncorrelated) sum of the filtered noise auto-correlation and the exponential thermal decay (black line). The effect of the low-pass filter on the thermal correlations is negligible due to their different time constants $\tau_\text{filter} = 1.62$ µs and $\tau_\text{th} = 2/\gamma_4 = 81$ µs. Cross-correlations between resonator signals at different frequencies do not suffer from the frequency-uncorrelated instrument noise.
	}
	\label{ctd:fig:noise-autocorr}

\end{figure*}

In the measurement of the resonators' thermal displacement, the dominant noise source is electronic noise in our instruments: a detector and a lock-in amplifier. The lock-in amplifier is used to demodulate the thermomechanical signals, and in doing so passes the voltage signal it has measured (including the instrument noise) through a digital low-pass filter. Assuming that the instrument noise is $\delta$-correlated with spectral density $S_{nn}$ (at least around the MHz frequencies of our resonators), this introduces time-correlations in the demodulated noise signal $y(t)$. Depicted in Fig.~\ref{ctd:fig:noise-autocorr}, this leads to a noise auto-correlation $R_{yy}(\tau) = S_{nn} (h(\tau) \star h(-\tau))$ that scales with the convolution of the low-pass filter response $h(t)$ with itself. The instrument noise gets added to the thermal auto-correlations $R_{jj}$ that we are interested in, as demonstrated in Fig.~\ref{ctd:fig:noise-autocorr:b}. In all auto-correlations shown in the manuscript, the noise contributions is corrected for.

\section*{Supplemental References}

\setcounter{NAT@ctr}{0}
\begin{list}{[\arabic{enumiv}]}{
  \usecounter{enumiv}
  \setlength{\leftmargin}{2.5em}
  \setlength{\labelwidth}{2em}
  \setlength{\labelsep}{0.5em}
  \setlength{\itemsep}{0pt}
  \setlength{\parsep}{0pt}
}
\makeatletter
\providecommand \@ifxundefined [1]{
 \@ifx{#1\undefined}
}
\providecommand \@ifnum [1]{
 \ifnum #1\expandafter \@firstoftwo
 \else \expandafter \@secondoftwo
 \fi
}
\providecommand \@ifx [1]{
 \ifx #1\expandafter \@firstoftwo
 \else \expandafter \@secondoftwo
 \fi
}
\providecommand \natexlab [1]{#1}
\providecommand \enquote  [1]{``#1''}
\providecommand \bibnamefont  [1]{#1}
\providecommand \bibfnamefont [1]{#1}
\providecommand \citenamefont [1]{#1}
\providecommand \href@noop [0]{\@secondoftwo}
\providecommand \href [0]{\begingroup \@sanitize@url \@href}
\providecommand \@href[1]{\@@startlink{#1}\@@href}
\providecommand \@@href[1]{\endgroup#1\@@endlink}
\providecommand \@sanitize@url [0]{\catcode `\\12\catcode `\$12\catcode `\&12\catcode `\#12\catcode `\^12\catcode `\_12\catcode `\%12\relax}
\providecommand \@@startlink[1]{}
\providecommand \@@endlink[0]{}
\providecommand \url  [0]{\begingroup\@sanitize@url \@url }
\providecommand \@url [1]{\endgroup\@href {#1}{\urlprefix }}
\providecommand \urlprefix  [0]{URL }
\providecommand \Eprint [0]{\href }
\providecommand \doibase [0]{https://doi.org/}
\providecommand \selectlanguage [0]{\@gobble}
\providecommand \bibinfo  [0]{\@secondoftwo}
\providecommand \bibfield  [0]{\@secondoftwo}
\providecommand \translation [1]{[#1]}
\providecommand \BibitemOpen [0]{}
\providecommand \bibitemStop [0]{}
\providecommand \bibitemNoStop [0]{.\EOS\space}
\providecommand \EOS [0]{\spacefactor3000\relax}
\providecommand \BibitemShut  [1]{\csname bibitem#1\endcsname}
\let\auto@bib@innerbib\@empty
\bibitem [{\citenamefont {Gardiner}\ and\ \citenamefont {Zoller}(2004{\natexlab{a}})}]{Gardiner2004}
  \BibitemOpen
  \bibfield  {author} {\bibinfo {author} {\bibfnamefont {C.~W.}\ \bibnamefont {Gardiner}}\ and\ \bibinfo {author} {\bibfnamefont {P.}~\bibnamefont {Zoller}},\ }\href@noop {} {\bibinfo {title} {{Quantum noise : a handbook of Markovian and non-Markovian quantum stochastic methods with applications to quantum optics}}} (\bibinfo {year} {2004}{\natexlab{a}})\BibitemShut {NoStop}
\bibitem [{\citenamefont {Reiter}\ and\ \citenamefont {S{\o}rensen}(2012)}]{Reiter2012effective}
  \BibitemOpen
  \bibfield  {author} {\bibinfo {author} {\bibfnamefont {F.}~\bibnamefont {Reiter}}\ and\ \bibinfo {author} {\bibfnamefont {A.~S.}\ \bibnamefont {S{\o}rensen}},\ }\href {https://doi.org/10.1103/PhysRevA.85.032111} {\bibfield  {journal} {\bibinfo  {journal} {Phys. Rev. A}\ }\textbf {\bibinfo {volume} {85}},\ \bibinfo {pages} {032111} (\bibinfo {year} {2012})}\BibitemShut {NoStop}
\bibitem [{\citenamefont {Gardiner}\ and\ \citenamefont {Zoller}(2004{\natexlab{b}})}]{Gardiner2004quantum}
  \BibitemOpen
  \bibfield  {author} {\bibinfo {author} {\bibfnamefont {C.~W.}\ \bibnamefont {Gardiner}}\ and\ \bibinfo {author} {\bibfnamefont {P.}~\bibnamefont {Zoller}},\ }\href@noop {} {\emph {\bibinfo {title} {Quantum noise: a handbook of {{Markovian}} and non-{{Markovian}} quantum stochastic methods with applications to quantum optics}}},\ \bibinfo {edition} {3rd}\ ed.,\ Springer series in synergetics\ (\bibinfo  {publisher} {{Springer}},\ \bibinfo {address} {{Berlin ; New York}},\ \bibinfo {year} {2004})\BibitemShut {NoStop}
\bibitem [{\citenamefont {Onsager}(1931)}]{Onsager1931reciprocal2}
  \BibitemOpen
  \bibfield  {author} {\bibinfo {author} {\bibfnamefont {L.}~\bibnamefont {Onsager}},\ }\href {https://doi.org/10.1103/PhysRev.38.2265} {\bibfield  {journal} {\bibinfo  {journal} {Phys. Rev.}\ }\textbf {\bibinfo {volume} {38}},\ \bibinfo {pages} {2265} (\bibinfo {year} {1931})}\BibitemShut {NoStop}
\bibitem [{\citenamefont {Meystre}\ and\ \citenamefont {Sargent}(2007)}]{Meystre2007}
  \BibitemOpen
  \bibfield  {author} {\bibinfo {author} {\bibfnamefont {P.}~\bibnamefont {Meystre}}\ and\ \bibinfo {author} {\bibfnamefont {M.}~\bibnamefont {Sargent}},\ }\href {https://doi.org/10.1007/978-3-540-74211-1} {\emph {\bibinfo {title} {Elements of Quantum Optics}}}\ (\bibinfo {year} {2007})\BibitemShut {NoStop}
\bibitem [{\citenamefont {Purkayastha}(2022)}]{Purkayastha2022}
  \BibitemOpen
  \bibfield  {author} {\bibinfo {author} {\bibfnamefont {A.}~\bibnamefont {Purkayastha}},\ }\href {https://doi.org/10.1103/PhysRevA.105.062204} {\bibfield  {journal} {\bibinfo  {journal} {Phys. Rev. A}\ }\textbf {\bibinfo {volume} {105}},\ \bibinfo {pages} {062204} (\bibinfo {year} {2022})}\BibitemShut {NoStop}
\bibitem [{\citenamefont {Clos}\ \emph {et~al.}(2016)\citenamefont {Clos}, \citenamefont {Porras}, \citenamefont {Warring},\ and\ \citenamefont {Schaetz}}]{Clos2016timeresolved}
  \BibitemOpen
  \bibfield  {author} {\bibinfo {author} {\bibfnamefont {G.}~\bibnamefont {Clos}}, \bibinfo {author} {\bibfnamefont {D.}~\bibnamefont {Porras}}, \bibinfo {author} {\bibfnamefont {U.}~\bibnamefont {Warring}},\ and\ \bibinfo {author} {\bibfnamefont {T.}~\bibnamefont {Schaetz}},\ }\href {https://doi.org/10.1103/PhysRevLett.117.170401} {\bibfield  {journal} {\bibinfo  {journal} {Phys. Rev. Lett.}\ }\textbf {\bibinfo {volume} {117}},\ \bibinfo {pages} {170401} (\bibinfo {year} {2016})}\BibitemShut {NoStop}
\bibitem [{\citenamefont {Verhagen}\ \emph {et~al.}(2012)\citenamefont {Verhagen}, \citenamefont {Del{\'e}glise}, \citenamefont {Weis}, \citenamefont {Schliesser},\ and\ \citenamefont {Kippenberg}}]{Verhagen2012quantumcoherent}
  \BibitemOpen
  \bibfield  {author} {\bibinfo {author} {\bibfnamefont {E.}~\bibnamefont {Verhagen}}, \bibinfo {author} {\bibfnamefont {S.}~\bibnamefont {Del{\'e}glise}}, \bibinfo {author} {\bibfnamefont {S.}~\bibnamefont {Weis}}, \bibinfo {author} {\bibfnamefont {A.}~\bibnamefont {Schliesser}},\ and\ \bibinfo {author} {\bibfnamefont {T.~J.}\ \bibnamefont {Kippenberg}},\ }\href {https://doi.org/10.1038/nature10787} {\bibfield  {journal} {\bibinfo  {journal} {Nature}\ }\textbf {\bibinfo {volume} {482}},\ \bibinfo {pages} {63} (\bibinfo {year} {2012})}\BibitemShut {NoStop}
\end{list}

\begin{thebibliography}{60}
\makeatletter
\providecommand \@ifxundefined [1]{
 \@ifx{#1\undefined}
}
\providecommand \@ifnum [1]{
 \ifnum #1\expandafter \@firstoftwo
 \else \expandafter \@secondoftwo
 \fi
}
\providecommand \@ifx [1]{
 \ifx #1\expandafter \@firstoftwo
 \else \expandafter \@secondoftwo
 \fi
}
\providecommand \natexlab [1]{#1}
\providecommand \enquote  [1]{``#1''}
\providecommand \bibnamefont  [1]{#1}
\providecommand \bibfnamefont [1]{#1}
\providecommand \citenamefont [1]{#1}
\providecommand \href@noop [0]{\@secondoftwo}
\providecommand \href [0]{\begingroup \@sanitize@url \@href}
\providecommand \@href[1]{\@@startlink{#1}\@@href}
\providecommand \@@href[1]{\endgroup#1\@@endlink}
\providecommand \@sanitize@url [0]{\catcode `\\12\catcode `\$12\catcode `\&12\catcode `\#12\catcode `\^12\catcode `\_12\catcode `\%12\relax}
\providecommand \@@startlink[1]{}
\providecommand \@@endlink[0]{}
\providecommand \url  [0]{\begingroup\@sanitize@url \@url }
\providecommand \@url [1]{\endgroup\@href {#1}{\urlprefix }}
\providecommand \urlprefix  [0]{URL }
\providecommand \Eprint [0]{\href }
\providecommand \doibase [0]{https://doi.org/}
\providecommand \selectlanguage [0]{\@gobble}
\providecommand \bibinfo  [0]{\@secondoftwo}
\providecommand \bibfield  [0]{\@secondoftwo}
\providecommand \translation [1]{[#1]}
\providecommand \BibitemOpen [0]{}
\providecommand \bibitemStop [0]{}
\providecommand \bibitemNoStop [0]{.\EOS\space}
\providecommand \EOS [0]{\spacefactor3000\relax}
\providecommand \BibitemShut  [1]{\csname bibitem#1\endcsname}
\let\auto@bib@innerbib\@empty
\bibitem [{\citenamefont {Seifert}(2012)}]{Seifert2012stochastic}
  \BibitemOpen
  \bibfield  {author} {\bibinfo {author} {\bibfnamefont {U.}~\bibnamefont {Seifert}},\ }\href {https://doi.org/10.1088/0034-4885/75/12/126001} {\bibfield  {journal} {\bibinfo  {journal} {Rep. Prog. Phys.}\ }\textbf {\bibinfo {volume} {75}},\ \bibinfo {pages} {126001} (\bibinfo {year} {2012})}\BibitemShut {NoStop}
\bibitem [{\citenamefont {Ciliberto}(2017)}]{Ciliberto2017experiments}
  \BibitemOpen
  \bibfield  {author} {\bibinfo {author} {\bibfnamefont {S.}~\bibnamefont {Ciliberto}},\ }\href {https://doi.org/10.1103/PhysRevX.7.021051} {\bibfield  {journal} {\bibinfo  {journal} {Phys. Rev. X}\ }\textbf {\bibinfo {volume} {7}},\ \bibinfo {pages} {021051} (\bibinfo {year} {2017})}\BibitemShut {NoStop}
\bibitem [{\citenamefont {Parrondo}\ \emph {et~al.}(2015)\citenamefont {Parrondo}, \citenamefont {Horowitz},\ and\ \citenamefont {Sagawa}}]{Parrondo2015thermodynamics}
  \BibitemOpen
  \bibfield  {author} {\bibinfo {author} {\bibfnamefont {J.~M.~R.}\ \bibnamefont {Parrondo}}, \bibinfo {author} {\bibfnamefont {J.~M.}\ \bibnamefont {Horowitz}},\ and\ \bibinfo {author} {\bibfnamefont {T.}~\bibnamefont {Sagawa}},\ }\href {https://doi.org/10.1038/nphys3230} {\bibfield  {journal} {\bibinfo  {journal} {Nat. Phys.}\ }\textbf {\bibinfo {volume} {11}},\ \bibinfo {pages} {131} (\bibinfo {year} {2015})}\BibitemShut {NoStop}
\bibitem [{\citenamefont {Rivas}(2020)}]{Rivas2020}
  \BibitemOpen
  \bibfield  {author} {\bibinfo {author} {\bibfnamefont {A.}~\bibnamefont {Rivas}},\ }\href {https://doi.org/10.1103/PhysRevLett.124.160601} {\bibfield  {journal} {\bibinfo  {journal} {Phys. Rev. Lett.}\ }\textbf {\bibinfo {volume} {124}},\ \bibinfo {pages} {160601} (\bibinfo {year} {2020})}\BibitemShut {NoStop}
\bibitem [{\citenamefont {Rademacher}\ \emph {et~al.}(2022)\citenamefont {Rademacher}, \citenamefont {Konopik}, \citenamefont {Debiossac}, \citenamefont {Grass}, \citenamefont {Lutz},\ and\ \citenamefont {Kiesel}}]{Rademacher2022nonequilibrium}
  \BibitemOpen
  \bibfield  {author} {\bibinfo {author} {\bibfnamefont {M.}~\bibnamefont {Rademacher}}, \bibinfo {author} {\bibfnamefont {M.}~\bibnamefont {Konopik}}, \bibinfo {author} {\bibfnamefont {M.}~\bibnamefont {Debiossac}}, \bibinfo {author} {\bibfnamefont {D.}~\bibnamefont {Grass}}, \bibinfo {author} {\bibfnamefont {E.}~\bibnamefont {Lutz}},\ and\ \bibinfo {author} {\bibfnamefont {N.}~\bibnamefont {Kiesel}},\ }\href {https://doi.org/10.1103/PhysRevLett.128.070601} {\bibfield  {journal} {\bibinfo  {journal} {Phys. Rev. Lett.}\ }\textbf {\bibinfo {volume} {128}},\ \bibinfo {pages} {070601} (\bibinfo {year} {2022})}\BibitemShut {NoStop}
\bibitem [{\citenamefont {Bermudez}\ \emph {et~al.}(2013)\citenamefont {Bermudez}, \citenamefont {Bruderer},\ and\ \citenamefont {Plenio}}]{Bermudez2013controlling}
  \BibitemOpen
  \bibfield  {author} {\bibinfo {author} {\bibfnamefont {A.}~\bibnamefont {Bermudez}}, \bibinfo {author} {\bibfnamefont {M.}~\bibnamefont {Bruderer}},\ and\ \bibinfo {author} {\bibfnamefont {M.~B.}\ \bibnamefont {Plenio}},\ }\href {https://doi.org/10.1103/PhysRevLett.111.040601} {\bibfield  {journal} {\bibinfo  {journal} {Phys. Rev. Lett.}\ }\textbf {\bibinfo {volume} {111}},\ \bibinfo {pages} {040601} (\bibinfo {year} {2013})}\BibitemShut {NoStop}
\bibitem [{\citenamefont {Huber}\ and\ \citenamefont {Rabl}(2019)}]{Huber2019active}
  \BibitemOpen
  \bibfield  {author} {\bibinfo {author} {\bibfnamefont {J.}~\bibnamefont {Huber}}\ and\ \bibinfo {author} {\bibfnamefont {P.}~\bibnamefont {Rabl}},\ }\href {https://doi.org/10.1103/PhysRevA.100.012129} {\bibfield  {journal} {\bibinfo  {journal} {Phys. Rev. A}\ }\textbf {\bibinfo {volume} {100}},\ \bibinfo {pages} {012129} (\bibinfo {year} {2019})}\BibitemShut {NoStop}
\bibitem [{\citenamefont {Benenti}\ \emph {et~al.}(2017)\citenamefont {Benenti}, \citenamefont {Casati}, \citenamefont {Saito},\ and\ \citenamefont {Whitney}}]{Benenti2017fundamental}
  \BibitemOpen
  \bibfield  {author} {\bibinfo {author} {\bibfnamefont {G.}~\bibnamefont {Benenti}}, \bibinfo {author} {\bibfnamefont {G.}~\bibnamefont {Casati}}, \bibinfo {author} {\bibfnamefont {K.}~\bibnamefont {Saito}},\ and\ \bibinfo {author} {\bibfnamefont {R.~S.}\ \bibnamefont {Whitney}},\ }\href {https://doi.org/10.1016/j.physrep.2017.05.008} {\bibfield  {journal} {\bibinfo  {journal} {Phys. Rep.}\ }\textbf {\bibinfo {volume} {694}},\ \bibinfo {pages} {1} (\bibinfo {year} {2017})}\BibitemShut {NoStop}
\bibitem [{\citenamefont {{Serra-Garcia}}\ \emph {et~al.}(2016)\citenamefont {{Serra-Garcia}}, \citenamefont {Foehr}, \citenamefont {Moler{\'o}n}, \citenamefont {Lydon}, \citenamefont {Chong},\ and\ \citenamefont {Daraio}}]{Serra-Garcia2016mechanical}
  \BibitemOpen
  \bibfield  {author} {\bibinfo {author} {\bibfnamefont {M.}~\bibnamefont {{Serra-Garcia}}}, \bibinfo {author} {\bibfnamefont {A.}~\bibnamefont {Foehr}}, \bibinfo {author} {\bibfnamefont {M.}~\bibnamefont {Moler{\'o}n}}, \bibinfo {author} {\bibfnamefont {J.}~\bibnamefont {Lydon}}, \bibinfo {author} {\bibfnamefont {C.}~\bibnamefont {Chong}},\ and\ \bibinfo {author} {\bibfnamefont {C.}~\bibnamefont {Daraio}},\ }\href {https://doi.org/10.1103/PhysRevLett.117.010602} {\bibfield  {journal} {\bibinfo  {journal} {Phys. Rev. Lett.}\ }\textbf {\bibinfo {volume} {117}},\ \bibinfo {pages} {010602} (\bibinfo {year} {2016})}\BibitemShut {NoStop}
\bibitem [{\citenamefont {Buddhiraju}\ \emph {et~al.}(2020)\citenamefont {Buddhiraju}, \citenamefont {Li},\ and\ \citenamefont {Fan}}]{Buddhiraju2020photonic}
  \BibitemOpen
  \bibfield  {author} {\bibinfo {author} {\bibfnamefont {S.}~\bibnamefont {Buddhiraju}}, \bibinfo {author} {\bibfnamefont {W.}~\bibnamefont {Li}},\ and\ \bibinfo {author} {\bibfnamefont {S.}~\bibnamefont {Fan}},\ }\href {https://doi.org/10.1103/PhysRevLett.124.077402} {\bibfield  {journal} {\bibinfo  {journal} {Phys. Rev. Lett.}\ }\textbf {\bibinfo {volume} {124}},\ \bibinfo {pages} {077402} (\bibinfo {year} {2020})}\BibitemShut {NoStop}
\bibitem [{\citenamefont {Giazotto}\ and\ \citenamefont {{Mart{\'i}nez-P{\'e}rez}}(2012)}]{Giazotto2012josephson}
  \BibitemOpen
  \bibfield  {author} {\bibinfo {author} {\bibfnamefont {F.}~\bibnamefont {Giazotto}}\ and\ \bibinfo {author} {\bibfnamefont {M.~J.}\ \bibnamefont {{Mart{\'i}nez-P{\'e}rez}}},\ }\href {https://doi.org/10.1038/nature11702} {\bibfield  {journal} {\bibinfo  {journal} {Nature}\ }\textbf {\bibinfo {volume} {492}},\ \bibinfo {pages} {401} (\bibinfo {year} {2012})}\BibitemShut {NoStop}
\bibitem [{\citenamefont {Zhu}\ and\ \citenamefont {Fan}(2014)}]{Zhu2014nearcomplete}
  \BibitemOpen
  \bibfield  {author} {\bibinfo {author} {\bibfnamefont {L.}~\bibnamefont {Zhu}}\ and\ \bibinfo {author} {\bibfnamefont {S.}~\bibnamefont {Fan}},\ }\href {https://doi.org/10.1103/PhysRevB.90.220301} {\bibfield  {journal} {\bibinfo  {journal} {Phys. Rev. B}\ }\textbf {\bibinfo {volume} {90}},\ \bibinfo {pages} {220301} (\bibinfo {year} {2014})}\BibitemShut {NoStop}
\bibitem [{\citenamefont {S\"utl\"uo\u{g}lu~Ege}\ \emph {et~al.}(2025)\citenamefont {S\"utl\"uo\u{g}lu~Ege}, \citenamefont {\"Ozdemir},\ and\ \citenamefont {Bulutay}}]{Ege2025exploring}
  \BibitemOpen
  \bibfield  {author} {\bibinfo {author} {\bibfnamefont {B.}~\bibnamefont {S\"utl\"uo\u{g}lu~Ege}}, \bibinfo {author} {\bibfnamefont {S.~K.}\ \bibnamefont {\"Ozdemir}},\ and\ \bibinfo {author} {\bibfnamefont {C.}~\bibnamefont {Bulutay}},\ }\href {https://doi.org/10.1103/9syn-mfn7} {\bibfield  {journal} {\bibinfo  {journal} {Phys. Rev. Res.}\ }\textbf {\bibinfo {volume} {7}},\ \bibinfo {pages} {043330} (\bibinfo {year} {2025})}\BibitemShut {NoStop}
\bibitem [{\citenamefont {Biehs}\ and\ \citenamefont {Agarwal}(2023)}]{Biehs2023breakdown}
  \BibitemOpen
  \bibfield  {author} {\bibinfo {author} {\bibfnamefont {S.-A.}\ \bibnamefont {Biehs}}\ and\ \bibinfo {author} {\bibfnamefont {G.~S.}\ \bibnamefont {Agarwal}},\ }\href {https://doi.org/10.1103/PhysRevLett.130.110401} {\bibfield  {journal} {\bibinfo  {journal} {Phys. Rev. Lett.}\ }\textbf {\bibinfo {volume} {130}},\ \bibinfo {pages} {110401} (\bibinfo {year} {2023})}\BibitemShut {NoStop}
\bibitem [{\citenamefont {Fern\'andez-Alc\'azar}\ \emph {et~al.}(2021)\citenamefont {Fern\'andez-Alc\'azar}, \citenamefont {Kononchuk}, \citenamefont {Li},\ and\ \citenamefont {Kottos}}]{Fernandez-Alcazar2021extreme}
  \BibitemOpen
  \bibfield  {author} {\bibinfo {author} {\bibfnamefont {L.~J.}\ \bibnamefont {Fern\'andez-Alc\'azar}}, \bibinfo {author} {\bibfnamefont {R.}~\bibnamefont {Kononchuk}}, \bibinfo {author} {\bibfnamefont {H.}~\bibnamefont {Li}},\ and\ \bibinfo {author} {\bibfnamefont {T.}~\bibnamefont {Kottos}},\ }\href {https://doi.org/10.1103/PhysRevLett.126.204101} {\bibfield  {journal} {\bibinfo  {journal} {Phys. Rev. Lett.}\ }\textbf {\bibinfo {volume} {126}},\ \bibinfo {pages} {204101} (\bibinfo {year} {2021})}\BibitemShut {NoStop}
\bibitem [{\citenamefont {Seif}\ \emph {et~al.}(2018)\citenamefont {Seif}, \citenamefont {DeGottardi}, \citenamefont {Esfarjani},\ and\ \citenamefont {Hafezi}}]{Seif2018thermal}
  \BibitemOpen
  \bibfield  {author} {\bibinfo {author} {\bibfnamefont {A.}~\bibnamefont {Seif}}, \bibinfo {author} {\bibfnamefont {W.}~\bibnamefont {DeGottardi}}, \bibinfo {author} {\bibfnamefont {K.}~\bibnamefont {Esfarjani}},\ and\ \bibinfo {author} {\bibfnamefont {M.}~\bibnamefont {Hafezi}},\ }\href {https://doi.org/10.1038/s41467-018-03624-y} {\bibfield  {journal} {\bibinfo  {journal} {Nat. Commun.}\ }\textbf {\bibinfo {volume} {9}},\ \bibinfo {pages} {1207} (\bibinfo {year} {2018})}\BibitemShut {NoStop}
\bibitem [{\citenamefont {Rivas}\ and\ \citenamefont {{Martin-Delgado}}(2017)}]{Rivas2017topological}
  \BibitemOpen
  \bibfield  {author} {\bibinfo {author} {\bibfnamefont {{\'A}.}~\bibnamefont {Rivas}}\ and\ \bibinfo {author} {\bibfnamefont {M.~A.}\ \bibnamefont {{Martin-Delgado}}},\ }\href {https://doi.org/10.1038/s41598-017-06722-x} {\bibfield  {journal} {\bibinfo  {journal} {Sci. Rep.}\ }\textbf {\bibinfo {volume} {7}},\ \bibinfo {pages} {6350} (\bibinfo {year} {2017})}\BibitemShut {NoStop}
\bibitem [{\citenamefont {Silveirinha}(2017)}]{Silveirinha2017topological}
  \BibitemOpen
  \bibfield  {author} {\bibinfo {author} {\bibfnamefont {M.~G.}\ \bibnamefont {Silveirinha}},\ }\href {https://doi.org/10.1103/PhysRevB.95.115103} {\bibfield  {journal} {\bibinfo  {journal} {Phys. Rev. B}\ }\textbf {\bibinfo {volume} {95}},\ \bibinfo {pages} {115103} (\bibinfo {year} {2017})}\BibitemShut {NoStop}
\bibitem [{\citenamefont {Zhu}\ and\ \citenamefont {Fan}(2016)}]{Zhu2016persistent}
  \BibitemOpen
  \bibfield  {author} {\bibinfo {author} {\bibfnamefont {L.}~\bibnamefont {Zhu}}\ and\ \bibinfo {author} {\bibfnamefont {S.}~\bibnamefont {Fan}},\ }\href {https://doi.org/10.1103/PhysRevLett.117.134303} {\bibfield  {journal} {\bibinfo  {journal} {Phys. Rev. Lett.}\ }\textbf {\bibinfo {volume} {117}},\ \bibinfo {pages} {134303} (\bibinfo {year} {2016})}\BibitemShut {NoStop}
\bibitem [{\citenamefont {Denis}\ \emph {et~al.}(2020)\citenamefont {Denis}, \citenamefont {Biella}, \citenamefont {Favero},\ and\ \citenamefont {Ciuti}}]{Denis2020permanent}
  \BibitemOpen
  \bibfield  {author} {\bibinfo {author} {\bibfnamefont {Z.}~\bibnamefont {Denis}}, \bibinfo {author} {\bibfnamefont {A.}~\bibnamefont {Biella}}, \bibinfo {author} {\bibfnamefont {I.}~\bibnamefont {Favero}},\ and\ \bibinfo {author} {\bibfnamefont {C.}~\bibnamefont {Ciuti}},\ }\href {https://doi.org/10.1103/PhysRevLett.124.083601} {\bibfield  {journal} {\bibinfo  {journal} {Phys. Rev. Lett.}\ }\textbf {\bibinfo {volume} {124}},\ \bibinfo {pages} {083601} (\bibinfo {year} {2020})}\BibitemShut {NoStop}
\bibitem [{\citenamefont {Biehs}\ \emph {et~al.}(2023)\citenamefont {Biehs}, \citenamefont {Rodriguez-Lopez}, \citenamefont {Antezza},\ and\ \citenamefont {Agarwal}}]{Biehs2023nonreciprocal}
  \BibitemOpen
  \bibfield  {author} {\bibinfo {author} {\bibfnamefont {S.-A.}\ \bibnamefont {Biehs}}, \bibinfo {author} {\bibfnamefont {P.}~\bibnamefont {Rodriguez-Lopez}}, \bibinfo {author} {\bibfnamefont {M.}~\bibnamefont {Antezza}},\ and\ \bibinfo {author} {\bibfnamefont {G.~S.}\ \bibnamefont {Agarwal}},\ }\href {https://doi.org/10.1103/PhysRevA.108.042201} {\bibfield  {journal} {\bibinfo  {journal} {Phys. Rev. A}\ }\textbf {\bibinfo {volume} {108}},\ \bibinfo {pages} {042201} (\bibinfo {year} {2023})}\BibitemShut {NoStop}
\bibitem [{\citenamefont {Biehs}\ and\ \citenamefont {Latella}(2025)}]{Biehs2025onpersistent}
  \BibitemOpen
  \bibfield  {author} {\bibinfo {author} {\bibfnamefont {S.-A.}\ \bibnamefont {Biehs}}\ and\ \bibinfo {author} {\bibfnamefont {I.}~\bibnamefont {Latella}},\ }\href {https://doi.org/https://doi.org/10.1016/j.jqsrt.2025.109660} {\bibfield  {journal} {\bibinfo  {journal} {J. Quant. Spectrosc. Radiat. Transf.}\ }\textbf {\bibinfo {volume} {347}},\ \bibinfo {pages} {109660} (\bibinfo {year} {2025})}\BibitemShut {NoStop}
\bibitem [{\citenamefont {Loos}\ and\ \citenamefont {Klapp}(2020)}]{Loos2020irreversibility}
  \BibitemOpen
  \bibfield  {author} {\bibinfo {author} {\bibfnamefont {S.~A.~M.}\ \bibnamefont {Loos}}\ and\ \bibinfo {author} {\bibfnamefont {S.~H.~L.}\ \bibnamefont {Klapp}},\ }\href {https://doi.org/10.1088/1367-2630/abcc1e} {\bibfield  {journal} {\bibinfo  {journal} {New J. Phys.}\ }\textbf {\bibinfo {volume} {22}},\ \bibinfo {pages} {123051} (\bibinfo {year} {2020})}\BibitemShut {NoStop}
\bibitem [{\citenamefont {Zhang}\ and\ \citenamefont {Garcia-Millan}(2023)}]{Zhang2023entropy}
  \BibitemOpen
  \bibfield  {author} {\bibinfo {author} {\bibfnamefont {Z.}~\bibnamefont {Zhang}}\ and\ \bibinfo {author} {\bibfnamefont {R.}~\bibnamefont {Garcia-Millan}},\ }\href {https://doi.org/10.1103/PhysRevResearch.5.L022033} {\bibfield  {journal} {\bibinfo  {journal} {Phys. Rev. Res.}\ }\textbf {\bibinfo {volume} {5}},\ \bibinfo {pages} {L022033} (\bibinfo {year} {2023})}\BibitemShut {NoStop}
\bibitem [{\citenamefont {Loos}\ \emph {et~al.}(2023)\citenamefont {Loos}, \citenamefont {Arabha}, \citenamefont {Hassanali},\ and\ \citenamefont {Rold{\'a}n}}]{Loos2023nonreciprocal}
  \BibitemOpen
  \bibfield  {author} {\bibinfo {author} {\bibfnamefont {S.~A.~M.}\ \bibnamefont {Loos}}, \bibinfo {author} {\bibfnamefont {S.}~\bibnamefont {Arabha}}, \bibinfo {author} {\bibfnamefont {A.}~\bibnamefont {Hassanali}},\ and\ \bibinfo {author} {\bibfnamefont {{\'E}.}~\bibnamefont {Rold{\'a}n}},\ }\href {https://doi.org/10.1038/s41598-023-31583-y} {\bibfield  {journal} {\bibinfo  {journal} {Sci. Rep.}\ }\textbf {\bibinfo {volume} {13}},\ \bibinfo {pages} {4517} (\bibinfo {year} {2023})}\BibitemShut {NoStop}
\bibitem [{\citenamefont {Yang}\ \emph {et~al.}(2024)\citenamefont {Yang}, \citenamefont {Liu}, \citenamefont {Zhao}, \citenamefont {Fan},\ and\ \citenamefont {Qiu}}]{Yang2024nonreciprocal}
  \BibitemOpen
  \bibfield  {author} {\bibinfo {author} {\bibfnamefont {S.}~\bibnamefont {Yang}}, \bibinfo {author} {\bibfnamefont {M.}~\bibnamefont {Liu}}, \bibinfo {author} {\bibfnamefont {C.}~\bibnamefont {Zhao}}, \bibinfo {author} {\bibfnamefont {S.}~\bibnamefont {Fan}},\ and\ \bibinfo {author} {\bibfnamefont {C.-W.}\ \bibnamefont {Qiu}},\ }\href {https://doi.org/10.1038/s41566-024-01409-y} {\bibfield  {journal} {\bibinfo  {journal} {Nat. Photon.}\ }\textbf {\bibinfo {volume} {18}},\ \bibinfo {pages} {412} (\bibinfo {year} {2024})}\BibitemShut {NoStop}
\bibitem [{\citenamefont {Shayegan}\ \emph {et~al.}(2023)\citenamefont {Shayegan}, \citenamefont {Biswas}, \citenamefont {Zhao}, \citenamefont {Fan},\ and\ \citenamefont {Atwater}}]{Shayegan2023direct}
  \BibitemOpen
  \bibfield  {author} {\bibinfo {author} {\bibfnamefont {K.~J.}\ \bibnamefont {Shayegan}}, \bibinfo {author} {\bibfnamefont {S.}~\bibnamefont {Biswas}}, \bibinfo {author} {\bibfnamefont {B.}~\bibnamefont {Zhao}}, \bibinfo {author} {\bibfnamefont {S.}~\bibnamefont {Fan}},\ and\ \bibinfo {author} {\bibfnamefont {H.~A.}\ \bibnamefont {Atwater}},\ }\href {https://doi.org/10.1038/s41566-023-01261-6} {\bibfield  {journal} {\bibinfo  {journal} {Nat. Photon.}\ }\textbf {\bibinfo {volume} {17}},\ \bibinfo {pages} {891} (\bibinfo {year} {2023})}\BibitemShut {NoStop}
\bibitem [{\citenamefont {Benenti}\ \emph {et~al.}(2011)\citenamefont {Benenti}, \citenamefont {Saito},\ and\ \citenamefont {Casati}}]{Benenti2011thermodynamic}
  \BibitemOpen
  \bibfield  {author} {\bibinfo {author} {\bibfnamefont {G.}~\bibnamefont {Benenti}}, \bibinfo {author} {\bibfnamefont {K.}~\bibnamefont {Saito}},\ and\ \bibinfo {author} {\bibfnamefont {G.}~\bibnamefont {Casati}},\ }\href {https://doi.org/10.1103/PhysRevLett.106.230602} {\bibfield  {journal} {\bibinfo  {journal} {Phys. Rev. Lett.}\ }\textbf {\bibinfo {volume} {106}},\ \bibinfo {pages} {230602} (\bibinfo {year} {2011})}\BibitemShut {NoStop}
\bibitem [{\citenamefont {Brandner}\ \emph {et~al.}(2013)\citenamefont {Brandner}, \citenamefont {Saito},\ and\ \citenamefont {Seifert}}]{Brandner2013strong}
  \BibitemOpen
  \bibfield  {author} {\bibinfo {author} {\bibfnamefont {K.}~\bibnamefont {Brandner}}, \bibinfo {author} {\bibfnamefont {K.}~\bibnamefont {Saito}},\ and\ \bibinfo {author} {\bibfnamefont {U.}~\bibnamefont {Seifert}},\ }\href {https://doi.org/10.1103/PhysRevLett.110.070603} {\bibfield  {journal} {\bibinfo  {journal} {Phys. Rev. Lett.}\ }\textbf {\bibinfo {volume} {110}},\ \bibinfo {pages} {070603} (\bibinfo {year} {2013})}\BibitemShut {NoStop}
\bibitem [{\citenamefont {Luo}\ \emph {et~al.}(2020)\citenamefont {Luo}, \citenamefont {Benenti}, \citenamefont {Casati},\ and\ \citenamefont {Wang}}]{Luo2020onsager}
  \BibitemOpen
  \bibfield  {author} {\bibinfo {author} {\bibfnamefont {R.}~\bibnamefont {Luo}}, \bibinfo {author} {\bibfnamefont {G.}~\bibnamefont {Benenti}}, \bibinfo {author} {\bibfnamefont {G.}~\bibnamefont {Casati}},\ and\ \bibinfo {author} {\bibfnamefont {J.}~\bibnamefont {Wang}},\ }\href {https://doi.org/10.1103/PhysRevResearch.2.022009} {\bibfield  {journal} {\bibinfo  {journal} {Phys. Rev. Res.}\ }\textbf {\bibinfo {volume} {2}},\ \bibinfo {pages} {022009} (\bibinfo {year} {2020})}\BibitemShut {NoStop}
\bibitem [{\citenamefont {Saryal}\ \emph {et~al.}(2022)\citenamefont {Saryal}, \citenamefont {Mohanta},\ and\ \citenamefont {Agarwalla}}]{Saryal2022bounds}
  \BibitemOpen
  \bibfield  {author} {\bibinfo {author} {\bibfnamefont {S.}~\bibnamefont {Saryal}}, \bibinfo {author} {\bibfnamefont {S.}~\bibnamefont {Mohanta}},\ and\ \bibinfo {author} {\bibfnamefont {B.~K.}\ \bibnamefont {Agarwalla}},\ }\href {https://doi.org/10.1103/PhysRevE.105.024129} {\bibfield  {journal} {\bibinfo  {journal} {Phys. Rev. E}\ }\textbf {\bibinfo {volume} {105}},\ \bibinfo {pages} {024129} (\bibinfo {year} {2022})}\BibitemShut {NoStop}
\bibitem [{\citenamefont {Lai}\ \emph {et~al.}(2020)\citenamefont {Lai}, \citenamefont {Huang}, \citenamefont {Yin}, \citenamefont {Hou}, \citenamefont {Li}, \citenamefont {Vitali}, \citenamefont {Nori},\ and\ \citenamefont {Liao}}]{Lai2020nonreciprocal}
  \BibitemOpen
  \bibfield  {author} {\bibinfo {author} {\bibfnamefont {D.-G.}\ \bibnamefont {Lai}}, \bibinfo {author} {\bibfnamefont {J.-F.}\ \bibnamefont {Huang}}, \bibinfo {author} {\bibfnamefont {X.-L.}\ \bibnamefont {Yin}}, \bibinfo {author} {\bibfnamefont {B.-P.}\ \bibnamefont {Hou}}, \bibinfo {author} {\bibfnamefont {W.}~\bibnamefont {Li}}, \bibinfo {author} {\bibfnamefont {D.}~\bibnamefont {Vitali}}, \bibinfo {author} {\bibfnamefont {F.}~\bibnamefont {Nori}},\ and\ \bibinfo {author} {\bibfnamefont {J.-Q.}\ \bibnamefont {Liao}},\ }\href {https://doi.org/10.1103/PhysRevA.102.011502} {\bibfield  {journal} {\bibinfo  {journal} {Phys. Rev. A}\ }\textbf {\bibinfo {volume} {102}},\ \bibinfo {pages} {011502} (\bibinfo {year} {2020})}\BibitemShut {NoStop}
\bibitem [{\citenamefont {Messina}\ \emph {et~al.}(2021)\citenamefont {Messina}, \citenamefont {Ott}, \citenamefont {Kathmann}, \citenamefont {Biehs},\ and\ \citenamefont {Ben-Abdallah}}]{Messina2021radiative}
  \BibitemOpen
  \bibfield  {author} {\bibinfo {author} {\bibfnamefont {R.}~\bibnamefont {Messina}}, \bibinfo {author} {\bibfnamefont {A.}~\bibnamefont {Ott}}, \bibinfo {author} {\bibfnamefont {C.}~\bibnamefont {Kathmann}}, \bibinfo {author} {\bibfnamefont {S.-A.}\ \bibnamefont {Biehs}},\ and\ \bibinfo {author} {\bibfnamefont {P.}~\bibnamefont {Ben-Abdallah}},\ }\href {https://doi.org/10.1103/PhysRevB.103.115440} {\bibfield  {journal} {\bibinfo  {journal} {Phys. Rev. B}\ }\textbf {\bibinfo {volume} {103}},\ \bibinfo {pages} {115440} (\bibinfo {year} {2021})}\BibitemShut {NoStop}
\bibitem [{\citenamefont {Wang}\ \emph {et~al.}(2022)\citenamefont {Wang}, \citenamefont {Wang}, \citenamefont {Wang}, \citenamefont {Chen},\ and\ \citenamefont {Jen}}]{Wang2022superior}
  \BibitemOpen
  \bibfield  {author} {\bibinfo {author} {\bibfnamefont {C.-C.}\ \bibnamefont {Wang}}, \bibinfo {author} {\bibfnamefont {Y.-C.}\ \bibnamefont {Wang}}, \bibinfo {author} {\bibfnamefont {C.-H.}\ \bibnamefont {Wang}}, \bibinfo {author} {\bibfnamefont {C.-C.}\ \bibnamefont {Chen}},\ and\ \bibinfo {author} {\bibfnamefont {H.~H.}\ \bibnamefont {Jen}},\ }\href {https://doi.org/10.1088/1367-2630/ac9ed5} {\bibfield  {journal} {\bibinfo  {journal} {New J. Phys.}\ }\textbf {\bibinfo {volume} {24}},\ \bibinfo {pages} {113020} (\bibinfo {year} {2022})}\BibitemShut {NoStop}
\bibitem [{\citenamefont {Wang}\ \emph {et~al.}(2024)\citenamefont {Wang}, \citenamefont {Niu}, \citenamefont {Liu}, \citenamefont {Wang}, \citenamefont {Zhang},\ and\ \citenamefont {Wang}}]{Wang2024enhancement}
  \BibitemOpen
  \bibfield  {author} {\bibinfo {author} {\bibfnamefont {L.}~\bibnamefont {Wang}}, \bibinfo {author} {\bibfnamefont {W.}~\bibnamefont {Niu}}, \bibinfo {author} {\bibfnamefont {S.}~\bibnamefont {Liu}}, \bibinfo {author} {\bibfnamefont {T.}~\bibnamefont {Wang}}, \bibinfo {author} {\bibfnamefont {S.}~\bibnamefont {Zhang}},\ and\ \bibinfo {author} {\bibfnamefont {H.-F.}\ \bibnamefont {Wang}},\ }\href {https://doi.org/10.1364/OE.541360} {\bibfield  {journal} {\bibinfo  {journal} {Opt. Express}\ }\textbf {\bibinfo {volume} {32}},\ \bibinfo {pages} {42149} (\bibinfo {year} {2024})}\BibitemShut {NoStop}
\bibitem [{\citenamefont {Chakraborty}\ \emph {et~al.}(2024)\citenamefont {Chakraborty}, \citenamefont {Chakraborty}, \citenamefont {Basu},\ and\ \citenamefont {Mukherjee}}]{Chakraborty2024quantum}
  \BibitemOpen
  \bibfield  {author} {\bibinfo {author} {\bibfnamefont {G.}~\bibnamefont {Chakraborty}}, \bibinfo {author} {\bibfnamefont {S.}~\bibnamefont {Chakraborty}}, \bibinfo {author} {\bibfnamefont {T.}~\bibnamefont {Basu}},\ and\ \bibinfo {author} {\bibfnamefont {M.}~\bibnamefont {Mukherjee}},\ }\href {https://doi.org/10.1103/PhysRevA.110.042216} {\bibfield  {journal} {\bibinfo  {journal} {Phys. Rev. A}\ }\textbf {\bibinfo {volume} {110}},\ \bibinfo {pages} {042216} (\bibinfo {year} {2024})}\BibitemShut {NoStop}
\bibitem [{\citenamefont {Roushan}\ \emph {et~al.}(2017)\citenamefont {Roushan}, \citenamefont {Neill}, \citenamefont {Megrant}, \citenamefont {Chen}, \citenamefont {Babbush}, \citenamefont {Barends}, \citenamefont {Campbell}, \citenamefont {Chen}, \citenamefont {Chiaro}, \citenamefont {Dunsworth}, \citenamefont {Fowler}, \citenamefont {Jeffrey}, \citenamefont {Kelly}, \citenamefont {Lucero}, \citenamefont {Mutus}, \citenamefont {O'Malley}, \citenamefont {Neeley}, \citenamefont {Quintana}, \citenamefont {Sank}, \citenamefont {Vainsencher}, \citenamefont {Wenner}, \citenamefont {White}, \citenamefont {Kapit}, \citenamefont {Neven},\ and\ \citenamefont {Martinis}}]{Roushan2017chiral}
  \BibitemOpen
  \bibfield  {author} {\bibinfo {author} {\bibfnamefont {P.}~\bibnamefont {Roushan}}, \bibinfo {author} {\bibfnamefont {C.}~\bibnamefont {Neill}}, \bibinfo {author} {\bibfnamefont {A.}~\bibnamefont {Megrant}}, \bibinfo {author} {\bibfnamefont {Y.}~\bibnamefont {Chen}}, \bibinfo {author} {\bibfnamefont {R.}~\bibnamefont {Babbush}}, \bibinfo {author} {\bibfnamefont {R.}~\bibnamefont {Barends}}, \bibinfo {author} {\bibfnamefont {B.}~\bibnamefont {Campbell}}, \bibinfo {author} {\bibfnamefont {Z.}~\bibnamefont {Chen}}, \bibinfo {author} {\bibfnamefont {B.}~\bibnamefont {Chiaro}}, \bibinfo {author} {\bibfnamefont {A.}~\bibnamefont {Dunsworth}}, \bibinfo {author} {\bibfnamefont {A.}~\bibnamefont {Fowler}}, \bibinfo {author} {\bibfnamefont {E.}~\bibnamefont {Jeffrey}}, \bibinfo {author} {\bibfnamefont {J.}~\bibnamefont {Kelly}}, \bibinfo {author} {\bibfnamefont {E.}~\bibnamefont {Lucero}}, \bibinfo {author} {\bibfnamefont {J.}~\bibnamefont {Mutus}}, \bibinfo {author} {\bibfnamefont {P.~J.~J.}\ \bibnamefont {O'Malley}}, \bibinfo {author} {\bibfnamefont {M.}~\bibnamefont {Neeley}}, \bibinfo {author} {\bibfnamefont {C.}~\bibnamefont {Quintana}}, \bibinfo {author} {\bibfnamefont {D.}~\bibnamefont {Sank}}, \bibinfo {author} {\bibfnamefont {A.}~\bibnamefont {Vainsencher}}, \bibinfo {author} {\bibfnamefont {J.}~\bibnamefont {Wenner}}, \bibinfo {author} {\bibfnamefont {T.}~\bibnamefont {White}}, \bibinfo {author} {\bibfnamefont {E.}~\bibnamefont {Kapit}}, \bibinfo {author} {\bibfnamefont {H.}~\bibnamefont {Neven}},\ and\ \bibinfo {author} {\bibfnamefont {J.}~\bibnamefont {Martinis}},\ }\href {https://doi.org/10.1038/nphys3930} {\bibfield  {journal} {\bibinfo  {journal} {Nat. Phys.}\ }\textbf {\bibinfo {volume} {13}},\ \bibinfo {pages} {146} (\bibinfo {year} {2017})}\BibitemShut {NoStop}
\bibitem [{\citenamefont {Xuereb}\ \emph {et~al.}(2015)\citenamefont {Xuereb}, \citenamefont {Imparato},\ and\ \citenamefont {Dantan}}]{Xuereb2015heat}
  \BibitemOpen
  \bibfield  {author} {\bibinfo {author} {\bibfnamefont {A.}~\bibnamefont {Xuereb}}, \bibinfo {author} {\bibfnamefont {A.}~\bibnamefont {Imparato}},\ and\ \bibinfo {author} {\bibfnamefont {A.}~\bibnamefont {Dantan}},\ }\href {https://doi.org/10.1088/1367-2630/17/5/055013} {\bibfield  {journal} {\bibinfo  {journal} {New J. Phys.}\ }\textbf {\bibinfo {volume} {17}},\ \bibinfo {pages} {055013} (\bibinfo {year} {2015})}\BibitemShut {NoStop}
\bibitem [{\citenamefont {Barzanjeh}\ \emph {et~al.}(2018)\citenamefont {Barzanjeh}, \citenamefont {Aquilina},\ and\ \citenamefont {Xuereb}}]{Barzanjeh2018manipulating}
  \BibitemOpen
  \bibfield  {author} {\bibinfo {author} {\bibfnamefont {S.}~\bibnamefont {Barzanjeh}}, \bibinfo {author} {\bibfnamefont {M.}~\bibnamefont {Aquilina}},\ and\ \bibinfo {author} {\bibfnamefont {A.}~\bibnamefont {Xuereb}},\ }\href {https://doi.org/10.1103/PhysRevLett.120.060601} {\bibfield  {journal} {\bibinfo  {journal} {Phys. Rev. Lett.}\ }\textbf {\bibinfo {volume} {120}},\ \bibinfo {pages} {060601} (\bibinfo {year} {2018})}\BibitemShut {NoStop}
\bibitem [{\citenamefont {Fong}\ \emph {et~al.}(2019)\citenamefont {Fong}, \citenamefont {Li}, \citenamefont {Zhao}, \citenamefont {Yang}, \citenamefont {Wang},\ and\ \citenamefont {Zhang}}]{Fong2019phonon}
  \BibitemOpen
  \bibfield  {author} {\bibinfo {author} {\bibfnamefont {K.~Y.}\ \bibnamefont {Fong}}, \bibinfo {author} {\bibfnamefont {H.-K.}\ \bibnamefont {Li}}, \bibinfo {author} {\bibfnamefont {R.}~\bibnamefont {Zhao}}, \bibinfo {author} {\bibfnamefont {S.}~\bibnamefont {Yang}}, \bibinfo {author} {\bibfnamefont {Y.}~\bibnamefont {Wang}},\ and\ \bibinfo {author} {\bibfnamefont {X.}~\bibnamefont {Zhang}},\ }\href {https://doi.org/10.1038/s41586-019-1800-4} {\bibfield  {journal} {\bibinfo  {journal} {Nature}\ }\textbf {\bibinfo {volume} {576}},\ \bibinfo {pages} {243} (\bibinfo {year} {2019})}\BibitemShut {NoStop}
\bibitem [{\citenamefont {Yang}\ \emph {et~al.}(2020)\citenamefont {Yang}, \citenamefont {Wei}, \citenamefont {Sheng},\ and\ \citenamefont {Wu}}]{Yang2020phonon}
  \BibitemOpen
  \bibfield  {author} {\bibinfo {author} {\bibfnamefont {C.}~\bibnamefont {Yang}}, \bibinfo {author} {\bibfnamefont {X.}~\bibnamefont {Wei}}, \bibinfo {author} {\bibfnamefont {J.}~\bibnamefont {Sheng}},\ and\ \bibinfo {author} {\bibfnamefont {H.}~\bibnamefont {Wu}},\ }\href {https://doi.org/10.1038/s41467-020-18426-4} {\bibfield  {journal} {\bibinfo  {journal} {Nat. Commun.}\ }\textbf {\bibinfo {volume} {11}},\ \bibinfo {pages} {4656} (\bibinfo {year} {2020})}\BibitemShut {NoStop}
\bibitem [{\citenamefont {Barzanjeh}\ \emph {et~al.}(2022)\citenamefont {Barzanjeh}, \citenamefont {Xuereb}, \citenamefont {Gr{\"o}blacher}, \citenamefont {Paternostro}, \citenamefont {Regal},\ and\ \citenamefont {Weig}}]{Barzanjeh2022optomechanics}
  \BibitemOpen
  \bibfield  {author} {\bibinfo {author} {\bibfnamefont {S.}~\bibnamefont {Barzanjeh}}, \bibinfo {author} {\bibfnamefont {A.}~\bibnamefont {Xuereb}}, \bibinfo {author} {\bibfnamefont {S.}~\bibnamefont {Gr{\"o}blacher}}, \bibinfo {author} {\bibfnamefont {M.}~\bibnamefont {Paternostro}}, \bibinfo {author} {\bibfnamefont {C.~A.}\ \bibnamefont {Regal}},\ and\ \bibinfo {author} {\bibfnamefont {E.~M.}\ \bibnamefont {Weig}},\ }\href {https://doi.org/10.1038/s41567-021-01402-0} {\bibfield  {journal} {\bibinfo  {journal} {Nat. Phys.}\ }\textbf {\bibinfo {volume} {18}},\ \bibinfo {pages} {15} (\bibinfo {year} {2022})}\BibitemShut {NoStop}
\bibitem [{\citenamefont {Sheng}\ \emph {et~al.}(2023)\citenamefont {Sheng}, \citenamefont {Yang},\ and\ \citenamefont {Wu}}]{Sheng2023nonequilibrium}
  \BibitemOpen
  \bibfield  {author} {\bibinfo {author} {\bibfnamefont {J.}~\bibnamefont {Sheng}}, \bibinfo {author} {\bibfnamefont {C.}~\bibnamefont {Yang}},\ and\ \bibinfo {author} {\bibfnamefont {H.}~\bibnamefont {Wu}},\ }\href {https://doi.org/10.1016/j.fmre.2022.09.005} {\bibfield  {journal} {\bibinfo  {journal} {Fund. Res.}\ }\textbf {\bibinfo {volume} {3}},\ \bibinfo {pages} {75} (\bibinfo {year} {2023})}\BibitemShut {NoStop}
\bibitem [{\citenamefont {Reisenbauer}\ \emph {et~al.}(2024)\citenamefont {Reisenbauer}, \citenamefont {Rudolph}, \citenamefont {Egyed}, \citenamefont {Hornberger}, \citenamefont {Zasedatelev}, \citenamefont {Abuzarli}, \citenamefont {Stickler},\ and\ \citenamefont {Deli\'{c}}}]{Reisenbauer2024nonhermitian}
  \BibitemOpen
  \bibfield  {author} {\bibinfo {author} {\bibfnamefont {M.}~\bibnamefont {Reisenbauer}}, \bibinfo {author} {\bibfnamefont {H.}~\bibnamefont {Rudolph}}, \bibinfo {author} {\bibfnamefont {L.}~\bibnamefont {Egyed}}, \bibinfo {author} {\bibfnamefont {K.}~\bibnamefont {Hornberger}}, \bibinfo {author} {\bibfnamefont {A.~V.}\ \bibnamefont {Zasedatelev}}, \bibinfo {author} {\bibfnamefont {M.}~\bibnamefont {Abuzarli}}, \bibinfo {author} {\bibfnamefont {B.~A.}\ \bibnamefont {Stickler}},\ and\ \bibinfo {author} {\bibfnamefont {U.}~\bibnamefont {Deli\'{c}}},\ }\href {https://doi.org/10.1038/s41567-024-02589-8} {\bibfield  {journal} {\bibinfo  {journal} {Nat. Phys.}\ }\textbf {\bibinfo {volume} {20}},\ \bibinfo {pages} {1629} (\bibinfo {year} {2024})}\BibitemShut {NoStop}
\bibitem [{\citenamefont {Xu}\ \emph {et~al.}(2019)\citenamefont {Xu}, \citenamefont {Jiang}, \citenamefont {Clerk},\ and\ \citenamefont {Harris}}]{Xu2019nonreciprocal}
  \BibitemOpen
  \bibfield  {author} {\bibinfo {author} {\bibfnamefont {H.}~\bibnamefont {Xu}}, \bibinfo {author} {\bibfnamefont {L.}~\bibnamefont {Jiang}}, \bibinfo {author} {\bibfnamefont {A.~A.}\ \bibnamefont {Clerk}},\ and\ \bibinfo {author} {\bibfnamefont {J.~G.~E.}\ \bibnamefont {Harris}},\ }\href {https://doi.org/10.1038/s41586-019-1061-2} {\bibfield  {journal} {\bibinfo  {journal} {Nature}\ }\textbf {\bibinfo {volume} {568}},\ \bibinfo {pages} {65} (\bibinfo {year} {2019})}\BibitemShut {NoStop}
\bibitem [{\citenamefont {Mathew}\ \emph {et~al.}(2020)\citenamefont {Mathew}, \citenamefont {del Pino},\ and\ \citenamefont {Verhagen}}]{Mathew2020synthetic}
  \BibitemOpen
  \bibfield  {author} {\bibinfo {author} {\bibfnamefont {J.~P.}\ \bibnamefont {Mathew}}, \bibinfo {author} {\bibfnamefont {J.}~\bibnamefont {del Pino}},\ and\ \bibinfo {author} {\bibfnamefont {E.}~\bibnamefont {Verhagen}},\ }\href {https://doi.org/10.1038/s41565-019-0630-8} {\bibfield  {journal} {\bibinfo  {journal} {Nat. Nanotechnol.}\ }\textbf {\bibinfo {volume} {15}},\ \bibinfo {pages} {198} (\bibinfo {year} {2020})}\BibitemShut {NoStop}
\bibitem [{\citenamefont {{Mercier de L{\'e}pinay}}\ \emph {et~al.}(2020)\citenamefont {{Mercier de L{\'e}pinay}}, \citenamefont {{Ockeloen-Korppi}}, \citenamefont {Malz},\ and\ \citenamefont {Sillanp{\"a}{\"a}}}]{MercierDeLepinay2020nonreciprocal}
  \BibitemOpen
  \bibfield  {author} {\bibinfo {author} {\bibfnamefont {L.}~\bibnamefont {{Mercier de L{\'e}pinay}}}, \bibinfo {author} {\bibfnamefont {C.~F.}\ \bibnamefont {{Ockeloen-Korppi}}}, \bibinfo {author} {\bibfnamefont {D.}~\bibnamefont {Malz}},\ and\ \bibinfo {author} {\bibfnamefont {M.~A.}\ \bibnamefont {Sillanp{\"a}{\"a}}},\ }\href {https://doi.org/10.1103/PhysRevLett.125.023603} {\bibfield  {journal} {\bibinfo  {journal} {Phys. Rev. Lett.}\ }\textbf {\bibinfo {volume} {125}},\ \bibinfo {pages} {023603} (\bibinfo {year} {2020})}\BibitemShut {NoStop}
\bibitem [{\citenamefont {{del Pino}}\ \emph {et~al.}(2022)\citenamefont {{del Pino}}, \citenamefont {Slim},\ and\ \citenamefont {Verhagen}}]{DelPino2022nonhermitian}
  \BibitemOpen
  \bibfield  {author} {\bibinfo {author} {\bibfnamefont {J.}~\bibnamefont {{del Pino}}}, \bibinfo {author} {\bibfnamefont {J.~J.}\ \bibnamefont {Slim}},\ and\ \bibinfo {author} {\bibfnamefont {E.}~\bibnamefont {Verhagen}},\ }\href {https://doi.org/10.1038/s41586-022-04609-0} {\bibfield  {journal} {\bibinfo  {journal} {Nature}\ }\textbf {\bibinfo {volume} {606}},\ \bibinfo {pages} {82} (\bibinfo {year} {2022})}\BibitemShut {NoStop}
\bibitem [{\citenamefont {Slim}\ \emph {et~al.}(2025)\citenamefont {Slim}, \citenamefont {del Pino},\ and\ \citenamefont {Verhagen}}]{Slim2025programmable}
  \BibitemOpen
  \bibfield  {author} {\bibinfo {author} {\bibfnamefont {J.~J.}\ \bibnamefont {Slim}}, \bibinfo {author} {\bibfnamefont {J.}~\bibnamefont {del Pino}},\ and\ \bibinfo {author} {\bibfnamefont {E.}~\bibnamefont {Verhagen}},\ }\href {https://doi.org/10.1038/s41467-025-62541-z} {\bibfield  {journal} {\bibinfo  {journal} {Nat. Commun.}\ }\textbf {\bibinfo {volume} {16}},\ \bibinfo {pages} {7471} (\bibinfo {year} {2025})}\BibitemShut {NoStop}
\bibitem [{\citenamefont {{Ockeloen-Korppi}}\ \emph {et~al.}(2019)\citenamefont {{Ockeloen-Korppi}}, \citenamefont {Gely}, \citenamefont {Damsk{\"a}gg}, \citenamefont {Jenkins}, \citenamefont {Steele},\ and\ \citenamefont {Sillanp{\"a}{\"a}}}]{Ockeloen-Korppi2019sideband}
  \BibitemOpen
  \bibfield  {author} {\bibinfo {author} {\bibfnamefont {C.~F.}\ \bibnamefont {{Ockeloen-Korppi}}}, \bibinfo {author} {\bibfnamefont {M.~F.}\ \bibnamefont {Gely}}, \bibinfo {author} {\bibfnamefont {E.}~\bibnamefont {Damsk{\"a}gg}}, \bibinfo {author} {\bibfnamefont {M.}~\bibnamefont {Jenkins}}, \bibinfo {author} {\bibfnamefont {G.~A.}\ \bibnamefont {Steele}},\ and\ \bibinfo {author} {\bibfnamefont {M.~A.}\ \bibnamefont {Sillanp{\"a}{\"a}}},\ }\href {https://doi.org/10.1103/PhysRevA.99.023826} {\bibfield  {journal} {\bibinfo  {journal} {Phys. Rev. A}\ }\textbf {\bibinfo {volume} {99}},\ \bibinfo {pages} {023826} (\bibinfo {year} {2019})}\BibitemShut {NoStop}
\bibitem [{\citenamefont {Huang}\ \emph {et~al.}(2022)\citenamefont {Huang}, \citenamefont {Lai}, \citenamefont {Liu}, \citenamefont {Huang}, \citenamefont {Nori},\ and\ \citenamefont {Liao}}]{Huang2022multimode}
  \BibitemOpen
  \bibfield  {author} {\bibinfo {author} {\bibfnamefont {J.}~\bibnamefont {Huang}}, \bibinfo {author} {\bibfnamefont {D.-G.}\ \bibnamefont {Lai}}, \bibinfo {author} {\bibfnamefont {C.}~\bibnamefont {Liu}}, \bibinfo {author} {\bibfnamefont {J.-F.}\ \bibnamefont {Huang}}, \bibinfo {author} {\bibfnamefont {F.}~\bibnamefont {Nori}},\ and\ \bibinfo {author} {\bibfnamefont {J.-Q.}\ \bibnamefont {Liao}},\ }\href {https://doi.org/10.1103/PhysRevA.106.013526} {\bibfield  {journal} {\bibinfo  {journal} {Phys. Rev. A}\ }\textbf {\bibinfo {volume} {106}},\ \bibinfo {pages} {013526} (\bibinfo {year} {2022})}\BibitemShut {NoStop}
\bibitem [{\citenamefont {Aranas}\ \emph {et~al.}(2017)\citenamefont {Aranas}, \citenamefont {Akram}, \citenamefont {Malz},\ and\ \citenamefont {Monteiro}}]{Aranas2017quantum}
  \BibitemOpen
  \bibfield  {author} {\bibinfo {author} {\bibfnamefont {E.~B.}\ \bibnamefont {Aranas}}, \bibinfo {author} {\bibfnamefont {M.~J.}\ \bibnamefont {Akram}}, \bibinfo {author} {\bibfnamefont {D.}~\bibnamefont {Malz}},\ and\ \bibinfo {author} {\bibfnamefont {T.~S.}\ \bibnamefont {Monteiro}},\ }\href {https://doi.org/10.1103/PhysRevA.96.063836} {\bibfield  {journal} {\bibinfo  {journal} {Phys. Rev. A}\ }\textbf {\bibinfo {volume} {96}},\ \bibinfo {pages} {063836} (\bibinfo {year} {2017})}\BibitemShut {NoStop}
\bibitem [{sup()}]{supmat}
  \BibitemOpen
  \href@noop {} {}\bibinfo {note} {See Supplemental Material at \url{http://link.aps.org/supplemental/10.1103/PhysRevLett.XXX.YYYYY} for details on the theoretical model, derivations, and additional figures.}\BibitemShut {Stop}
\bibitem [{\citenamefont {Koch}\ \emph {et~al.}(2010)\citenamefont {Koch}, \citenamefont {Houck}, \citenamefont {Hur},\ and\ \citenamefont {Girvin}}]{Koch2010timereversalsymmetry}
  \BibitemOpen
  \bibfield  {author} {\bibinfo {author} {\bibfnamefont {J.}~\bibnamefont {Koch}}, \bibinfo {author} {\bibfnamefont {A.~A.}\ \bibnamefont {Houck}}, \bibinfo {author} {\bibfnamefont {K.~L.}\ \bibnamefont {Hur}},\ and\ \bibinfo {author} {\bibfnamefont {S.~M.}\ \bibnamefont {Girvin}},\ }\href {https://doi.org/10.1103/PhysRevA.82.043811} {\bibfield  {journal} {\bibinfo  {journal} {Phys. Rev. A}\ }\textbf {\bibinfo {volume} {82}},\ \bibinfo {pages} {043811} (\bibinfo {year} {2010})}\BibitemShut {NoStop}
\bibitem [{\citenamefont {Horowitz}\ and\ \citenamefont {Gingrich}(2020)}]{Horowitz2020thermodynamic}
  \BibitemOpen
  \bibfield  {author} {\bibinfo {author} {\bibfnamefont {J.~M.}\ \bibnamefont {Horowitz}}\ and\ \bibinfo {author} {\bibfnamefont {T.~R.}\ \bibnamefont {Gingrich}},\ }\href {https://doi.org/10.1038/s41567-019-0702-6} {\bibfield  {journal} {\bibinfo  {journal} {Nat. Phys.}\ }\textbf {\bibinfo {volume} {16}},\ \bibinfo {pages} {15} (\bibinfo {year} {2020})}\BibitemShut {NoStop}
\bibitem [{\citenamefont {Jiang}\ \emph {et~al.}(2021)\citenamefont {Jiang}, \citenamefont {Liu},\ and\ \citenamefont {Sillanp\"a\"a}}]{Jiang2020energylevelattraction}
  \BibitemOpen
  \bibfield  {author} {\bibinfo {author} {\bibfnamefont {C.}~\bibnamefont {Jiang}}, \bibinfo {author} {\bibfnamefont {Y.-L.}\ \bibnamefont {Liu}},\ and\ \bibinfo {author} {\bibfnamefont {M.~A.}\ \bibnamefont {Sillanp\"a\"a}},\ }\href {https://doi.org/10.1103/PhysRevA.104.013502} {\bibfield  {journal} {\bibinfo  {journal} {Phys. Rev. A}\ }\textbf {\bibinfo {volume} {104}},\ \bibinfo {pages} {013502} (\bibinfo {year} {2021})}\BibitemShut {NoStop}
\bibitem [{\citenamefont {Peterson}\ \emph {et~al.}(2017)\citenamefont {Peterson}, \citenamefont {Lecocq}, \citenamefont {Cicak}, \citenamefont {Simmonds}, \citenamefont {Aumentado},\ and\ \citenamefont {Teufel}}]{Peterson2017demonstration}
  \BibitemOpen
  \bibfield  {author} {\bibinfo {author} {\bibfnamefont {G.~A.}\ \bibnamefont {Peterson}}, \bibinfo {author} {\bibfnamefont {F.}~\bibnamefont {Lecocq}}, \bibinfo {author} {\bibfnamefont {K.}~\bibnamefont {Cicak}}, \bibinfo {author} {\bibfnamefont {R.~W.}\ \bibnamefont {Simmonds}}, \bibinfo {author} {\bibfnamefont {J.}~\bibnamefont {Aumentado}},\ and\ \bibinfo {author} {\bibfnamefont {J.~D.}\ \bibnamefont {Teufel}},\ }\href {https://doi.org/10.1103/PhysRevX.7.031001} {\bibfield  {journal} {\bibinfo  {journal} {Phys. Rev. X}\ }\textbf {\bibinfo {volume} {7}},\ \bibinfo {pages} {031001} (\bibinfo {year} {2017})}\BibitemShut {NoStop}
\bibitem [{\citenamefont {Bernier}\ \emph {et~al.}(2017)\citenamefont {Bernier}, \citenamefont {T{\'o}th}, \citenamefont {Koottandavida}, \citenamefont {Ioannou}, \citenamefont {Malz}, \citenamefont {Nunnenkamp}, \citenamefont {Feofanov},\ and\ \citenamefont {Kippenberg}}]{Bernier2017nonreciprocal}
  \BibitemOpen
  \bibfield  {author} {\bibinfo {author} {\bibfnamefont {N.~R.}\ \bibnamefont {Bernier}}, \bibinfo {author} {\bibfnamefont {L.~D.}\ \bibnamefont {T{\'o}th}}, \bibinfo {author} {\bibfnamefont {A.}~\bibnamefont {Koottandavida}}, \bibinfo {author} {\bibfnamefont {M.~A.}\ \bibnamefont {Ioannou}}, \bibinfo {author} {\bibfnamefont {D.}~\bibnamefont {Malz}}, \bibinfo {author} {\bibfnamefont {A.}~\bibnamefont {Nunnenkamp}}, \bibinfo {author} {\bibfnamefont {A.~K.}\ \bibnamefont {Feofanov}},\ and\ \bibinfo {author} {\bibfnamefont {T.~J.}\ \bibnamefont {Kippenberg}},\ }\href {https://doi.org/10.1038/s41467-017-00447-1} {\bibfield  {journal} {\bibinfo  {journal} {Nat. Commun.}\ }\textbf {\bibinfo {volume} {8}},\ \bibinfo {pages} {604} (\bibinfo {year} {2017})}\BibitemShut {NoStop}
\bibitem [{\citenamefont {Slim}\ \emph {et~al.}(2024)\citenamefont {Slim}, \citenamefont {Wanjura}, \citenamefont {Brunelli}, \citenamefont {Del~Pino}, \citenamefont {Nunnenkamp},\ and\ \citenamefont {Verhagen}}]{Slim2024optomechanical}
  \BibitemOpen
  \bibfield  {author} {\bibinfo {author} {\bibfnamefont {J.~J.}\ \bibnamefont {Slim}}, \bibinfo {author} {\bibfnamefont {C.~C.}\ \bibnamefont {Wanjura}}, \bibinfo {author} {\bibfnamefont {M.}~\bibnamefont {Brunelli}}, \bibinfo {author} {\bibfnamefont {J.}~\bibnamefont {Del~Pino}}, \bibinfo {author} {\bibfnamefont {A.}~\bibnamefont {Nunnenkamp}},\ and\ \bibinfo {author} {\bibfnamefont {E.}~\bibnamefont {Verhagen}},\ }\href {https://doi.org/10.1038/s41586-024-07174-w} {\bibfield  {journal} {\bibinfo  {journal} {Nature}\ }\textbf {\bibinfo {volume} {627}},\ \bibinfo {pages} {767} (\bibinfo {year} {2024})}\BibitemShut {NoStop}
\bibitem [{\citenamefont {Meystre}\ and\ \citenamefont {Sargent}(2007)}]{Meystre2007elements}
  \BibitemOpen
  \bibfield  {author} {\bibinfo {author} {\bibfnamefont {P.}~\bibnamefont {Meystre}}\ and\ \bibinfo {author} {\bibfnamefont {M.}~\bibnamefont {Sargent}},\ }\href@noop {} {\emph {\bibinfo {title} {Elements of quantum optics}}},\ \bibinfo {edition} {4th}\ ed.\ (\bibinfo  {publisher} {{Springer}},\ \bibinfo {address} {{Berlin ; New York}},\ \bibinfo {year} {2007})\BibitemShut {NoStop}

\end{thebibliography}
\end{document}